\documentclass{msuphddissertation}
\author{Dillon Kyle Frame} %% Put your name in full as it is officially recognized by Michigan State University here.
\title{Ab Initio Simulations of Light Nuclear Systems Using Eigenvector Continuation and Auxiliary Field Monte Carlo} %% Put the title of your dissertation here.
\unit{Physics --- Doctor of Philosophy} %% Copy and paste these two items here 
\usepackage{amsmath}
\usepackage[utf8]{inputenc}
\usepackage{float}
\usepackage{graphicx}
\usepackage{lipsum}
\usepackage{wrapfig}
\usepackage[section]{placeins}
\makeatletter
\AtBeginDocument{%
  \expandafter\renewcommand\expandafter\subsection\expandafter{%
    \expandafter\@fb@secFB\subsection
  }%
}
\makeatother

\begin{document}

\maketitlepage %%This command will produce the title page of your thesis.

%% If you wish to include a "public abstract" (i.e.; in layman's terms), remove the "%" 
%% in from of the command \begin{pub abstract} and remove the "%" in front of
%% \end{pub abstract} below. A public abstract isn't required, but might be useful
%% for some readers.
\begin{pubabstract}
%%Type the text of your public abstract here. A public abstract is optional.
Since the beginning of the 20th century, the field of nuclear physics has had a lot of major developments, from Rutherford's discovery of the nucleus in 1911, to Chadwick's discovery of the neutron in 1932, to the development of quantum mechanics in to 1920's through the 1940's, and to the development of quantum chromodynamics (QCD) in the 1970's. 

While there has been a large experimental effort to measure the properties of the nuclear forces, most models that describe these properties are phenomenological. We know that QCD is the fundamental theory that describes the interactions in a nucleus, but accurately simulating even a light nucleus in this framework is beyond the capability of modern supercomputers. Clearly, a middle-ground is the approach one needs. One kind of \textit{ab initio} method used by our research group is called "chiral effective field theory" ($\chi$-EFT), where the nuclear forces are broken down into interactions between only protons, neutrons, and a force-carrying meson, the pion.

In this work, I will describe two techniques that we have used to simulate the properties of light nuclear systems. The first is called auxiliary field Monte Carlo, in which we re-write the interactions between particles as an interaction of a single particle with a fluctuating background field. The second is a new method called eigenvector continuation, in which we can compute properties in some simple or easy-to-compute configuration, then accurately extrapolate to configurations that are not able to be computed directly.
\end{pubabstract}

\begin{abstract}
%% Type your abstract here. An abstract is REQUIRED and limited to two pages.
%% The abstract must not include any figures.
In this work, we discuss a new method for calculation of extremal eigenvectors and eigenvalues in systems or regions of parameter space where direct calculation is problematic. This technique relies on the analytic continuation of the power series expansion for the eigenvector around a point in the complex plane.

We start this document by introducing the background material relevant to understand the basics of quantum mechanics and quantum field theories on the lattice, how we perform our numerical simulations, and how this relates to the nuclear physics we probe. We then move to the mathematical formalism of the eigenvector continuation. 

Here, we present the foundations of the method, which is rooted in analytic function theory and linear algebra. We then discuss how these techniques are implemented numerically (with a discussion about the computational costs), and the systematic errors associated with this method. 

Finally, we discuss applications of this method to full, quantum many-body systems. These include neutron matter, the Bose-Hubbard model, the Lipkin model, and the Coulomb interaction in light nuclei with LO chiral forces. These systems cover two categories of interest to the field: systems with a substantial sign problem, or systems that exhibit quantum phase transitions.

\end{abstract}

%% If you wish to have a copyright page, remove the "%" 
%% in front of \begin{copyrt}
%% and remove the "%" in front of \end{copyrt}.
%% An acceptable form of a copyright page  
%% will be generated automatically. 
%% TO INCLUDE A COPYRIGHT, YOU MUST REGISTER
%% IT. See the Formatting Guide for instructions. 
%\begin{copyrt}
%\end{copyrt}

%% If you wish to have a dedication, remove the "%" in front of
%% \begin{dedication} and remove the "%" in front of
%% \end{dedication} below.
%% A dedication must be single-spaced and 
%% centered on the page. Both will be done automatically. 
\begin{dedication} 
%% Type your dedication here. A dedication is optional.
\noindent To my family, for being supportive, and for encouraging me to give it my all. To my friends back home, for keeping me from going insane. To my colleagues and classmates, for helping me when books weren't enough. And to my academic advisor, Dean Lee, for always pushing me to be my best, and for the criticisms that helped me learn from my mistakes. None of this would have been possible without you all. Thank you.
\end{dedication}

%% If you wish to have an acknowledgment, remove the "%" in front of  \begin{acknowledgment}
%% and remove the "%" in front of  \end{acknowledgment} below.  
%\begin{acknowledgment}
%% Type your acknowledgment here. An acknowledgment is optional.
%\end{acknowledgment}

%% If you wish to have a preface, remove the "%" in front of \begin{preface}
%% and remove the "%" in front of \end{preface} below. The formatting of
%% a preface isn't specified, but it is included in the TOC.
%\begin{preface}
%% Type your preface here. A preface is optional.
%\end{preface}

\TOC %% This command produces the Table of Contents. DO NOT REMOVE IT!

%% If your document contains tables, remove the "%" in front of 
%%  the following line.
\LOT

%% If your document contains figures, remove the "%" in front of
%% the following line.
\LOF

%%%% LIST OF SYMBOLS OR LIST OF ABBREVIATIONS %%%%
%% If you wish to have a list of symbols or a list of abbreviations, 
%% it should be here. For a list of symbols remove the "%" in front of 
%% \begin{symbols} and remove the "%" in front of \end{symbols} below.
%\begin{symbols}
%% Type your list using a list environment here.
%\end{symbols}
%% Similarly for a list of abbreviations remove the "%" in front of 
%% \begin{abbrev} and remove the "%" in front of \end{abbrev} below.
%\begin{abbrev}
%% Type your list using a list environment here.
%\end{abbrev}
%% The list will be included in the TOC as
%% KEY TO SYMBOLS or KEY TO ABBREVIATIONS
%%%%%%%%%%

\newpage
\pagenumbering{arabic}
\begin{doublespace}

%% Put the body of your dissertation here. 
%% DO NOT include the bibliography or any appendices.
%% These topics will be discussed later.

%%%%%%%%%%%%%%%%%%%%%%%%%%%%%%%%%%%%%%%%%%%%%%%%%%%%%%%%%%%%%%%%%%
%%%%%%%%%%%%%%%%%%%%%%%%%%%%%%%%%%%%%%%%%%%%%%%%%%%%%%%%%%%%%%%%%%
%%%%%%%%%%%%%%%%%%%%%%%%%%%%%%%%%%%%%%%%%%%%%%%%%%%%%%%%%%%%%%%%%%
\chapter{Introduction}

Eigenvector continuation \cite{Frame:2018ecs} is a new tool that allows calculations of the properties of quantum many-body systems that are otherwise inaccessible; either through the Monte Carlo sign problem, or by sheer cost in computation. This technique relies on the analytic continuation of the power series expansion for the eigenvector around a point in the complex plane. We assume we have a Hamiltonian that can be written as
\begin{equation}
H = H_0 + c H_1
\end{equation}
\noindent where $H_0$ is some baseline, and $H_1$ is the interacting term, with tunable, complex coupling $c$ that we wish to compute matrix elements of. In describing the mathematics and numerical techniques associated with this method, we will be assuming that $H$ is a Hermitian matrix, and that its eigenvalues and eigenvectors can be accurately calculated for at least a few values of $c$.

This work is organized into three main sections. In the first chapter, we begin by introducing the background material relevant to understand the basics of quantum mechanics and quantum field theories on the lattice, how we perform our numerical simulations, and how this can be tied back to the nuclear physics we are interested in probing. The most fundamental QFT, quantum chromodynamics, is the most reasonable place to start this discussion. The QCD Lagrangian \cite{Tanabashi:2018rpp} is given by
\begin{equation}
\mathcal{L} = \sum_q \bar{\psi}_{q,a} \left( i \gamma^\mu \partial_\mu \delta_{ab} - g_s \gamma^\mu t_{ab}^C A_\mu^C - m_q \delta_{ab} \right) \psi_{q,b} - \frac{1}{4}F_{\mu\nu}^A F^{A \mu\nu}
\end{equation}
\noindent and it describes an SU(3) color gauge field interacting with quarks and gluons; the building blocks of nucleons and other fundamental particles. We then move down in energy scale to something more suited for nuclear physics; chiral EFT. Chiral EFT \cite{Machleidt:2011cef}\cite{Scherer:2003icp} is an effective field theory built around the chiral symmetry of QCD, which is a symmetry that arises from the limit of zero quark masses. The mass of the up and down quarks are quite small, so this limit can be described well by an effective field theory. To do simulations using chiral EFT, one first needs to decide how many terms in the expansion to keep, and then, one needs to find a way to constrain the low-energy-constants, or LECs, that each term is responsible for. We do this by fitting low-energy n-n, n-p, and p-p scattering data, as well as ground state properties of the deuteron \cite{Li:2018nps}\cite{Lu:2018een}\cite{Lu:2015riz}. All of this input can provide sufficient constraints to use this framework for heavier nuclear systems.

In Chapter 2, we move to the primary topic of this work; the presentation of the mathematical formalism of eigenvector continuation. Here, we present the foundations of the method, which is rooted in analytic function theory and linear algebra. 
We take as our starting point the perturbative expansion of the eigenvectors of $H$, denoted $|\psi_j(c)\rangle$
\begin{equation}
|\psi_j(c)\rangle = \sum_{n=0}^{\infty} |\psi^{(n)}_j(0)\rangle c^n /n!
\end{equation}
\noindent were $c$ is assumed to be in a region in which it converges. Suppose that we are interested in a value of $c$ outside of this region. Using just perturbation theory, we would be out of luck. However, suppose that we construct a new series about a point $|w| < |c|$. Then, we could write
\begin{align}
|\psi_j(c)\rangle &= \sum_{n=0}^{\infty} |\psi^{(n)}_j(w)\rangle \left(c-w\right)^n /n! \\
|\psi^{(n)}_j(w)\rangle &= \sum_{m=0}^{\infty} |\psi^{(n+m)}_j(0)\rangle w^m /m!
\end{align}
\noindent Combining these, we can obtain
\begin{equation}
|\psi_j(c)\rangle = \sum_{n=0}^{\infty} \sum_{m=0}^{\infty} \left(c-w\right)^n w^m |\psi^{(n+m)}_j(0)\rangle / m! n!
\end{equation}

\noindent This multi-series expansion is the heart of the eigenvector continuation method. In a sense, we now have an expression for $|\psi_j(c)\rangle$ with a wider area of convergence. Now, from a numerical standpoint, the analogue of derivatives of the eigenvectors at $c=0$ are the finite differences, i.e. computing the eigenvectors at multiple small values for the couplings. 

Once we obtain these eigenvectors, we can form them into a projection operator $P$ and project the Hamiltonian $H$ at some target coupling $C_\odot$ into this lower-dimensional subspace. We call the projected Hamiltonian $M = P^\dagger H P$, and the norm matrix $N = P^\dagger P$. Then, we solve the generalized eigenvalue problem
\begin{equation}
M \vec{v} = \lambda_i N \vec{v}
\end{equation}
\noindent The lowest eigenvalue of this system is called the EC estimate for the ground state energy, and the ground state eigenvector in the full space can be reconstructed from the projection operator
\begin{equation}
|\psi_0(C_\odot)\rangle = P \vec{v}_0
\end{equation}

\noindent From here, we briefly discuss how these techniques are implemented numerically, with a discussion about the computational costs.

In Chapter 3, we discuss applications of this method to full, quantum many-body systems. The first of these is the Bose-Hubbard model, which consists of a system of N bosons interacting on a 3D lattice. The Hamiltonian of this system is given by 
\begin{equation}
H = -t\sum_{\langle \textbf{n},\textbf{n'} \rangle} a^\dagger (\textbf{n'}) a(\textbf{n}) + \frac{U}{2}\sum_{\textbf{n}} \rho(\textbf{n}) \left[ \rho(\textbf{n}) - \textbf{1} \right] - \mu \sum_{\textbf{n}} \rho(\textbf{n})
\end{equation}
\noindent where $a^\dagger (\textbf{n})$ and $a(\textbf{n})$ are the creation and annihilation operators, and $\rho(\textbf{n})$ are the density operators. This system exhibits a phase transition from a weakly-bound Bose gas to a strongly-bound cluster around the point $U/t = -3.8$. This makes perturbative methods difficult, and a good test for the eigenvector continuation. We also looked at neutron matter. This difficulty in direct calculations here is that for large projection times, the Monte Carlo sign problem begins to dominate, so reasonable extraction of the ground state energy is not possible. 

The Lipkin model is another such model were perturbation theory has difficulty. This model considers a N-fold degenerate two-level system, with N spin-1/2 fermions. The Hamiltonian consists of the angular momentum operators $J_z$,$J_+$, and $J_-$.
\begin{equation}
H= \epsilon J_z + \frac{V}{2} ( J_{+}^2 + J_{-}^2 ) + \frac{W}{2} ( J_+ J_- + J_- J_+ )
\end{equation}
\noindent The states of the system $| J, m \rangle$ are described by their total angular momentum $J$ and the projection of that momentum onto the z-axis $m$. Since no term in the Hamiltonian mixes different J states, we can focus on a single part of $H$ that has the maximum angular momentum $J = J_\text{max}$. For small $NV/\epsilon$, the energy levels of the system are ordered by the $m$ of that state. For larger $NV/\epsilon$, the energy levels pair up, giving an even and odd parity state with the same energy. 

Finally, we discuss adding the Coulomb interaction into our group's lattice EFT calculations. This normally is plagued by the sign problem, due to the repulsive nature of the Coulomb interaction, so this is somewhere that eigenvector continuation would help. These simulations are done at LO in Chiral EFT, and the Coulomb interaction is added in non-perturbatively using the Hubbard-Stratonivich transformation
\begin{equation}
\exp \left[ -\frac{C\alpha_t}{2} (a^\dagger(n) a(n))^2 \right] = \sqrt{\frac{1}{2\pi}}\int_{-\infty}^{\infty} ds \exp \left[-\frac{1}{2} s(n)^2 + \sqrt{-C \alpha_t} s(n) a^\dagger(n) a(n) \right]
\end{equation}
\noindent which allows us to write a two-body interaction in terms of a one-body density interacting with a fluctuating background field. Using this ``Coulomb" field, we can simulate the Coulomb interaction for multiple values of the fine-structure constant $\alpha_em$. Then, the EC method can give us estimates for the ground state energy at the physical value for $\alpha_em$.

In Chapter 4, we briefly summarize our results, and highlight a few plans that we have moving forward. The eigenvector continuation technique is proving to be an invaluable tool for quantum many-body simulations, and we would like to investigate applications to many other systems and problems.

%%%%%%%%%%%%%%%%%%%%%%%%%%%%%%%%%%%%%%%%%%%%%%%%%%%%%%%%%%%%%%%%%%
%%%%%%%%%%%%%%%%%%%%%%%%%%%%%%%%%%%%%%%%%%%%%%%%%%%%%%%%%%%%%%%%%%
%%%%%%%%%%%%%%%%%%%%%%%%%%%%%%%%%%%%%%%%%%%%%%%%%%%%%%%%%%%%%%%%%%
\chapter{Background Material}

The purpose of this chapter is to provide the information that is essential to the work presented in this document. Experts in this area should skip ahead to the next chapter for the new content described in this work.

First, an overview of the essential parts of quantum mechanics are presented. This covers the discussion of the Schr\"odinger equation and how wavefuctions are described in terms of bras and kets. Then, we jump into a review of quantum field theory, focusing on the the lattice methods and effective fields theories used in this work.

Afterwards, we will begin laying down the foundations of lattice field theory. These will describe a general framework for discretizing the path integrals present in QFT into a form that can be easily expressed on a computer.

The Monte Carlo methods used by our research group will then be covered in detail, with accompanying pseduocode when necessary. We use three primary Monte Carlo methods: Euclidean time projection Monte Carlo, which involves computing powers of a transfer matrix, auxiliary field Monte Carlo, which allows us to re-write the transfer matrix in terms of an integral over background fields, and hybrid Monte Carlo, which provides a sophisticated method for updating these auxiliary fields.

%%%%%%%%%%%%%%%%%%%%%%%%%%%%%%%%%%%%%%%%%%%%%%%%%%%%%%%%%%%%%%%%%%
\section{Elements of Quantum Mechanics}

When compared to other fields of physics, like classical mechanics or electromagnetism, the field of quantum mechanics is fairly young, having been developed around the 1920s. 

The principles of quantum mechanics can be summarized by two key observations. The first is that on the smallest scales, charge, light, and energy are all quantized; that is, they can only come in discrete packets, called ``quanta". This fact was observed during early experiments like the Millikan oil drop experiment, J.J. Thompson's e/m experiment, and measurements of the photoelectric effect.

The second observation was the wave-particle duality exhibited by fundamental particles, like the electron. This was observed by de Broglie in the double slit experiment, showing that electrons could constructively and destructively interfere with each other.

We will now begin summarizing the essential information from this field. This will not even remotely be a complete treatment, so for further reference, see \cite{Peebles:1992qm}, \cite{Sakurai:2011mqm}, or \cite{Tipler:2002nrc}.

\subsection{The Schr\"odinger Equation}

Since it was determined that quantum particles acted like waves, it was a natural assumption that the equation of motion for these particles was a wave equation. This is what Erwin Schr\"odinger postulated in 1926.

For light, the wave equation in one dimension is given by
\begin{equation}
\frac{d\mathcal{E}^2}{dx^2} = \frac{1}{c^2}\frac{d\mathcal{E}}{dt^2}
\end{equation}
\noindent with the electric field denoted $\mathcal{E}(x,t)$, and with $c$ being the speed of light. This equation has the well-known solution
\begin{equation}
\mathcal{E}(x,t) = \mathcal{E}_0 \cos (kx -\omega t)
\end{equation}
with $k = p/\hbar$ and $\omega = E/\hbar$, where $p$ is the momentum of the photon, and $E$ is its energy.

Now, we wish to look at a particle with mass $m$, and some interaction potential $V$. From the de Broglie relations for an electron, we have
\begin{equation}
\hbar \omega = \frac{\hbar^2 k^2}{2m} + V
\end{equation}
which suggests the equations of motion will only have one time derivative for the $\omega$, two space derivatives for the $k^2$ term, and no derivatives for the $V$ term. 

The Schr\"odinger equation can then be written as
\begin{equation}
-\frac{\hbar^2}{2m}\frac{\partial^2 \Psi(x,t)}{\partial^2x^2} + V(x,t) \Psi(x,t) = i\hbar\frac{\partial \Psi(x,t)}{\partial t}
\end{equation}
\noindent The presence of the $i$ in this equation suggests that the solutions to this equation will in general be complex-valued functions. These solutions, $\Psi(x,t)$ are referred to as wavefunctions.

It is often easier to factor $\Psi(x,t)$ into two pieces, one that is dependent on $x$ and one that is dependent on $t$.
\begin{equation}
\Psi(x,t) = \psi(x) \phi(t)
\end{equation}
\noindent Substituting this into the Schr\"odinger equation, we get
\begin{equation}
-\frac{\hbar^2}{2m} \phi(t) \frac{d^2 \psi(x)}{dx^2} + V(x)\psi(x)\phi(t) = i\hbar\psi(x)\frac{d\phi(t)}{dt}
\end{equation}
\noindent and dividing by $\psi(x) \phi(t)$ on both sides, we get
\begin{equation}
-\frac{\hbar^2}{2m} \frac{1}{\psi(x)} \frac{d^2 \psi(x)}{dx^2} + V(x) = i\hbar \frac{1}{\phi(t)} \frac{d\phi(t)}{dt}
\end{equation}
\noindent We notice that each side is a function of only a single variable, so we can treat each side as a separate differential equation, with a common ``separation constant" $E$, which is the total energy of the particle. We then get two equations
\begin{align}
\left[ -\frac{\hbar^2}{2m} \frac{d^2}{dx^2} + V(x) \right] \psi(x) &= E \psi(x) \\
\left[ i\hbar \frac{d}{dt} \right] \phi(t) = E \phi(t)
\end{align} 

The first equation is referred to as the time-independent Schr\"odinger equation, often simplified to $H\psi = E\psi$, with $H$ being the Hamiltonian of the system. The second equation will then govern the time evolution of the wavefunction. Just looking at the time-dependent piece, we see that it is only a 1-D ODE, so it can be solved easily, resulting in
\begin{equation}
\phi(t) = e^{-iEt/\hbar}
\end{equation}
\noindent This is referred to as the time-evolution operator.

\subsection{Operators, States \& Measurements}
Now the we have built up a description of the dynamics of a quantum particle in terms of wavefunctions, it is time to step back and look at how we use these wavefunctions to calculate properties of the particle.

First, we note that the states of the quantum particle live in a complex vector space. The size and dimension of this space depends on the system being studied; i.e. for a 1-D particle in a harmonic oscillator potential can be labelled by $n$, noting which energy level the particle is in, and by its $x$-position.

To describe these vector states, we will use bra-ket notation, developed by Dirac, in which the state labels are in a ket $|\alpha \rangle$ and the complex conjugate is in a bra $\langle \beta |$. The bras and kets are vectors in the full complex vector space, and the labels $\alpha$ and $\beta$ are the quantum numbers needed to completely describe a quantum state.

Normal operations of linear algebra can be used when working with these objects: the inner product of a bra and ket
$\langle \beta | \alpha \rangle$ will be a complex number, the norm of a vector is given by $\langle \alpha| \alpha \rangle$, and the outer product is given by $| \alpha \rangle\langle \beta |$. In the language of linear algebra, the kets are column vectors, and the bras are row vectors.

If the label on the state is continuous, like say the position $x$, we still label the states in terms of bras and kets. In this case, the vector space has an infinite number of dimensions. We call such a space a Hilbert space.

In general, we assume the different basis states of a given system are orthonormal, meaning they are orthogonal to each other
\begin{equation}
\langle \alpha | \beta \rangle = \delta_{\alpha\beta}
\end{equation}
\noindent and normalized
\begin{equation}
\langle \alpha | \alpha \rangle  = 1
\end{equation}
\noindent Since we are working in a linear space, any arbitrary quantum state of a system can be expressed as a linear combination of basis states.
\begin{equation}
| \psi \rangle = \sum_{n} c_n | n \rangle
\end{equation}
\noindent This set of basis states $|n\rangle$ is called a complete set if
\begin{equation}
\sum_{n} | n \rangle \langle n | = \textbf{1}
\end{equation}
\noindent The advantage of using this machinery is that we can view operators as matrices that multiply a ket and give us a new ket. 
\begin{equation}
\hat{A} | \alpha \rangle = a_i | \alpha \rangle
\end{equation}
\noindent That is, the whole of quantum mechanics can be turned into solutions of eigenvalue problems.

If we want to perform a measurement, and take the expectation value of some operator $\hat{A}$ on some state $|\alpha\rangle$, then we compute the quantity
\begin{equation}
\langle \beta | \hat{A} | \alpha \rangle
\end{equation}
\noindent These quantities are usually referred to as ``matrix elements", for the reason that for $\alpha,\beta = 0,\cdots,N$, there will be $N \times N$ results of this expectation value.

In our simulations, we consider states with 3 spatial labels, $n_x$, $n_y$, $n_z$, one temporal label, $n_t$, a label for spin and isospin, $S$, $I$, and a label for particle number $n_f$. Then, our kets will be $|n_x, n_y, n_z, n_t, S, I, n_f>$. This is bulky to carry around, so often just gets simplified to $|\Psi\rangle$.

%%%%%%%%%%%%%%%%%%%%%%%%%%%%%%%%%%%%%%%%%%%%%%%%%%%%%%%%%%%%%%%%%%
\section{Lattice Field Theory}

\subsection{Path Integral Formulation}

One of the fundamental problems in quantum mechanics is the evaluation of how a quantum state evolves in time. In the Schr\"{o}dinger picture, any quantum state can be written in terms of energy eigenstates of the Hamiltonian of the system, i.e. solutions of the Schr\"{o}dinger equation.
\begin{equation}
\hat{H} |n\rangle = E_n |n \rangle
\end{equation}
\begin{equation}
| \alpha, t \rangle = e^{-i\hat{H}(t-t_0)} | \alpha, t=t_0 \rangle = \sum_{n} e^{-i E_n (t-t_0)} |n\rangle \langle n | \alpha, t=t_0 \rangle
\end{equation}

\noindent We set $t_0 = 0$, $\hbar = c = 1$. The wavefunction is this state $| \alpha, t >$ projected into position eigenstates.
\begin{align}
\psi_\alpha \left( x, t\right) = \langle x | \alpha, t \rangle &= \sum_{n} e^{-i E_n t} \langle x |n\rangle \langle n | \alpha, 0 \rangle \\ & \equiv \sum_{n} e^{-i E_n t} u_n(x) c_n
\end{align}

\noindent where $u_n(x) = \langle x | n \rangle$ and
\begin{align}
c_n &= \langle n | \alpha, 0 \rangle \\ &= \int d^3 x \langle x | n \rangle \langle x | \alpha, 0 \rangle \\ &=  \int d^3 x u^*_n(x) \psi_\alpha \left( x , 0\right)
\end{align}

In quantum field theory, systems are described in terms of fields, $\phi\left(\vec{x},t\right)$, which are functions of the space-time coordinates. The most fundamental object that describes a system is the path integral, which is defined as
\begin{equation}
\mathcal{Z} = \int D\phi \exp \left\lbrace i \int_{0}^{t_f} dt \text{ }\mathcal{S}\left(\phi,\partial_\mu \phi \right) \right\rbrace
\end{equation}  

\noindent with $D\phi$ defined as
\begin{equation}
D\phi = \prod_{i=0}^{\infty} d\phi_i\left(\vec{x},t\right)
\end{equation}

\noindent and $\mathcal{S}$ is the action associated with the system
\begin{equation}
\mathcal{S}\left(\phi,\partial_\mu \phi \right)  = \int d^3 x \text{ }\mathcal{L}\left(\phi,\partial_\mu \phi \right)
\end{equation}

\noindent and $\mathcal{L}$ is the Lagrangian density (often just called the Lagrangian.) Physically, this path integral represents a sum over all possible paths that a particle (field) can take, travelling from one point in time to another, with each path having an associated weight. 

It is often easier to work in imaginary time, $\tau = it$, also called Euclidean time. The reason behind this is that we would like to interpret the weights in Eqn. (2.24) as probabilities, instead of complex phases. With this substitution, our path integral becomes
\begin{equation}
\mathcal{Z} = \int D\phi \exp \left\lbrace - \int_{0}^{\tau_f} d\tau \int d^3 x \text{ }\mathcal{L}\left(\phi,\partial_\mu \phi \right) \right\rbrace
\end{equation}

Lattice field theory refers to a specific formulation of this path integral, developed by Wilson \cite{Wilson:1974coq}, in which space-time is discretized on a N+1 dimensional lattice. This is done by replacing the integrals by sums and by having the fields $\phi_\mu \left(\vec{x},t\right)$ be defined at a discrete set of points $\phi_\mu \left(a\vec{n},a_t n_t\right)$, where $a$ is the spacing between lattice sites in the N spatial dimensions, and $a_t$ is the spacing between lattice sites in the time dimension.

The advantage of this method is that it reduces the infinite dimensional path integral into a finite dimensional, ordinary integral. 
\begin{equation}
Z = \prod_{i=0}^{N} \left[ \int d\phi_i \right] \exp \left\lbrace - \frac{1}{L^3 L_t} \sum_{n_t} \sum_{\vec{n}} \text{ }\mathcal{L}\left(\phi_i\left(\vec{n},n_t\right), \partial_\mu \phi_i\left(\vec{n},n_t\right) \right) a^3 a_t \right\rbrace
\end{equation}

\noindent Thus, using suitable integration methods, this integral can be evaluated, and the dynamics of the system can be studied.

A free field theory is defined by any involved fields having no interactions; that is, the lattice action consists of only a kinetic energy term. On the lattice, this corresponds to a finite difference operation, or a hopping term that allows particles to travel from one site on the lattice to another. Considering only one spatial dimension, this free lattice action is defined as:
\begin{equation}
\mathcal{S} = a^4 \sum_{i = 0}^{N} \sum_{x} m \dot{\phi}^2_i(x) = a^4 \sum_{i = 0}^{N} \frac{1}{4a} \sum_{x} m \left(\phi_i(x) - \phi_i(x-a) \right)^2
\end{equation}
\noindent An interacting field theory adds potential energy terms between the fields $V(|x-x'|)$, which depends on the distance between lattice sites. In this case, the action has the form
\begin{equation}
\mathcal{S} = a^4 \sum_{i = 0}^{N} \left[ \sum_{x} \frac{1}{4a} m \left(\phi_i(x) - \phi_i(x-a)\right)^2 + \frac{a}{2} \sum_{x,x'} \phi_i(x) V(x - x') \phi_i(x') \right]
\end{equation}

\noindent If we take the interaction potential as a delta function $V(x-x') = C\delta(x-x')$, then this reduces to

\begin{equation}
\mathcal{S} = a^4 \sum_{i = 0}^{N} \sum_{x} \left[  \frac{1}{4a} m \left(\phi_i(x) - \phi_i(x-a)\right)^2 + \frac{C a}{2}\phi^2_i(x) \right]
\end{equation}

\subsection{Grassmann Path Integral}

In the previous section, we described the path integral for a scalar field. However, in nuclear systems, we are concerned with anti-commuting fermionic fields. These are described using a Grassmann algebra, in which its elements $\eta_1$, $\eta_2$, ..., anti-commute
\begin{equation}
\left\lbrace \eta_1, \eta_1\right\rbrace = 0
\end{equation}
\noindent These variables have a few interesting properties:

\begin{itemize}
\item \textbf{Linearity}
\begin{equation}
\int \left[ a f(\eta) + b g (\eta) \right] d\eta = a \int f(\eta) d\eta + b \int g(\eta) d\eta
\end{equation}

\item \textbf{Integration/Differentiation}
\begin{align}
& \int \left[ \frac{\partial}{\partial \eta} f(\eta) \right] d\eta = 0 \nonumber \\
&\int 1 d\eta = 0 \\
&\int \eta d\eta = 1 \nonumber
\end{align} 

\end{itemize}

\noindent Let $c_i$ and $c^*_i$ be anti-commuting Grassmann fields for spin $i$. The Grassmann fields are periodic with respect to the spatial lengths of the $L^3$ lattice,
\begin{equation}
c_i(\vec{n},n_t) = c_i(\vec{n}+L\hat{e_1},n_t) = c_i(\vec{n}+L\hat{e_2},n_t) = c_i(\vec{n}+L\hat{e_3},n_t)
\end{equation}

\noindent and anti-periodic along the temporal direction,
\begin{equation}
c_i(\vec{n},n_t) = -c_i(\vec{n},n_t + L_t)
\end{equation}

\noindent The Grassmann path integral is given by
\begin{equation}
\mathcal{Z} = \int Dc Dc^* \exp \left[ -\mathcal{S}(c,c^*) \right]
\end{equation}
\begin{equation}
\mathcal{S}(c,c^*) = \mathcal{S}_\text{free} (c,c^*) + C\alpha_t \sum_{\vec{n},n_t} \rho_{\uparrow}(\vec{n},n_t) \rho_{\downarrow}(\vec{n},n_t)
\end{equation}

\noindent The action $\mathcal{S}(c,c^*)$ consists of the free fermion action
\begin{equation}
\begin{split}
\mathcal{S}_\text{free} (c,c^*) &= \sum_{\vec{n},n_t,i=\uparrow,\downarrow} \left[ c_i^* (\vec{n},n_t) c_i(\vec{n},n_t + 1) - (1-6h)  c_i^* (\vec{n},n_t) c_i(\vec{n},n_t) \right] \\ 
& -h \sum_{\vec{n},n_t,i=\uparrow,\downarrow} \sum_{l = 1,2,3} \left[ c_i^* (\vec{n} + \hat{e}_l,n_t) c_i(\vec{n},n_t) + c_i^* (\vec{n},n_t) c_i(\vec{n}-\hat{e}_l,n_t) \right]
\end{split}
\end{equation}

\noindent and an attractive interaction between up and down spins. The Grassmann spin densities $\rho_\uparrow$ and $\rho_\downarrow$ are defined as
\begin{equation}
\rho_\uparrow (\vec{n},n_t) = c_\uparrow^* (\vec{n},n_t) c_\uparrow(\vec{n},n_t)
\end{equation}
\begin{equation}
\rho_\downarrow (\vec{n},n_t) = c_\uparrow^* (\vec{n},n_t) c_\downarrow(\vec{n},n_t)
\end{equation}

Now, we wish to extend the definition of the Grassmann path integral to add in a real-valued, auxiliary field $s(\vec{n},n_t)$. We do this by defining the Grassmann actions
\begin{equation}
\mathcal{S}_j (c,c^*,s) = \mathcal{S}_{\text{free}}(c,c^*) - \sum_{\vec{n},n_t} A_j\left[s(\vec{n},n_t)\right]\cdot\left[ \rho_{\uparrow}(\vec{n},n_t) + \rho_{\downarrow}(\vec{n},n_t) \right]
\end{equation}

\noindent and Grassmann path integrals,
\begin{equation}
\mathcal{Z}_j = \prod_{\vec{n},n_t}\left[\int d_j s(\vec{n},n_t) \right] \int Dc Dc^* \exp\left[-\mathcal{S}_j (c,c^*,s) \right]
\end{equation}

\noindent Here, $A_j$ is a particular functional form involving the auxiliary field. The first possibility is a Gaussian-integral transformation similar to the original Hubbard-Stratonovich transformation, denoted by $j=1$.
\begin{equation}
\int d_1 s(\vec{n},n_t) \equiv \frac{1}{\sqrt{2\pi}} \int_{-\infty}^{+\infty} ds(\vec{n},n_t) e^{-\frac{1}{2}s^2 (\vec{n},n_t)}
\end{equation}
\begin{equation}
A_1 \left[s(\vec{n},n_t)\right] = \sqrt{-C_1 \alpha_t}s(\vec{n},n_t)
\end{equation}

\noindent Another possibility is a bounded, continuous auxiliary field transformation, denoted by $j=4$.
\begin{equation}
\int d_4 s(\vec{n},n_t) \equiv \frac{1}{2\pi} \int_{-\pi}^{+\pi} ds(\vec{n},n_t)
\end{equation}
\begin{equation}
A_4 \left[s(\vec{n},n_t)\right] = \sqrt{-C_4 \alpha_t}\sin\left(s(\vec{n},n_t)\right)
\end{equation}

\noindent Other transformations are described in \cite{Lee:2008gse}, but only the two transformations detailed here are used in our simulations.

\subsection{Transfer Matrix Formalism}

Relating the Grassmann path integral to the transfer matrix, we have
\begin{equation}
\mathcal{Z} = \text{Tr} \left( M^{L_t} \right)
\end{equation}
\noindent That is, the path integral can be re-written as the trace of a product of transfer matrix operators. Let us define the free lattice Hamiltonian as
\begin{equation}
H_{\text{free}} = \frac{3}{m} \sum_{\vec{n},i=\uparrow,\downarrow} a_i^{\dagger}(\vec{n}) a_i(\vec{n}) - \frac{1}{2m} \sum_{\vec{n},i=\uparrow,\downarrow} \sum_{l=1,2,3} \left[a_i^{\dagger}(\vec{n}) a_i(\vec{n} + \hat{e}_l) + a_i^{\dagger}(\vec{n}) a_i(\vec{n} - \hat{e}_l) \right]
\end{equation}
\noindent and the density operators as
\begin{equation}
\rho_{\uparrow} (\vec{n}) = a_\uparrow^\dagger (\vec{n}) a_\uparrow(\vec{n})
\end{equation}
\begin{equation}
\rho_{\downarrow} (\vec{n}) = a_\downarrow^\dagger (\vec{n}) a_\downarrow(\vec{n})
\end{equation}
\noindent Then, the normal ordered transfer matrix operator is given by
\begin{equation}
M = : \exp \left[ -H_{\text{free}}\alpha_t - C\alpha_t \sum_{\vec{n}} \rho_{\uparrow} (\vec{n}) \rho_{\downarrow} (\vec{n}) \right] :
\end{equation}

Like before, we wish to express the Grassmann path integral as a trace of products of transfer matrix operators. Now, the transfer matrix depends on the auxiliary fields,
\begin{equation}
\mathcal{Z} = \prod_{\vec{n},n_t}\left[\int d_j s(\vec{n},n_t) \right] \text{Tr} \left[ M_j (s,Lt-1) \cdots M_j (s,0) \right]
\end{equation}
\noindent where
\begin{equation}
M = : \exp \left\lbrace -H_{\text{free}}\alpha_t + \sum_{\vec{n}} A_j \left[ s(\vec{n},n_t) \right] \cdot \left[ \rho_{\uparrow}(\vec{n},n_t) + \rho_{\downarrow}(\vec{n},n_t) \right] \right\rbrace :
\end{equation}
\noindent Note that in this transformation, the interaction part of the lattice  Hamiltonian in Eqn. (2.52) involved two-body operators, but in Eq. (2.54), only one-body operators are involved.

Now the the groundwork is laid for generating a baseline interaction using an auxiliary field, we discuss a method for adding in extra interactions with a known potential or propagator. We add in a SU(2N)-invariant potential $V(\vec{n}-\vec{n}')$ to the lattice Hamiltonian:
\begin{equation}
H_{\text{new}} = H_{\text{base}}  + \frac{1}{2}\sum_{\vec{n},\vec{n}'}\sum_{i=\uparrow,\downarrow} : \rho_i(\vec{n}) V(\vec{n} -\vec{n}') \rho_i(\vec{n}'):
\end{equation}
\noindent where
\begin{equation}
H_{\text{base}} = H_{\text{free}} + C \sum_{\vec{n}} : \rho_{\uparrow}(\vec{n}) \rho_{\downarrow} (\vec{n}) :
\end{equation}
\noindent as described earlier in this section. We can then define the normal ordered transfer matrix $M$ as:
\begin{equation}
M = : \exp \left[ -H_{\text{free}}\alpha_t - C\alpha_t \sum_{\vec{n}} \rho_{\uparrow} (\vec{n}) \rho_{\downarrow} (\vec{n}) - \frac{\alpha_t}{2}\sum_{\vec{n},\vec{n}'}\sum_{i=\uparrow,\downarrow}  \rho_i(\vec{n}) V(\vec{n} -\vec{n}') \rho_i(\vec{n}') \right] :
\end{equation}
\noindent We will denote the Fourier Transforms of quantities (or their momentum space versions) in this section with a tilde. Thus, the Fourier Transform of the potential $V(\vec{n}-\vec{n}')$ is given by
\begin{equation}
\tilde{V}(2\pi\vec{k}_s / L) = \sum_{\vec{n}} V(\vec{n}) e^{2\pi i \vec{n}\cdot\vec{k}/L}
\end{equation}
\noindent We assume that $\tilde{V}(2\pi\vec{k}_s / L)$ is negative definite. The inverse of the potential, $V^{-1}(\vec{n})$ is
\begin{equation}
V^{-1}(\vec{n}) = \frac{1}{L^d} \sum_{\vec{k}} \frac{e^{-2\pi i \vec{n}\cdot\vec{k} / L}}{\tilde{V}(2\pi\vec{k}_s / L)}
\end{equation}
\noindent Now, we can add in an auxiliary field $s_1$, resulting in the lattice action
\begin{equation}
\mathcal{S}_{\text{new}} = \mathcal{S}_{\text{base}} - \frac{\alpha_t}{2} \sum_{n_t} \sum_{\vec{n},\vec{n}'} s_1(\vec{n},n_t) V^{-1}(\vec{n} - \vec{n}') s_1(\vec{n}',n_t)
\end{equation}
\noindent where $\mathcal{S}_{\text{base}}$ is given in Eqn. (2.42). The modified transfer matrix, with both auxiliary fields, is then given by
\begin{equation}
M(s,s_1,t) = : \exp \left\lbrace -H_{\text{free}}\alpha_t + \sum_{\vec{n}} A_j \left[ s \right] \cdot \left[ \rho_{\uparrow} + \rho_{\downarrow} \right] + \sum_{\vec{n}} A_{j'} \left[ s_1\right] \cdot \left[ \rho_{\uparrow} + \rho_{\downarrow} \right]  \right\rbrace :
\end{equation}
\noindent where the space and time indices have been suppressed to save line space. For our simulations, we use $j = 1$ and $j' = 4$. These $A_j \left[ s(\vec{n},n_t) \right]$ functionals are the same as described in Section 2.2.2. Substituting in the appropriate $A_j[s]$, our transfer matrix has the form:
\begin{align}
M(s,s_1,t) = : \exp \lbrace & -H_{\text{free}}\alpha_t + \sqrt{-C \alpha_t} \sum_{\vec{n}} \sin s(\vec{n},n_t) \left[ \rho_{\uparrow} + \rho_{\downarrow} \right] + \\ & \sqrt{-C_1 \alpha_t} \sum_{\vec{n}}  s_1(\vec{n},n_t) \cdot \left[ \rho_{\uparrow} + \rho_{\downarrow} \right]  \rbrace :
\end{align}

\subsection{Calculating Observables}

Mentioned earlier, we wish to take as our observable the powers of the transfer matrix. This Euclidean time projection amplitude $Z_{N,N}(t)$ is defined as

\begin{equation}
Z_{N,N}(t) \equiv \prod_{\vec{n},n_t} \left[ \int d_j s(\vec{n},n_t) \right] \langle \Psi^{0,\text{free}}_{N,N} | M_j (s,t) \cdots M_j (s,0) | \Psi^{0,\text{free}}_{N,N} \rangle
\end{equation}
\noindent As a result of the normal ordering, the transfer matrix only consists of single-particle operators acting with the background auxiliary field. There are no direct interactions between particles. Thus,
\begin{equation}
\langle \Psi^{0,\text{free}}_{N,N} | M_j (s,t) \cdots M_j (s,0) | \Psi^{0,\text{free}}_{N,N} \rangle = \left[ \det \textbf{M}_j (s,t) \right]^2
\end{equation}
\noindent where
\begin{equation}
\left[ \textbf{M}_j (s,t) \right]_{k,k'} = \langle \vec{p}_k' | M_j (s,t) \cdots M_j (s,0) | \vec{p}_k \rangle
\end{equation}
\noindent for matrix indices $k,k' = 1,\cdots,N$. $| \vec{p}_k \rangle$ and $| \vec{p}_k' \rangle$ re the single-particle momentum states comprising the initial Slater determinant state. The single particle interactions in the transfer matrix are the same for up and down spin particles, which results in the determinant being squared in Eqn. (2.65). This matrix is real-valued, so the square of the determinant is non-negative and there is no sign problem.

To extract the ground state energy, we compute $Z_{N,N}(t)$ and $Z_{N,N}(t - \alpha_t)$, and take the ratio:
\begin{equation}
E_{N,N}(t) = \frac{\Lambda}{\alpha_t}\log \left( \frac{Z_{N,N}(t - \alpha_t)}{Z_{N,N}(t)} \right)
\end{equation}
\noindent For large $t$, this quantity converges to the ground state energy:
\begin{equation}
E_{N,N}^{0} = \lim_{t \to \infty} E_{N,N}(t)
\end{equation}

%%%%%%%%%%%%%%%%%%%%%%%%%%%%%%%%%%%%%%%%%%%%%%%%%%%%%%%%%%%%%%%%%%
\section{Nuclear Forces from Effective Field Theory}

Nuclear physics is  field that spans multiple scales; from the level of quarks and gluons, to the level of neutrons and protons, to the scale of collectivity in large nuclei. This hierarchy is summarized in Fig. [2.1]. This hierarchy can be broken into energy scales, which represent the ``resolution" of the associated model. At each level, the properties being investigated don't tend to depend on the higher energy scales; if you care about the rotational bands in Lead, it doesn't really matter what quark \#534 is doing. The higher energy degrees of freedom are ``integrated out", such that they don't have any influence in the dynamics of the system.

In this section, we will try to summarize the stepping stones we need to get from the fundamental interactions in quantum chromodynamics to chiral EFT, which is what we use in our lattice simulations. Apart from some beyond-the-standard-model theories, QCD is the most fundamental description of nucleons, in terms of their constituent particles, the quarks and gluons. However, systems of single protons or neutrons are at the cutting-edge for lattice QCD calculations. It will be a long time before supercomputers and the numerical algorithms are able to push the limits into nuclear systems.

\begin{figure}	% FILLER PLOT
\begin{center}
\makebox[\textwidth][c]{\includegraphics[width=0.9\textwidth]{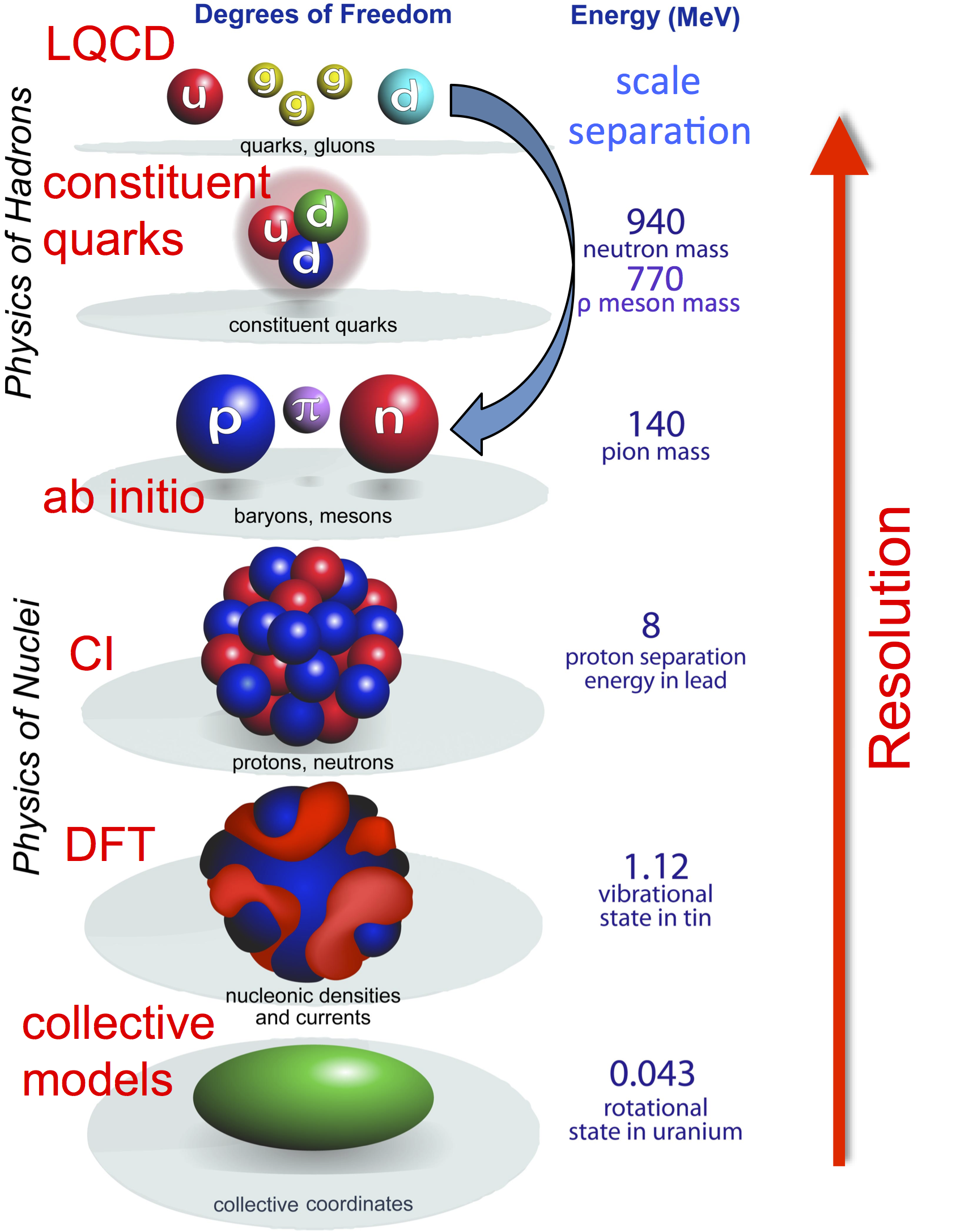}}
\caption[Scales in Nuclear Physics]{Energy scales in nuclear physics. At each step, our resolution changes, and so do the relevant degrees of freedom. This work follows the blue arrow; describing the jump from QCD to \textit{ab initio} methods which rely on EFTs. Figure from G. F. Bertsch et al \cite{Bertsch:2007can}.}
\end{center}
\end{figure}

\subsection{Quantum Chromodynamics (QCD)}

Quantum Chromodynamics (QCD) is an SU(3) gauge field theory that describes the strong interactions between quarks and gluons. It was developed throughout the 1970s, as an application of Yang-Mills theory.

The QCD Lagrangian is given by
\begin{equation}
\mathcal{L} = \sum_q \bar{\psi}_{q,a} \left( i \gamma^\mu \partial_\mu \delta_{ab} - g_s \gamma^\mu t_{ab}^C A_\mu^C - m_q \delta_{ab} \right) \psi_{q,b} - \frac{1}{4}F_{\mu\nu}^A F^{A \mu\nu}
\end{equation}
\noindent where repeated indices are summed over. The $\gamma^\mu$ are the Dirac $\gamma$-matrices. The $\psi_{q,a}$ is the quark field spinors, for quark flavour $q = u, d, s, c, t, \text{or} b$, with corresponding masses $m_q$, and a color index $a = 1,\cdots,N_c=3$, i.e. quarks come with three ``color charges." These quark fields are therefore elements of fundamental representation of SU(3). The $A_\mu^C$ are the gluon fields, where $\mu$ is once again the Dirac index, and $C = 1,\cdots,N_c^2-1=8$ runs over the 8 kinds of gluons. These gluon fields transform under the adjoint representation of SU(3). The $t_{ab}^C$ are the Gell-Mann matrices, defined by the relation
\begin{equation}
\left[ t^A, t^B \right] = i f_{ABC} t^C
\end{equation}
\noindent where $f_{ABC}$ are the structure factors of $SU(3)$. Finally, the field strength tensor $F_{\mu\nu}^A$ is defined by
\begin{equation}
F_{\mu\nu}^A = \partial_\mu A_{\nu}^A - \partial_\nu A_{\mu}^A - g_s f_{ABC} A_{\mu}^B A_{\nu}^C
\end{equation}

The coupling of the strong interaction is $g_s$. These enter into the three fundamental diagrams of QCD: a quark-antiquark-gluon vertex, a three-gluon vertex, and a four-gluon vertex. These diagrams contribute factors of $g_s$, $g_s$, and $g_s^2$, respectively.

One peculiarity of QCD is that quarks and gluons cannot be observed as free particles; instead they must be observed as color-neutral (color singlet) states. These can form mesons (a quark-antiquark) and baryons (three quarks). This property is known as ``confinement."

\begin{figure}
\begin{center}
\makebox[\textwidth][c]{\includegraphics[width=\textwidth]{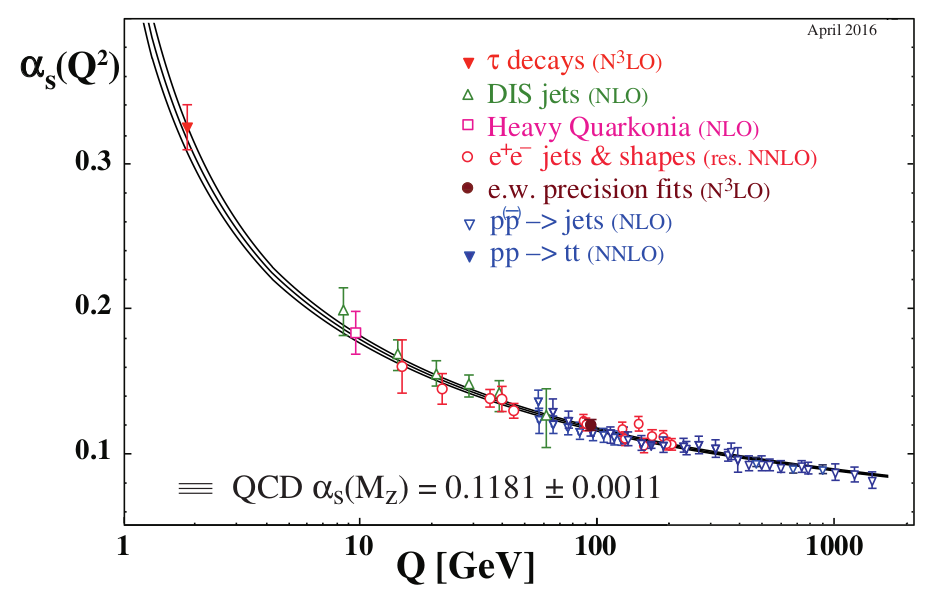}}
\caption[Running coupling in QCD]{The running coupling of QCD. Shown is a plot of the renormalized coupling $\alpha_s$ as a function of momentum transfer $Q$. Shown are the numerous experimental and numerical calculations of $\alpha_s$, at various $Q$, and the best fit value at the Z-boson mass $M_Z$ is presented \cite{Tanabashi:2018rpp}}
\end{center}
\end{figure}

Another interesting feature of QCD is that of asymptotic freedom; the strength of the strong interaction gets weaker at higher energies, and stronger at lower energies. This running coupling is shown in Fig. [2.2]. For low energies, QCD is non-perturbative, so it is often the case that computations are done at higher values for the coupling, the extrapolated down to the ``physical point".

\subsection{Chiral Effective Field Theory}

Chiral effective field theory ($\chi$-EFT) and chiral perturbation theory ($\chi$-PT) are built from a low energy effective field theory that was designed to be consistent with the chiral symmetry of the QCD Lagrangian. This chiral symmetry, simply put, is the limit where the quark masses go to zero.

Construction of the chiral Lagrangian starts with considering all terms that are consistent with the approximate chiral symmetry of QCD. This symmetry can be spontaneously broken, giving rise to the Goldstone bosons, the pions.

In this low-energy EFT, the relevant degrees of freedom move from the quarks and gluons to just the hadrons. In $\chi$-EFT, these are taken to be the nucleons. Then, one can work in terms of a power counting scheme in terms of some momentum scale.

If one looks at a given diagram, the behaviour can be analysed by rescaling the external momentum lines $p_i \to t p_i$ and the quark masses $m_i \to t^2 m_i$. Then, Weinberg's power counting shows that a given digram scales as $t^D$ for $D \geq 2$ is the number of vertices in the diagram. For a given $D$, this tells us to what order in the expansion of $\mathcal{L}_{\chi}$ we need.

These expansions tend to work for momentum smaller than the chiral symmetry breaking scale $\Lambda_{\chi SB} \approx 4\pi f_0$, where $f_0 = 93$ MeV is the pion decay constant. Additionally, the other scale where $\chi$-PT tends to break down is near the $\rho$ mass, which is $\approx 780$ MeV.

\begin{figure}	% FILLER PLOT
\begin{center}
\makebox[\textwidth][c]{\includegraphics[width=\textwidth]{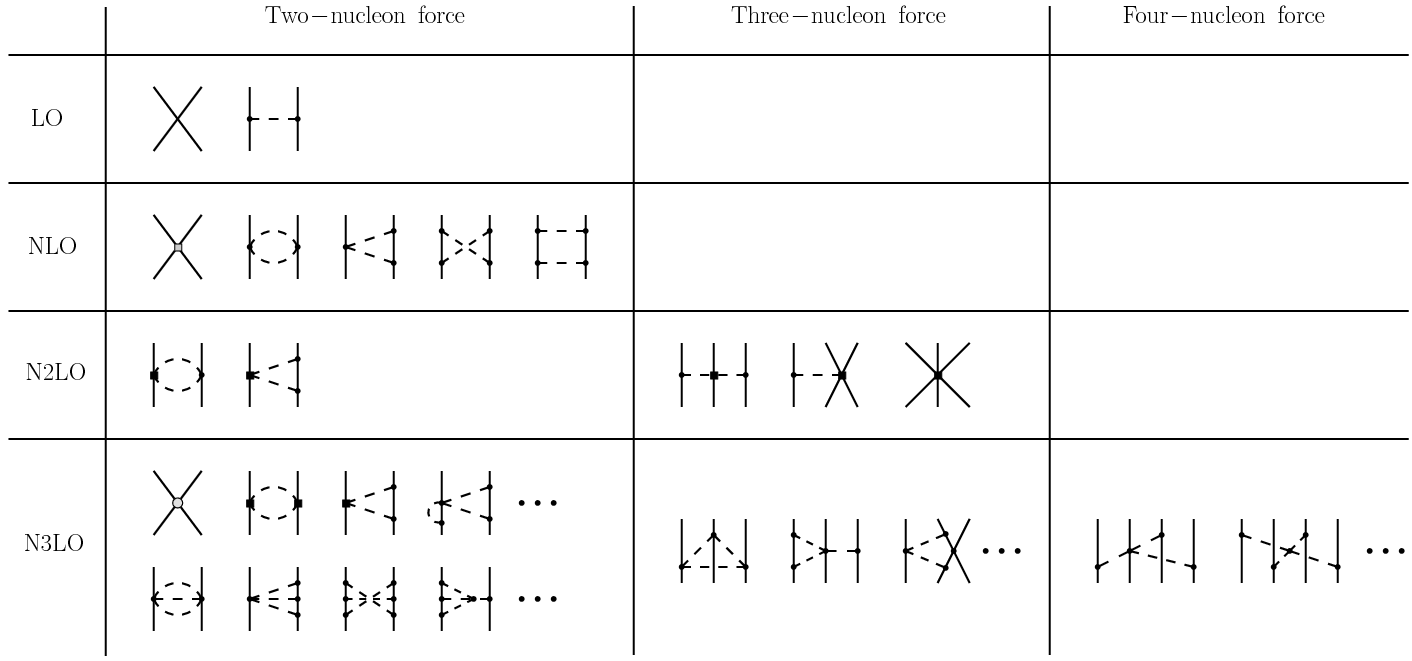}}
\caption[Diagrams in Chiral EFT]{An organization of the diagrams in Chiral EFT. Each row indicates the order of the diagram, i.e. how many powers of $(Q/\Lambda)$ the diagram contributes. The solid lines indicate nucleon lines, and the dashed lines indicate pion lines. The columns represent how many nucleons are involved in the process.}
\end{center}
\end{figure}

Shown in Fig. [2.3] is the organisation of diagrams in terms of this power counting scheme. In the rows are all diagrams which contribute a given power of $(Q/\Lambda)$. In the columns are diagrams which consist of $N$ external nucleon lines.

From this, we can see that the three-nucleon force does not make an appearance until N$^2$LO, which contributes at the $(Q/\Lambda)^3$ level.

For more information on chiral perturbation theory or chiral effective field theory, see \cite{Scherer:2003icp} and \cite{Machleidt:2011cef}.

\subsection{Fitting the LECs to NN Scattering Data}

Chiral EFT has several low-energy constants, or LECs, that need to be constrained before any calculations can be done. These are usually done by looking at low-energy scattering experiments between nucleons, such as n-n, n-p, or p-p scattering, or by binding energies of light systems, like the deuteron or alpha particle.

The first step in determining the LECs is to express the relevant operators in terms of angular momentum states. Then, the operators that act on a given partial wave can have their own particular LEC. By looking at the phase shifts, these can be fitted directly with lattice calculations. This section follows the discussion in \cite{Li:2018nps}. For a full discussion, please see this reference.

To start determining the phase shifts on the lattice, we need to re-write the full wavefunction in terms of spherical harmonics. This is done by
\begin{equation}
| r \rangle^{L,L_z} = \sum_{\textbf{r}'} Y_{L,L_z}(\hat{r}')\delta(|\textbf{r}'| - r)|\textbf{r}'\rangle
\end{equation}
\noindent where $r$ is a fixed distance from the origin, and the sum runs over all lattice points. The delta function counts only the lattice points that are a distance $r$ from the origin.

With this definition of the radial wavefunction, the Hamiltonian can be reduce to a form that contains only a single variable, $r$. Then, the phase shifts and mixing angles can be extracted in the region of vanishing potentials. Here, the radial wavefunction can be written in terms of spherical Bessel functions
\begin{equation}
| r \rangle^{L,L_z} \approx A_L h^-_L(kr) + B_L h^+_L (kr)
\end{equation}
\noindent where $k = \sqrt{2\mu E}$, $\mu$ is the reduced mass, and $E$ is the energy. The scattering coefficients $A_L$ and $B_L$ are related to the S-matrix by
\begin{equation}
B_L = S_L A_L
\end{equation}
\noindent with $S_L = \exp(2i\delta_L)$, where $\delta_L$ is the phase shift. This phase shift is determined by
\begin{equation}
\delta_L = \frac{1}{2i}\log \left( \frac{B_L}{A_L} \right)
\end{equation}

Fitting the phase shifts is done using a least-squares minimization. Shown in Fig. [2.4] are the results of fits of the phase shifts and mixing angles for several partial waves, done on a lattice with lattice spacing $a = 1.97 \text{ fm}$, for the energy range $E_{\text{lab}} \leq 50 \text{ MeV}$.

\begin{figure}
\begin{center}
\makebox[\textwidth][c]{\includegraphics[width=\textwidth]{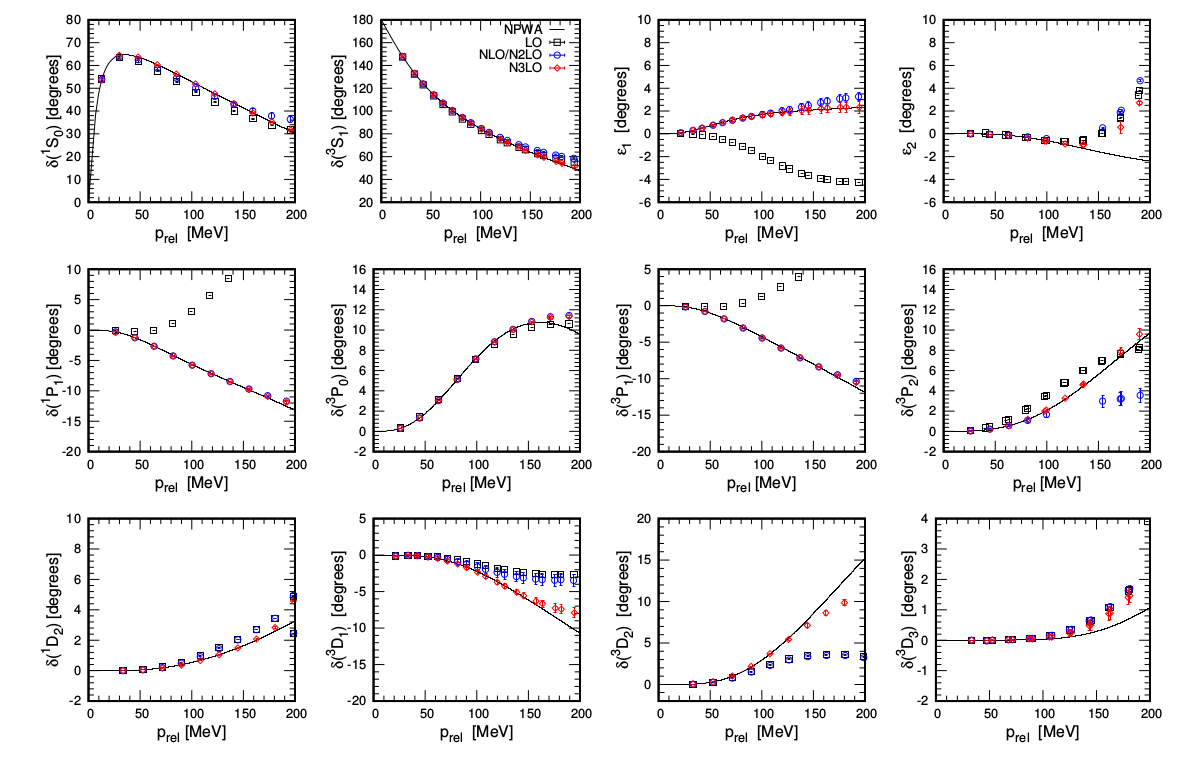}}
\caption[Neutron-proton phase shifts calculated on the lattice]{Neutron-proton phase shifts and mixing angles, calculated on the lattice, and shown as a function of the relative momenta $p_{\text{rel}}$. Figure courtesy of N. Li \cite{Li:2018nps}.}
\end{center}
\end{figure}

%%%%%%%%%%%%%%%%%%%%%%%%%%%%%%%%%%%%%%%%%%%%%%%%%%%%%%%%%%%%%%%%%%
\section{Monte Carlo Methods}

In this section, we will discuss the main methods that we use to evaluate these discretized path integrals.

\subsection{What is Monte Carlo?}

Monte Carlo algorithms are a type of algorithm that involves repeatedly generating random numbers, and using them in some meaningful way. They were named after the city of Monte Carlo in Monaco, since these algorithms have the gambling/rolling dice feel to them.

One of the primary uses of Monte Carlo is in numerically integrating a function. If we consider a function of one variable $f(x)$, over some closed interval $[a b]$, then we can view the integral of this function over this interval
\begin{equation}
\int_{a}^{b} f(x) dx
\end{equation}
as the area under the curve. All that we need to do is pick points in the domain and range of $f(x)$, and check to see if the point is above or below $f(x)$. Then, the integral can be evaluated as
\begin{equation}
\int_{a}^{b} f(x) dx = \lim_{N \to \infty} \left( \frac{N_\text{below}}{N} \right)
\end{equation}

The other primary use of Monte Carlo algorithm is in sampling probability distributions. For a given continuous probability distribution $\mathcal{P}(x)$ and operator $\mathcal{O}(x)$
\begin{equation}
\left\langle \mathcal{O} \right\rangle = \int \mathcal{P}(x) \mathcal{O}(x) dx = \lim_{N \to \infty} \frac{1}{N} \sum_{i=1}^N \mathcal{P}(x_i) \mathcal{O}(x_i)
\end{equation}
\noindent That is, any expectation value which involves integrating a quantity over some weight distribution can be expressed as a discrete sum over a finite number of samples pulled from that distribution.

In lattice field theory, this probability distribution is the Boltzmann distribution $e^{-\mathcal{S}}$, and the integration variables are the fermion fields. Now, since there is an infinite sum for each variable being integrated, direct sampling would never work for these lattice field theories. For this, we need more specialized algorithms.

\subsection{The Sign Problem}

Before we hop into the discussion of the Monte Carlo algorithms we use in our simulations, we first make a few remarks about the sign problem.

The sign problem/fermion sign problem/signal-to-noise problem is a common issue that arises in the simulations of fermions, and it is due to the fact that the probability distribution is not necessarily positive-definite. If we consider the expectation values of some operator
\begin{equation}
\langle A \rangle = \frac{\int D\phi A[\phi] p[\phi]}{\int D\phi p[\phi]}
\end{equation}
\noindent with bosonic fields $\phi$, and the probability weight defined by
\begin{equation}
p[\phi] = \det\left(M[\phi]\right) \exp \left(-\mathcal{S}[\phi]\right)
\end{equation}
\noindent which contains the fermion determinant and the Boltzmann weighting factor. When $p[\phi]$ is positive for all $\phi$, this integral is well-defined, and can be evaluated numerically.

However, if $p[\phi]$ is ever non-positive (usually due to a complex $\mathcal{S}$), then the integral becomes highly oscillatory, can cannot reliably be determined numerically. This can be shown if we break the probability weights into a modulus and a phase
\begin{equation}
p[\phi] = \rho[\phi] \exp(i\theta[\phi])
\end{equation}
\noindent and then we move the phase into the definition of the expectation value.
\begin{equation}
\langle A \rangle =  \frac{\int D\phi \rho[\phi] \exp(i\theta[\phi])}{\int D\phi \rho[\phi] \exp(i\theta[\phi])} = \frac{\langle A[\phi] \exp(i\theta[\phi]) \rangle}{\langle \exp(i\theta[\phi]) \rangle}
\end{equation}
\noindent This $\rho[\phi]$ is now strictly real and positive-definite, so can always be interpreted as a probability weight. This phase factor now encodes all of the complex behaviour of the observables and fields.
For $\exp(i\theta[\phi]) \approx 1$, there is no sign problem, and this expectation value is well-defined. If, however,  $\exp(i\theta[\phi]) \to 0$, both the numerator and denominator become small, leading to wildly fluctuating estimates for the expectation value.
 
This problem is an NP-Hard problem, meaning that a general solution to it is likely not possible (see P = NP). Instead, the best we can do is to find ways to circumvent the problem. There are multiple ways to minimize this problem in the context of lattice simulations; meron cluster algorithms \cite{COX2000777}, the complex Langevin method \cite{Aarts:2008rr}, a density of states method \cite{Langfeld:2016mct}, pseudofermions \cite{Fucito:1980fh}, Canonical methods \cite{Alexandru:2005ix,deForcrand:2006ec}, analytic continuation from imaginary chemical potentials \cite{deForcrand:2006pv}, fermion bags \cite{Chandrasekharan:2013rpa}, re-weighting methods \cite{Fodor:2001au}, sign-optimized manifolds \cite{Alexandru:2018fqp}, or thimble methods \cite{Cristoforetti:2013qaa,Cristoforetti:2012su}. Each has its own particular formulation, advantages, and disadvantages. In this work, we will show how the eigenvector continuation method can avoid this problem from becoming too much of a problem.

\subsection{Euclidean Time Projection Monte Carlo}

If we are interested in studying the dynamics of a quantum system on the lattice, then it would be natural to look at the time evolution of a system. However, in quantum mechanics, the time evolution operator is simply a phase factor
\begin{equation}
\hat{T}(t_i - t_j) = e^{-i\hat{H}(t_i - t_j)}
\end{equation}
\noindent It is then a ``natural" extension to consider the case $t \to i\tau$, which is a Wick rotation for the time variable. This new ``time" variable $\tau$ is referred to as the Euclidean time, or projection time. The effect of this transformation is it turns the time projection operator into one that can be calculated easily, as it is a simple exponential decay
\begin{equation}
\hat{T}_E (\tau_i - \tau_j) = e^{-\hat{H}(\tau_i - \tau_j)}
\end{equation}
\noindent  In Monte Carlo simulations, we keep track of the Euclidean time projection amplitude
\begin{equation}
Z_{N,N}(\tau) \equiv \prod_{\vec{n},n_t} \left[ \int d_j s(\vec{n},n_t) \right] \langle \Psi^{0,\text{free}}_{N,N} | M_j (s,t) \cdots M_j (s,0) | \Psi^{0,\text{free}}_{N,N} \rangle
\end{equation}

\begin{figure}
\begin{center}
\makebox[\textwidth][c]{\includegraphics[width=0.7\textwidth]{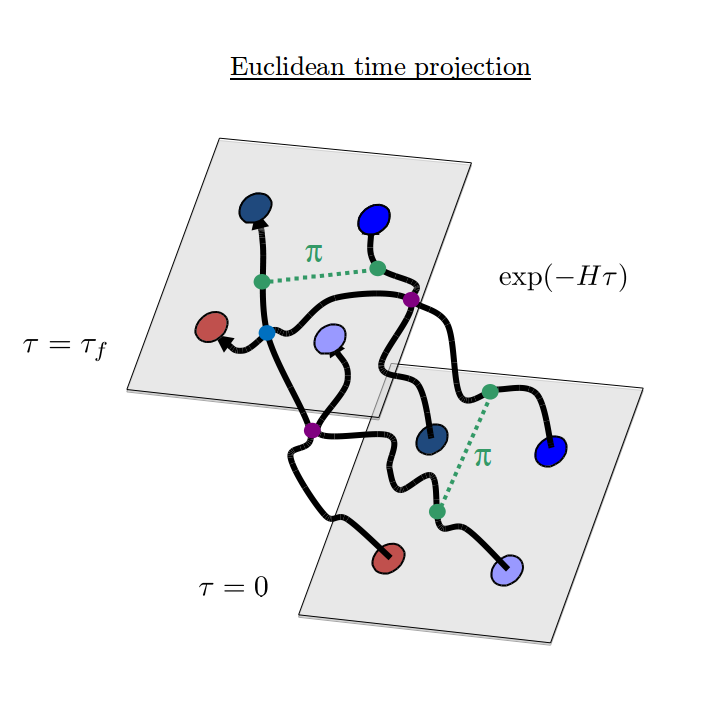}}
\caption[Euclidean time propagation of many-body state]{Propagation of the many-body state from ($\tau = 0$) to a final state ($\tau = \tau_f$). During the propagation, particles can interact via any of the chiral forces (LO contact and one-pion exchange shown here.)}
\end{center}
\end{figure}

Combining this method with the auxiliary fields, we can break the propagation involving many-body interactions (shown in Fig. [2.5]) into the propagation of single particle states through interactions with fluctuating background fields. This is shown in Fig. [2.6].

\begin{figure}
\begin{center}
\makebox[\textwidth][c]{\includegraphics[width=0.7\textwidth]{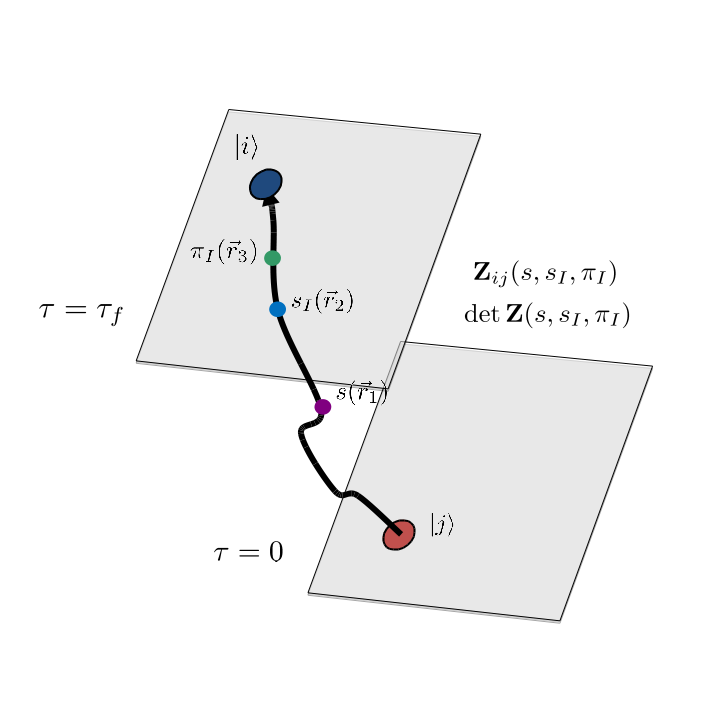}}
\caption[Euclidean time projection of single-particle state with auxiliary fields]{Propagation of some initial single-particle state ($\tau = 0$) to a final state ($\tau = \tau_f$). Along the path, the state interacts with the background auxiliary fields.}
\end{center}
\end{figure}

We now define an online energy expectation value that depends on the Euclidean time $\tau$
\begin{equation}
E_{N,N}(\tau) = \frac{1}{\alpha_t}\log \frac{Z_{N,N}(\tau - a_t)}{Z_{N,N}(\tau)}
\end{equation}
\noindent Looking at the spectral decomposition of $Z_{N,N}$, we see
\begin{equation}
Z_{N,N}(\tau) = \sum_k |c_{N,N}^k |^2 e^{-E_{N,N}^k \tau} 
\end{equation}

For the limit of $\tau \to \infty$, the lowest energy eigenstate is the primary contributor to the amplitude, and all the higher energy states are exponentially suppressed.
\begin{equation}
E_{N,N}(\tau) \approx E_{N,N}^0 + \sum_{k \neq 0} \left( \frac{|c_{N,N}^k|^2}{|c_{N,N}^0|^2} \right) \left( \frac{e^{(E_{N,N}^k - E_{N,N}^0) a_t} - 1}{\alpha_t} \right) e^{-(E_{N,N}^k - E_{N,N}^0)\tau}
\end{equation}
\noindent This can be further simplified by noting that the difference between the energy eigenstates $(E_{N,N}^k - E_{N,N}^0)$ is usually smaller than the energy cutoff scale $1/\alpha_t$. This can simplify then to
\begin{equation}
E_{N,N}(\tau) \approx E_{N,N}^0 + \sum_{k \neq 0} \left( \frac{|c_{N,N}^k|^2}{|c_{N,N}^0|^2} \right) \left( E_{N,N}^k - E_{N,N}^0 \right) e^{-(E_{N,N}^k - E_{N,N}^0)\tau}
\end{equation}
\noindent With this, we say that the ground state energy $E_{N,N}^0$ is the large $\tau$ limit
\begin{equation}
E_{N,N}^0 = \lim_{\tau \to \infty} E_{N,N}(\tau)
\end{equation}

\subsection{Auxiliary Field Monte Carlo}

Now that we have shown how to evaluate the ground state energy by propagating single-particle states through Euclidean time, we now wish to show how the auxiliary fields for the transfer matrices are generated.

The auxiliary fields are defined used a Gaussian integral transformation, known as the Hubbard-Stratonovich \cite{PhysRevLett.3.77} transformation.
\begin{equation}
\exp \left[ -\frac{C\alpha_t}{2} (a^\dagger(n) a(n))^2 \right] = \sqrt{\frac{1}{2\pi}}\int_{-\infty}^{\infty} ds \exp \left[-\frac{1}{2} s(n)^2 + \sqrt{-C \alpha_t} s(n) a^\dagger(n) a(n) \right]
\end{equation}

This transformation allows us to decompose two-body interactions into the integral over one-body interactions with a normally-distributed, background field. The only price is the addition of an integral at each lattice site for each interaction to be written in this manner. However, this is where Monte Carlo is well-suited.  

\begin{figure}
\begin{center}
\makebox[\textwidth][c]{\includegraphics[width=0.7\textwidth]{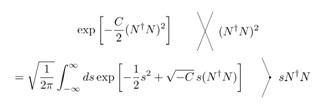}}
\caption[Hubbard-Stratonivich transformation]{A diagram of the Hubbard-Stratonivich transformation. Here, we can express the two-body interaction in terms of a one-body interaction with a fluctuating background field.}
\end{center}
\end{figure}

For simple contact interactions, the auxiliary field can be directly sampled from a normal distribution
\begin{equation}
P = exp\left( -\frac{1}{2} s(n)^2 \right)
\end{equation}
\noindent However, for more complicated interactions, like Coulomb or one-pion exchange, we have to work harder. Here, we take as our distribution
\begin{equation}
P = \exp \left( -\mathcal{S} \right) = \exp \left( - \sum_{\vec{n},\vec{n}'} s_1(\vec{n},n_t) V^{-1}(\vec{n} - \vec{n}') s_1(\vec{n}',n_t) \right)
\end{equation}
\noindent where this action is the same as in Eqn. (2.60), except we have removed the constant factor in front as well as the sum over time steps.

This form of the action is difficult to sample by itself, so we factor it into a Gaussian-like form. Defining a $t(\vec{n},n_t)$ variable, we would like
\begin{equation}
\sum_{\vec{n},\vec{n}'} s_1(\vec{n},n_t) V^{-1}(\vec{n} - \vec{n}') s_1(\vec{n}',n_t) = \frac{\left[ t(\vec{n},n_t) \right]^2}{2}
\end{equation}
\noindent so that the probability distribution simplifies to
\begin{equation}
P = \exp \left( -\mathcal{S} \right) = \exp \left( \frac{-\left[t(\vec{n},n_t) \right]^2}{2} \right)
\end{equation}
\noindent By inspection, we see that if we define $t(\vec{n},n_t) $ as
\begin{equation}
t(\vec{n},n_t) = \sum_{\vec{n}'}V^{-1/2}(\vec{n} - \vec{n}') s_1(\vec{n}',n_t)
\end{equation}
\noindent we reproduce the original distribution. If we invert this equation to solve for $s_1(\vec{n}',n_t)$, we have
\begin{equation}
s_1(\vec{n},n_t) = \sum_{\vec{n}'}V^{+1/2}(\vec{n} - \vec{n}') t(\vec{n}',n_t)
\end{equation}

This equation provides a simple prescription for generating a new $s_1(\vec{n},n_t)$. We just need to generate a set of normally-distributed numbers $t(\vec{n},n_t)$, and plug those in. However, this is a slow process, so using a convolution can speed this updating up tremendously.

If we define $\tilde{u}(\vec{k})$ to be,
\begin{equation}
\tilde{u}(\vec{k},n_t) = \left( \tilde{V}^{+1/2}(\vec{k}) \right) * \left( \tilde{t}(\vec{k},n_t) \right)
\end{equation}

\noindent where $*$ indicates a convolution, then the auxiliary field $s_1(\vec{n},n_t)$ is simply
\begin{equation}
s_1(\vec{n},n_t) = u(\vec{n},n_t)
\end{equation}

\noindent where the tilde represents the momentum space representations of each variable. These are obtained through the Fourier transform. With these formulas, the algorithm for updating the field is then: 

\noindent \underline{Step \#1}: Generate $\tilde{V}^{+1/2}(\vec{k})$. This only needs to be done once, at the beginning of the code.

\noindent \underline{Step \#2}: For each $\vec{n},n_t$, generate $t(\vec{n},n_t)$ according to Eqn. (2.96).
\begin{equation}
P \left[ t(\vec{n},n_t) \right] \propto \exp \left\lbrace -\frac{1}{2} \left[   t(\vec{n},n_t) \right] ^2 \right\rbrace
\end{equation}

\noindent \underline{Step \#3}: For each time step $n_t$, generate the quantities
\begin{equation}
\tilde{t}(\vec{k},n_t) = \frac{1}{L^{3/2}} \sum_{\vec{n}} t(\vec{n},n_t) e^{2\pi i \vec{n}\cdot\vec{k}/L}
\end{equation}
\begin{equation}
\tilde{u}(\vec{k},n_t) = \tilde{V}^{+1/2}(\vec{k}) \tilde{t}(\vec{k},n_t)
\end{equation}
\begin{equation}
s_1(\vec{n},n_t) = \frac{1}{L^{3/2}} \sum_{\vec{k}} \tilde{u}(\vec{k},n_t) e^{-2\pi i \vec{n}\cdot\vec{k}/L}
\end{equation}

\noindent Once observables are computed using this auxiliary field configuration, go back to step \#2 and generate a new configuration. Since the fields are being directly sampled from the respective probability distribution, there is no need for any importance sampling or Metropolis accept/reject stages.

This technique is very straight-forward, but usually not the most optimal way to generate field configurations. For this, we turn to more sophisticated updating algorithms, like Hybrid Monte Carlo.

\subsection{Hybrid Monte Carlo}

Recall, our goal is to calculate the ratio
\begin{equation}
\frac{Z_{N,N}(t - a_t)}{Z_{N,N}(t)}
\end{equation}
\noindent using a Markov chain Monte Carlo method. For $t = L_t a_t$,  auxiliary field configurations for the base interaction are sampled according to the weight
\begin{equation}
\exp \left\lbrace -U_j(s) + 2 \log \left[ | \det \textbf{M}_j (s,t) |\right]  \right\rbrace
\end{equation}
\noindent where
\begin{equation}
U_1(s) = \frac{1}{2} \sum_{\vec{n},n_t}\left[s(\vec{n},n_t)^2\right]
\end{equation}
\noindent and
\begin{equation}
U_4(s) = 0
\end{equation}
This importance sampling is handled by the Hybrid Monte Carlo (HMC) method \cite{DUANE1987216}. This involves computing molecular dynamics trajectories using the Hamiltonian
\begin{equation}
H_j(s,p) = \frac{1}{2}\sum_{\vec{n},n_t} \left[ p(\vec{n},n_t) \right]^2 + V_j(s)
\end{equation}
\noindent where $p(\vec{n},n_t)$ is the conjugate momentum of $s(\vec{n},n_t)$ and
\begin{equation}
V_j(s) = U_j(s) - 2 \log \left[ | \det \textbf{M}_j (s,t) |\right]
\end{equation}
\noindent The algorithm for updating the field is as follows:

\noindent \underline{Step \#1}: Select an arbitrary initial configuration $s^0$. 

\noindent \underline{Step \#2}: Select a configuration $p^0$ according to a normal distribution.

\begin{equation}
P \left[ p^0(\vec{n},n_t) \right] \propto \exp \left\lbrace -\frac{1}{2} \left[   p^0(\vec{n},n_t) \right] ^2 \right\rbrace
\end{equation}
\noindent \underline{Step \#3}: For each $\vec{n},n_t$, let
\begin{equation}
\tilde{p}^{0}(\vec{n},n_t) = p^{0}(\vec{n},n_t) - \frac{\epsilon_{\text{step}}}{2} \left[ \frac{\partial V_j(s)}{\partial s(\vec{n},n_t)} \right]_{s=s^0}
\end{equation}
\hspace{1.9cm}for some small positive $\epsilon_{\text{step}}$. The derivative of $V_j(s)$ is computed using
\begin{equation}
\begin{split}
\frac{\partial V_j(s)}{\partial s(\vec{n},n_t)} &= \frac{\partial U_j(s)}{\partial s(\vec{n},n_t)} - \frac{2}{\det \textbf{M}_j} \sum_{k,l} \frac{\partial \det \textbf{M}_j}{\partial \left[ \textbf{M}_j \right]_{kl}} \frac{\partial \left[ \textbf{M}_j \right]_{kl}}{\partial s(\vec{n},n_t)} \\
&= \frac{\partial U_j(s)}{\partial s(\vec{n},n_t)} - 2 \sum_{k,l} \left[ \textbf{M}_{j}^{-1} \right]_{lk} \frac{\partial \left[ \textbf{M}_j \right]_{kl}}{\partial s(\vec{n},n_t)}
\end{split}
\end{equation}
\noindent \underline{Step \#4}: For steps $i=0,1,\cdots,N_{\text{step}}-1$, let
\begin{align}
&s^{i+1}(\vec{n},n_t) = s^{i}(\vec{n},n_t) + \epsilon_{\text{step}} \tilde{p}^{i}(\vec{n},n_t) \\
&\tilde{p}^{i+1}(\vec{n},n_t) = \tilde{p}^{i}(\vec{n},n_t) - \epsilon_{\text{step}} \left[ \frac{\partial V_j(s)}{\partial s(\vec{n},n_t)} \right]_{s=s^{i+1}}
\end{align}
\hspace{1.9cm}for each $\vec{n},n_t$. \\
\noindent \underline{Step \#5}: For each $\vec{n},n_t$, let
\begin{equation}
p^{N_{\text{step}}}(\vec{n},n_t) = \tilde{p}^{N_{\text{step}}}(\vec{n},n_t) - \frac{\epsilon_{\text{step}}}{2} \left[ \frac{\partial V_j(s)}{\partial s(\vec{n},n_t)} \right]_{s=s^{\text{step}}}
\end{equation}
\noindent \underline{Step \#6}: Select a random number $r \in [0,1)$. If,
\begin{equation}
r < \exp \left[ -H(s^{N_{\text{step}}},p^{N_{\text{step}}}) + H(s^0,p^0)\right]
\end{equation}
\hspace{1.9cm}then set $s^0 = s^{N_{\text{step}}}$. Otherwise, leave $s^0$ as is. \\
Once observables are computed using this auxiliary field configuration, go back to step \#2 and start a new trajectory. The observable we calculate at each configuration is
\begin{equation}
O_j(s,L_t a_t) = \frac{\left[\det \textbf{M}_j(s,(L_t-1) \alpha_t) \right]^2}{\left[\det \textbf{M}_j(s,L_t \alpha_t) \right]^2}
\end{equation}
\noindent by taking the ensemble average of this observable over many HMC trajectories, we obtain
\begin{equation}
\frac{Z_{N,N}(t - \alpha_t)}{Z_{N,N}(t)}
\end{equation}

\subsection{Shuttle Algorithm}

Now that we have algorithms for generating auxiliary field configurations and propagating the states through Euclidean time, we are ready to put everything together. In older codes, we used to generate all of the auxiliary fields first, then propagate the initial wavefunctions through Euclidean time, and then calculate the expectation values of observables at the middle time-step. This order is not very efficient, and wastes a lot of computational time.

A new algorithm we began to use is the Shuttle algorithm, discussed briefly in \cite{Lu:2018een}. This algorithm instead updates the time steps one at the time, sweeping back and forth from $\tau = 0$ to $\tau = L_t a_t$. The schematic of this method is shown in Fig. (2.8).

\begin{figure}
\begin{center}
\makebox[\textwidth][c]{\includegraphics[width=0.8\textwidth]{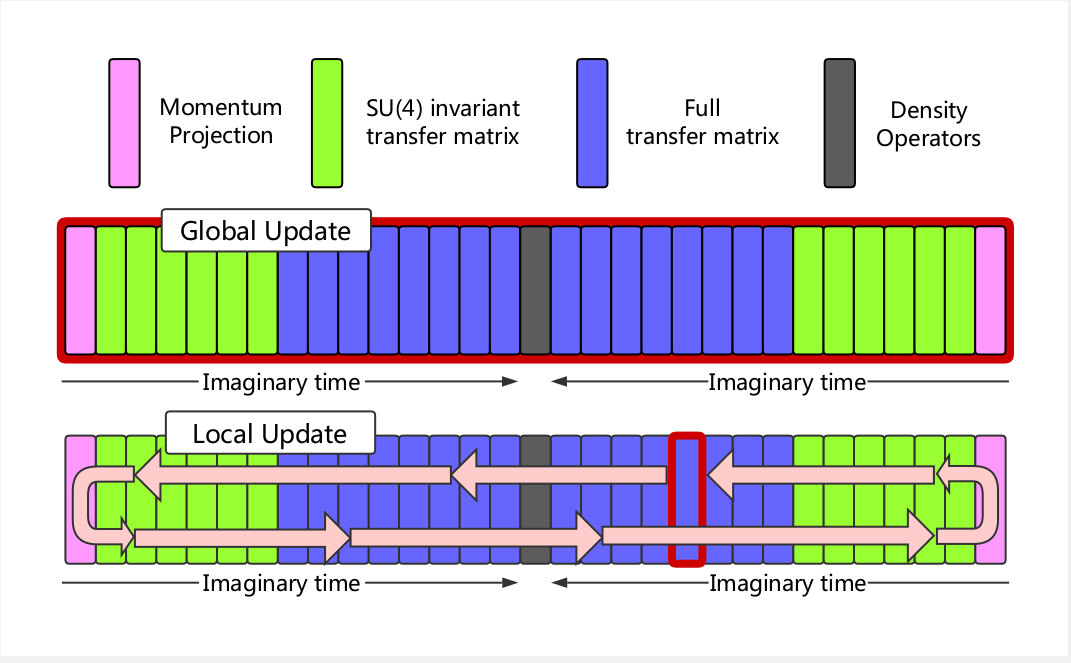}}
\caption[Schematic of the Shuttle Algorithm]{Schematic of Shuttle Algorithm. Courtesy of Bingnan Lu (MSU)}
\end{center}
\end{figure}

The global update here is the older algorithm, and the local update is this new algorithm. On the diagram, the red box shows the "current" time-step being updated. As this red box moves left and right, new auxiliary fields are generated, using either HMC for the full transfer matrix steps, or the simpler Metropolis updates for the SU(4)-invariant time steps.

When the red box reaches the middle time-step, the expectation value of all of the operators are computed. And finally, when the red box reaches the first or last time step, the initial states are re-updated, and the sweep direction is reversed. Using this algorithm, we can minimize the wasted time during the Monte Carlo simulations.

%%%%%%%%%%%%%%%%%%%%%%%%%%%%%%%%%%%%%%%%%%%%%%%%%%%%%%%%%%%%%%%%%%
%%%%%%%%%%%%%%%%%%%%%%%%%%%%%%%%%%%%%%%%%%%%%%%%%%%%%%%%%%%%%%%%%%
%%%%%%%%%%%%%%%%%%%%%%%%%%%%%%%%%%%%%%%%%%%%%%%%%%%%%%%%%%%%%%%%%%

\chapter{Eigenvector Continuation}

In this chapter, we discuss the main topic of this dissertation; the Eigenvector Continuation technique. This method begins with some Hermitian Hamiltonian matrix $H$
\begin{equation}
H = H_0 + c H_1
\end{equation}
\noindent with the eigenvectors denoted $|\psi_j(c)\rangle$. Generating a power series expansion for $|\psi_j(c)\rangle$, we get
\begin{equation}
|\psi_j(c)\rangle = \sum_{n=0}^{\infty} |\psi^{(n)}_j(0)\rangle c^n /n!
\end{equation}
\noindent were $c$ is assumed to be in a region in which this series converges. Suppose that we are interested in a value of $c$ outside of this region. Using just perturbation theory, we would be out of luck. However, suppose that we construct a new series about a point $|w| < |c|$. Then, we could write
\begin{align}
|\psi_j(c)\rangle &= \sum_{n=0}^{\infty} |\psi^{(n)}_j(w)\rangle \left(c-w\right)^n /n! \\
|\psi^{(n)}_j(w)\rangle &= \sum_{m=0}^{\infty} |\psi^{(n+m)}_j(0)\rangle w^m /m!
\end{align}
\noindent Combining these, we can obtain
\begin{equation}
|\psi_j(c)\rangle = \sum_{n=0}^{\infty} \sum_{m=0}^{\infty} \left(c-w\right)^n w^m |\psi^{(n+m)}_j(0)\rangle / m! n!
\end{equation}

\noindent This multi-series expansion is the heart of the eigenvector continuation method. In a sense, we now have an expression for $|\psi_j(c)\rangle$ with a wider area of convergence. Now, from a numerical standpoint, the analogue of derivatives of the eigenvectors at $c=0$ are the finite differences, i.e. computing the eigenvectors at multiple small values for the couplings. In this work, we will refer to these eigenvectors as EC eigenvectors or training eigenvectors.

\noindent Once we obtain these eigenvectors, we can form them into a projection operator $P$ and project the Hamiltonian $H$ at some target coupling $C_\odot$ into this lower-dimensional subspace. We call the projected Hamiltonian $M = P^\dagger H P$, and the norm matrix $N = P^\dagger P$. Then, we solve the generalized eigenvalue problem
\begin{equation}
M \vec{v} = \lambda_i N \vec{v}
\end{equation}
\noindent The lowest eigenvalue of this system is called the EC estimate for the ground state energy, and the ground state eigenvector in the full space can be reconstructed from the projection operator
\begin{equation}
|\psi_0(C_\odot)\rangle = P \vec{v}_0
\end{equation}

\noindent From here, we briefly discuss how these techniques are implemented numerically, with a side discussion about the computational costs. The computation associated with the eigenvector continuation method hinges on the computation of the training eigenvectors for a few values of the coupling $c$. If these can be accurately obtained, and if the overlaps between any two EC training eigenvectors are not nearly 1, then the method will be applicable.

%%%%%%%%%%%%%%%%%%%%%%%%%%%%%%%%%%%%%%%%%%%%%%%%%%%%%%%%%%%%%%%%%%
\section{Mathematical Formalism}

We start this discussion by considering a finite-dimensional linear space and a single parameter family of Hamiltonian matrices 
\begin{equation}
H(c) = H_0 + c H_1
\end{equation}

\noindent where $H_0$ and $H_1$ are Hermitian. Let $|\psi_j(c)\rangle$ denote the eigenvectors of $H(c)$ with corresponding eigenvalues $E_j(c)$. Since $H(c)$ is Hermitian for real $c$, each $E_j(c)$ is strictly real, and thus there are no singularities on the real axis.

Now, we can expand $E_j(c)$ and $|\psi_j(c)\rangle$ as power series about $c=0$.
\begin{align}
E_j(c) &= \sum_{n=0}^{\infty} E_j^{(n)}(0) c^n /n! \\
|\psi_j(c)\rangle &= \sum_{n=0}^{\infty} |\psi^{(n)}_j(0)\rangle c^n /n!
\end{align}

\noindent where the superscript $(n)$ refers to the n-th derivative. This is the approach of perturbation theory. These series are said to converge for all $|c| < |z|$, where $z$ and its complex conjugate are the nearest singularities to $c=0$ in the complex plane.

\begin{figure}
\begin{center}
\makebox[\textwidth][c]{\includegraphics[width=0.8\textwidth]{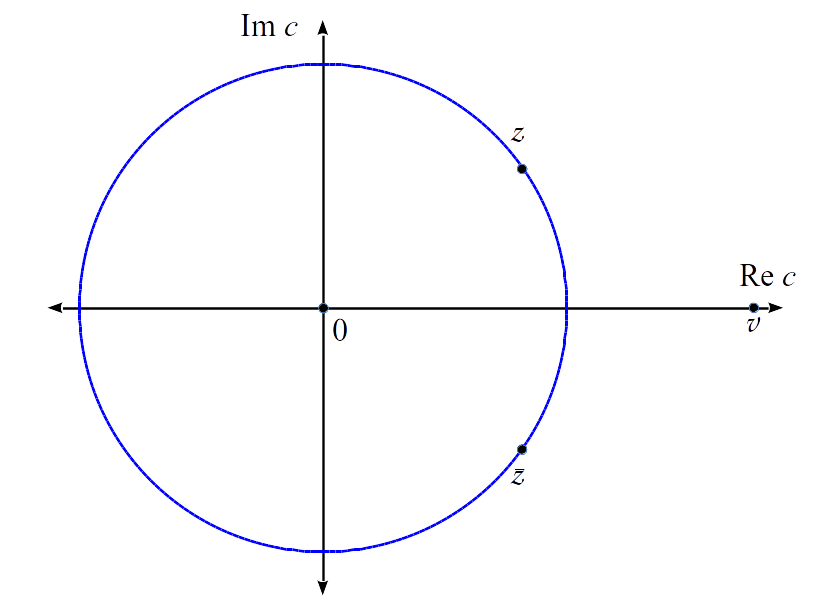}}
\caption[Diagram of convergence region for Eqn. (3.2)]{Diagram of convergence region for Eqn. (3.2)}
\end{center}
\end{figure}

If we are interested in points $|c| \geq |z|$, perturbation theory is no longer applicable. We can instead construct an analytic extension of Eqn. (3.2) by constructing a new series about the point $c = w$, where $w$ is real and $|w| < |z|$.

\begin{align}
|\psi_j(c)\rangle &= \sum_{n=0}^{\infty} |\psi^{(n)}_j(w)\rangle \left(c-w\right)^n /n! \\
|\psi^{(n)}_j(w)\rangle &= \sum_{m=0}^{\infty} |\psi^{(n+m)}_j(0)\rangle w^m /m!
\end{align}

\noindent Combining these, we can obtain
\begin{equation}
|\psi_j(c)\rangle = \sum_{n=0}^{\infty} \sum_{m=0}^{\infty} \left(c-w\right)^n w^m |\psi^{(n+m)}_j(0)\rangle / m! n!
\end{equation}

\noindent that is, the eigenvector $|\psi_j(c)\rangle$, can be expressed as a linear combination of the eigenvectors $|\psi_j(0)\rangle$ in the region $|c| < |z|$ \textbf{and} the circular region $|c-w| < |z-w|$ centered at $w$. 

\begin{figure}
\begin{center}
\makebox[\textwidth][c]{\includegraphics[width=0.8\textwidth]{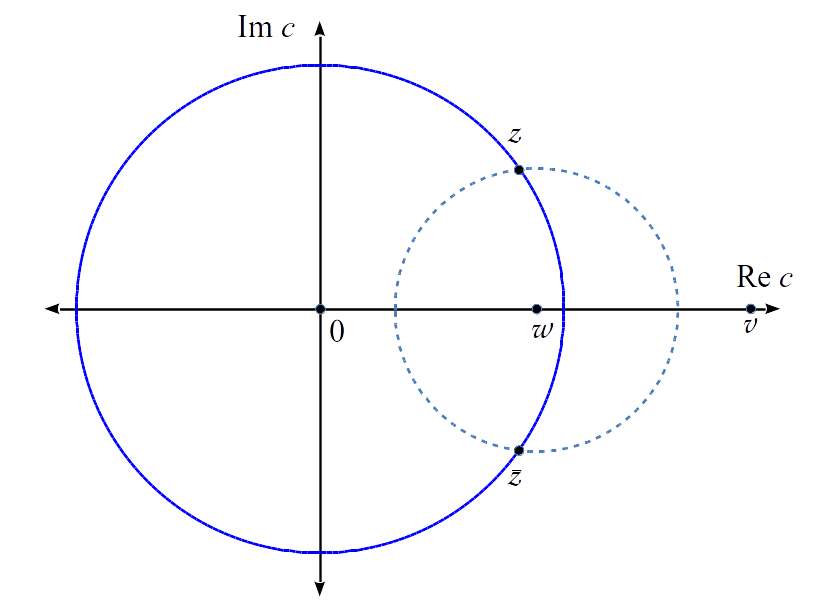}}
\caption[Diagram of extended convergence region, Eqn. (3.5)]{Diagram of Extended Convergence Region, Eqn. (3.5)}
\end{center}
\end{figure}

\begin{figure}
\begin{center}
\makebox[\textwidth][c]{\includegraphics[width=0.8\textwidth]{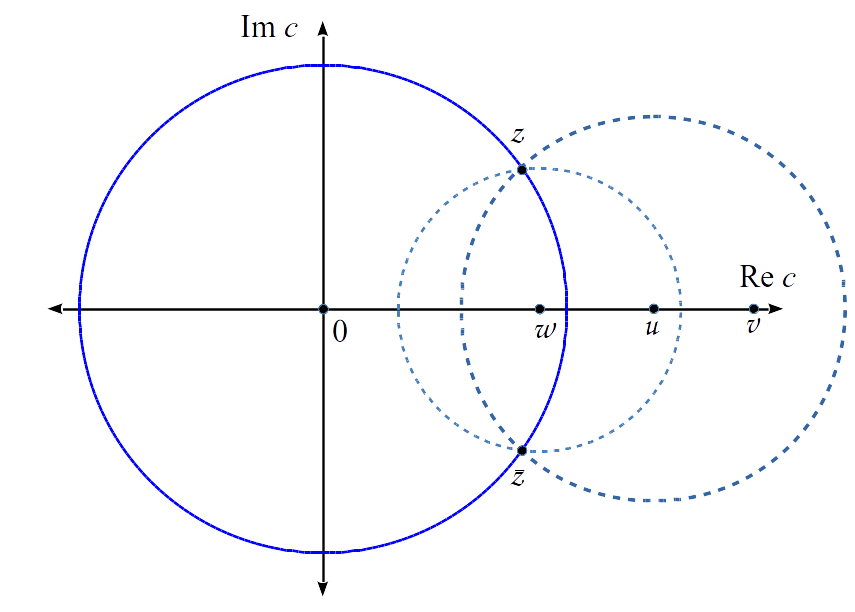}}
\caption[Extending the convergence region again to include $c=v$]{Extending the Convergence region to include the point $c=v$}
\end{center}
\end{figure}

\begin{figure}
\begin{center}
\makebox[\textwidth][c]{\includegraphics[width=0.8\textwidth]{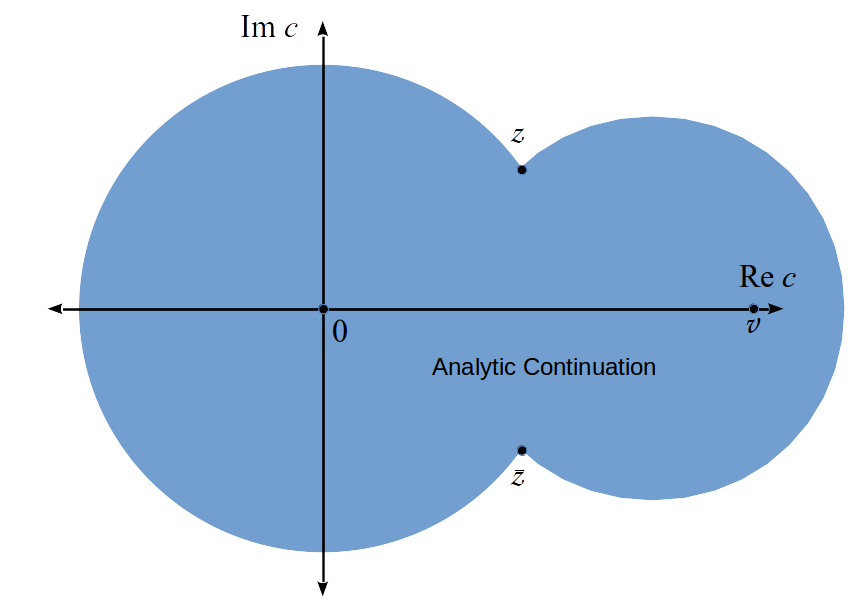}}
\caption[Final convergence region]{Final region of convergence}
\end{center}
\end{figure}

Analytic function theory help us understand why this technique works. Although in the original expansion for $|\psi_j(c)\rangle$ failed to converge for $|c| < |z|$, using the analytic extension Eqn. (3.5) allowed us to express a new series in terms of the original series, to arbitrary accuracy. This process can be repeated, until the target value of the coupling $c = v$ is within the analytically-extended convergence region. The number of vectors required is determined by the number of expansions needed in the analytic continuation, and the order-by-order convergence of each series expansion. 

In cases where there are singularities near the real axis, the convergence of these series can be accelerated by including multiple eigenvectors at each value of the coupling. Let us focus on a single eigenvector and eigenvalue of $H(c)$, which we label as $|\psi_1(c)\rangle$ and $E_1$, respectively. Furthermore, let D be the Riemann surface associated with $|\psi_1(c)\rangle$. We will assume that $|\psi_1(c)\rangle$ is analytic everywhere except for branch point singularities at $c = z$ and $c = \bar{z}$.

\begin{figure}
\begin{center}
\makebox[\textwidth][c]{\includegraphics[width=0.8\textwidth]{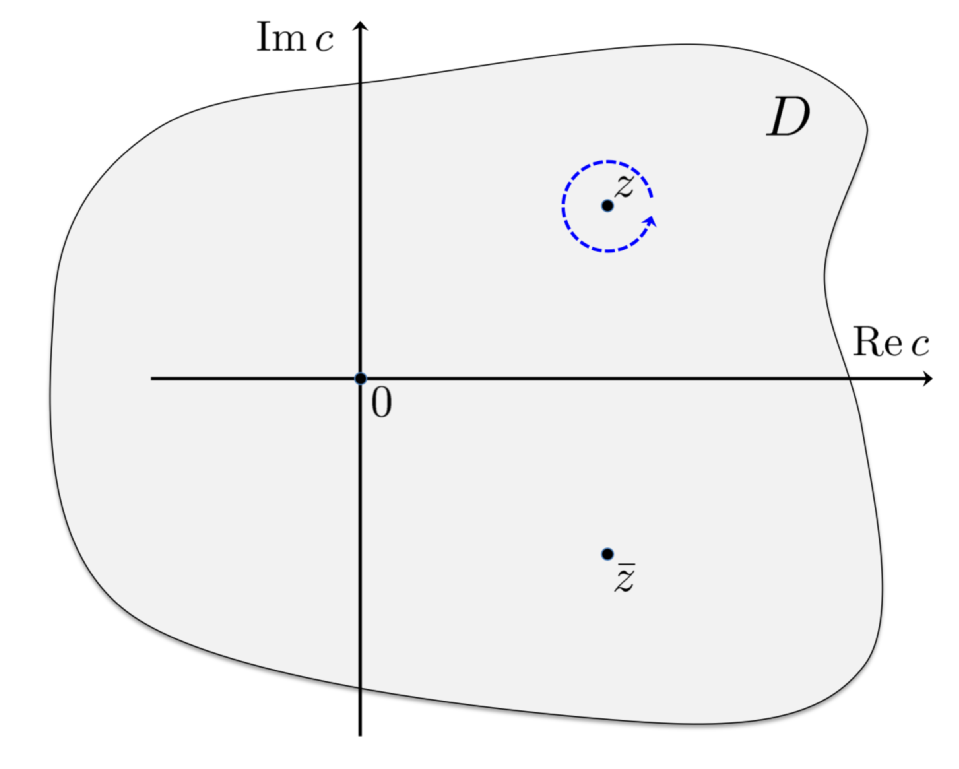}}
\caption[Riemann Surface of EC Subspace]{D is a finite region of the Riemann surface associated with the vector $|\psi_1(c)\rangle$. We assume $|\psi_1(c)\rangle$ is analytic everywhere except for branch point singularities at $z$ and $\bar{z}$. The dashed blue line shows the path traced around $z$.}
\end{center}
\end{figure}

Since the matrix elements of $H(c)$ are analytic everywhere, the characteristic polynomial of $H(c)$ is also analytic everywhere. We now consider what happens when traversing a closed loop around one of these branch points, as in Fig. [3.5]. The effect this process has on the eigenvectors and eigenvalues we will call the monodromy transformation $T(z)$.

We consider the case in which the $E_1(z)$ is a unique eigenvalue of $H(z)$. In this case, $T(z)$ transforms the eigenvector $|\psi_1(c)\rangle$ to a vector which is proportional to $|\psi_1(c)\rangle$. To remove the proportionality constant, we pick some state $|f>$ in the linear space such that $\langle f|\psi_1(c)\rangle \neq 0$ for all points $ c \in D$. This renormalized eigenvector can then be defined as
\begin{equation}
|\phi_1(c)\rangle = \frac{1}{\langle f|\psi_1(c)\rangle} |\psi_1(c)\rangle
\end{equation}
Since $\langle f|\phi_1(c)\rangle = 1$ for all $c \in D$, the transformation $T(z)$ transforms $\langle f|\phi_1(c)\rangle$ into itself. Therefore, $|\phi_1(c)\rangle$ is also an invariant of the transformation. While $|\psi_1(c)\rangle$ has a branch point singularity at $c = z$, the renormalized eigenvector now is analytic at $c = z$. Since the linear space spanned by $|\phi_1(c)\rangle$ is identical to the linear space spanned by $|\psi_1(c)\rangle$, it is clear that the singularities in the eigenvector normalization have no effect on the eigenvector continuation method.

We now consider the case in which $E_1(z)$ is not a unique eigenvalue of $H(z)$. In this case, the transformation $T(z)$ can result in a permutation of eigenvalues that are degenerate at $c = z$. Without loss of generality, we will assume that $T(z)$ induces a permutation cycle involving $k$ eigenvalues that we label $E_1(c), E_2(c), \cdots, E_k(c)$. With this notation, the monodromy transformation produces the cyclic permutation
\begin{equation}
T(z) : E_1(z) \to E_2(z) \to \cdots \to E_k(z) \to E_1(z)
\end{equation}
The case $k=1$ will lead to the same conclusion as the unique eigenvalue case. We will focus on the case $k > 1$. Let us label the image of $|\phi_1(c)\rangle$ under this transformation as $|\phi_2(c)\rangle$, an eigenvector of $H(z)$ with eigenvalue $E_2(z)$. We continue the labelling in this manner such that the transformation on the eigenvectors has the form
\begin{equation}
T(z) : \phi_1(c)\rangle \to \phi_2(c)\rangle \to \cdots \to \phi_k(c)\rangle \to \phi_1(c)\rangle
\end{equation}
The last entry in the cycle is $\phi_1(c)\rangle$ due to the fact that $\langle f|\phi_1(c)\rangle$ is single-valued under the transformation $\left[ T(z)\right]^k$.

We note that the linear combinations
\begin{equation}
|\gamma_n(c,z)\rangle = \sum_{j=0}^{k-1} e^{i2\pi nj/k}|\phi_j(k)\rangle
\end{equation}
for $n = 0,\cdots,k-1$ are the eigenvectors of the monodromy transformation with eigenvalues $e^{-i2\pi n/k}$. We can now renormalize each of these eigenvectors so that the branch point singularity at $c=z$ is removed for each n
\begin{equation}
|\gamma'_n(c,z)\rangle = e^{n\log(c-z)/k}|\gamma_n(c,z)\rangle
\end{equation}
These $|\gamma'_n(c,z)\rangle$ vectors are now analytic at $c=z$. In essence, what we have done is made linear combinations of the eigenvectors in which all of the non-integer fractional powers of $c-z$ have been removed. In a similar fashion, we can construct the vectors $|\gamma'_n(c,\bar{z})\rangle$ that are analytic at $c=\bar{z}$.

Let us now return to the discussion of performing eigenvector continuation with an EC subspace consisting of the k eigenvectors, $|\psi_1(c_i)\rangle,|\psi_2(c_i)\rangle, \cdots, |\psi_k(c_i)\rangle$ for each sampling point $c_i$. We note that the singularities at $c=z$ and $c=\bar{z}$ do not cause convergence problems when when determining $|\psi_1(c)\rangle$. This is because $|\psi_1(c)\rangle$ can be written as a linear combination of vectors $|\gamma'_n(c,z)\rangle$ that are analytic at $c=z$. Thus, to accelerate convergence near a branch point singularity, the EC subspace should contain all the eigenvectors whose Riemann sheets become intertwined at the branch point.

%%%%%%%%%%%%%%%%%%%%%%%%%%%%%%%%%%%%%%%%%%%%%%%%%%%%%%%%%%%%%%%%%%
\section{Algorithms}

The strategy of eigenvector continuation is to construct a set of wavefunctions for different values of the couplings (called the training data or training vectors,) which are used to form a ``low-dimensional" subspace that constrains the trajectory of $|\psi_j(c)\rangle$. We start with the smallest eigenvalue $E_1$ and corresponding eigenvector $|\psi_1(c)\rangle$ of the Hamiltonian $H(c)$. The approach will be to sample several values of the coupling $c = c_i$, $i = 1, \cdots, \mu$, where we use $\mu$ to denote the ``order" of the EC method. 

The reason for this is that in the limit of small steps between the $c_i$'s, the derivatives in the multi-series EC expansion can be approximated by finite differences, and the space spanned by the eigenvectors at various couplings will be approximately identical to the derivatives of the eigenvectors at zero coupling.

This approach gives us $\mu$ eigenvectors $|\psi_1(c_i)\rangle$ with which to form the EC subspace. The couplings $c_i$ are chosen to be such that $|\psi_1(c_i)\rangle$ can be accurately computed. If it is the case that $|\psi_1(c_i)\rangle$ cannot be accurately computed for any $c_i$, then eigenvector continuation will not be possible.

We assume that $c=c_\odot$ is the target value of the coupling, where we wish to determine $E_1(c_\odot)$ and $|\psi_1(c_\odot)\rangle$, and that direct computation of these quantities is not feasible.

The core algorithm of eigenvector continuation starts by forming a $\mu \times \mu$ norm matrix N by taking the inner products between each of the EC wavefunctions
\begin{equation}
N_{i,i'} = \left\langle \psi_1(c_{i'}) | \psi_1(c_i) \right\rangle
\end{equation}
and the matrix elements of $H(c_\odot)$
\begin{equation}
M_{i,i'} =  \left\langle \psi_1(c_{i'}) | H(c_\odot) | \psi_1(c_i) \right\rangle
\end{equation}
Then, we can solve the generalized eigenvalue problem, which consists of solving the equation
\begin{equation}
\hat{M} \vec{v} = \lambda \hat{N} \vec{v}
\end{equation}
The smallest eigenvalue $\lambda_0$ is the EC estimate for $E_1(c_\odot)$, and the estimate for $|\psi_1(c_\odot)\rangle$ can be constructed by the equation
\begin{equation}
|\psi_1(c_\odot)\rangle = P \vec{v}_1
\end{equation}
\noindent where $P$ is the projection operator, an $N \times \mu$ matrix whose columns are the training eigenvectors.

In the Monte Carlo simulations, it is sometimes advantageous to split $M$ into two pieces; one containing the base interactions $M_0$, and the other containing the term that has the coupling being varied $M_{\text{EC}}$. Then, we can compute the EC estimate for the energy for arbitrary target coupling $C_\odot$ by computing the generalized eigenvalues of 
\begin{equation}
(M_0 + C_\odot M_{\text{EC}}) \cdot \vec{v} = u \cdot N \cdot \vec{v}
\end{equation}
This can be done as a post-processing step, saving time during the simulation itself.

One issue that we run into while implementing the eigenvector continuation algorithm is in the inversion of the norm matrix required for the generalized eigenvalue problem. The norm matrix is constructed by taking the overlaps between each of the EC eigenvectors, a la Eqn. (3.19). If these overlaps are too close to unity, i.e. if the EC eigenvectors are too similar, then this norm matrix will become singular, and thus non-invertible. This provides a strict numerical bound to this technique.

To illustrate this problem, we take as an example the Lipkin model (discussed in Section 4.4). We perform a computation to 3rd order EC, computing 3 training eigenvectors with which we form our norm matrix. These three eigenvectors are taken to have couplings that are integer multiples of some step size, i.e. $C_i = 0\Delta C, 1\Delta C, ..., i\Delta C$. As we vary $\Delta C$, we expect the  overlaps between these wavefunctions to become smaller, and the norm matrix to become less singular. In Fig [3.6], we show the 2nd eigenvalue of this norm matrix.

\begin{figure}
\begin{center}
\makebox[\textwidth][c]{\includegraphics[width=\textwidth]{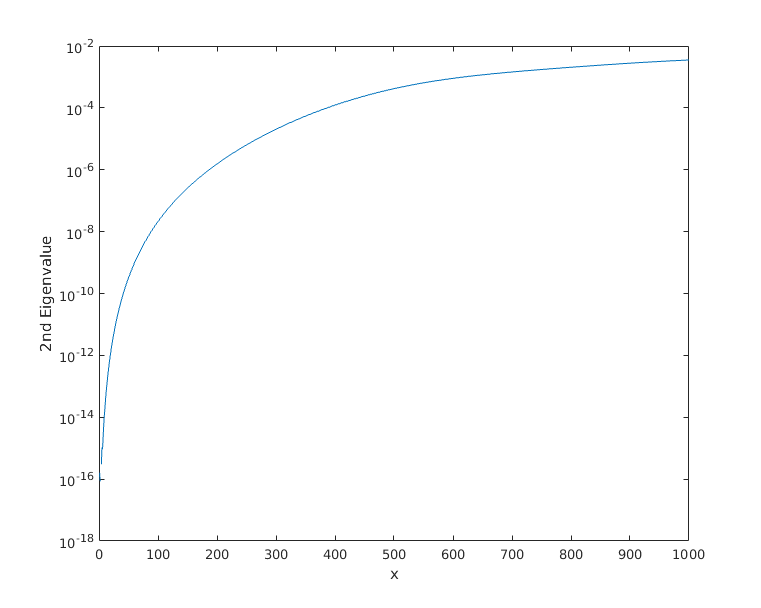}}
\caption[2nd eigenvalue of EC norm matrix ]{2nd Eigenvalue of EC norm matrix, as a function of the step size between values of the couplings used to generate the EC eigenvectors. Here, the coupling step size is $\Delta C = 0.001x$.}
\end{center}
\end{figure}

This second eigenvalue is a diagnostic about the invertibility of the norm matrix. It will tend towards zero if the matrix becomes singular, and will become larger as the overlaps between the EC eigenvectors become smaller. It is difficult to tell how small of an eigenvalue is ``too small"; in our experience, it depends on the system.

To demonstrate the effect this problem has on the extracted ground state wavefunctions and energies, we show a plot of the EC ground state energy estimate at a target coupling of $NV/\epsilon = 1.0$, s well as the overlaps of the exact ground state wavefunction with the EC estimate for the wavefunction. In Fig. [3.7], we show the overlaps of the two wavefunctions. In order to plot it in a readable way, we choose to plot one minus the overlap, $1 - <\psi_0|\psi_{EC}>$ on a log scale, so the estimate can be shown to get exponentially better as the norm matrix becomes less singular.

\begin{figure}
\begin{center}
\makebox[\textwidth][c]{\includegraphics[width=\textwidth]{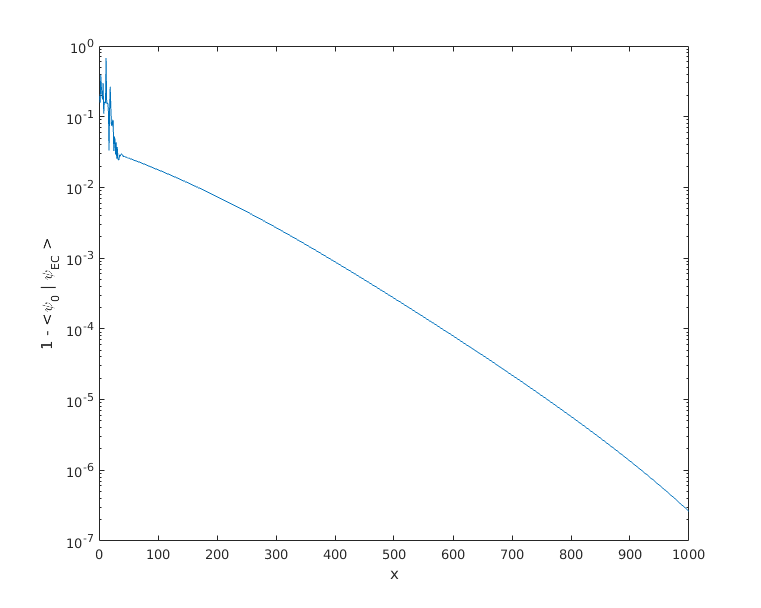}}
\caption[Overlap of EC wavefunction with exact ground state]{Overlap of EC wavefunction with exact ground state.}
\end{center}
\end{figure}

In Fig. [3.8], we show the estimates for the ground state energy of this system. If the norm matrix is singular or ill-conditioned, this estimate will be impossible or very noisy. Here, we see that the small $\Delta C$ estimates are very chaotic, and it takes a while to converge to a steady value.

\begin{figure}
\begin{center}
\makebox[\textwidth][c]{\includegraphics[width=\textwidth]{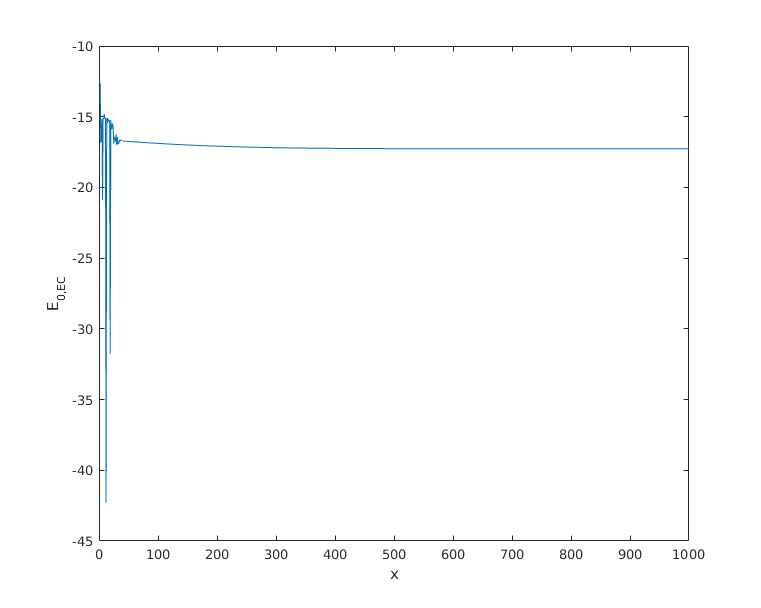}}
\caption[EC estimate for ground state energy vs EC coupling step size]{EC estimate for ground state energy vs EC coupling step size}
\end{center}
\end{figure}

\begin{figure}
\begin{center}
\makebox[\textwidth][c]{\includegraphics[width=\textwidth]{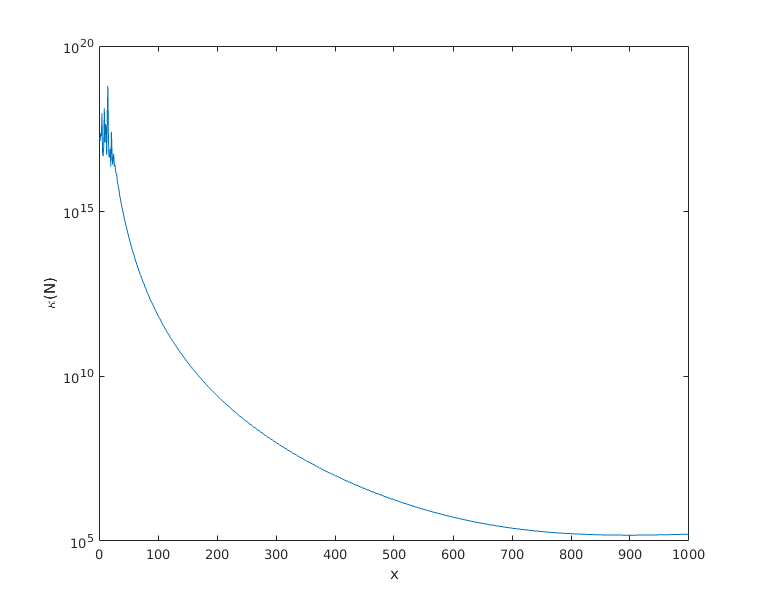}}
\caption[Condition Number of EC Norm Matrix vs $\Delta C$]{Here, we show the condition number of the norm matrix for 3rd order EC, in the Lipkin model. A condition number $\approx 1$ means the matrix can be inverted with no loss of precision, whereas a large condition number indicates numerical instability during the inversion procedure.}
\end{center}
\end{figure}

Another diagnostic that is useful in determining the stability of the algorithm is the condition number of the norm matrix. This condition number is defined as the ratio of the norms of the largest eigenvalue to the smallest eigenvalue
\begin{equation}
\kappa(N) = \frac{||\max{\lambda_i}||}{||\min{\lambda_i}||}
\end{equation}
\noindent which is denoted $\kappa$. This condition number usually indicates the loss of precision that any algorithm relying on inverting the matrix would have. $\kappa \approx 1$ would be able to be inverted with no loss of precision, whereas $\kappa \to \infty$ would lose all information. This definition is rather loose, but still can be helpful. In Fig. [3.9], we show the condition number of the same Lipkin model system, at 3rd order EC, plotted as a function of the EC coupling step size $\Delta C$.

Welford's online algorithm \cite{Welford1962:nmc} was used to extract more stable estimates for the average and error, compared to the standard Jackknife analysis. This algorithm is an online, iterative method in which running values for the mean $\bar{x}$ and variance $\sigma^2$ are kept track of, and updated for new samples $x$ are obtained
\begin{align}
\bar{x}_n &= \bar{x}_{n-1} + \frac{x_n - \bar{x}_{n-1}}{n} \nonumber \\
M_{2,n} &= M_{2,n-1} + (x_n + \bar{x}_{n-1})(x_n - \bar{x}_n) \\
\sigma^2_{n} &= \frac{M_{2,n}}{n-1} \nonumber
\end{align}
\noindent This algorithm is well-suited to run alongside Monte Carlo simulations, since averages and errors can be updated on the fly, rather than completely re-evaluated after each sample.

We now demonstrate how this algorithm is implemented in our Monte Carlo simulations. This is shown diagrammatically in Figs. [3.10] and [3.11]. First, we start by running the code in parallel, using MPI, such that we have $N$ processors (or cores) each running a copy of the code with different initial random seeds. As we generate Monte Carlo configurations, we group together a large number of configurations, and compute an average on each core (Step \#1). Assuming each Monte Carlo configuration is sampled from the same probability distribution, these averages is guaranteed by the central limit theorem to be normally distributed. We would then be able to compute a global average and variance/standard deviation from the data across all cores.

In the eigenvector continuation method, we need to do some computation before we average over the different cores, however. Here, we perform the needed eigenvalue solving, and extract the EC estimates for the ground states. Now, we have an estimate for the energy on each core. To begin applying the online algorithm, we will first average all of the EC estimates on each core into a single value (Step \#3). Then, Eqn. (3.25) tell us how to update the global moving average and population variance (Step \#4). After this step, we go back to Step \#1, and repeat until the average is converged and the errors are low.

\begin{figure}
\begin{center}
\makebox[\textwidth][c]{\includegraphics[width=\textwidth]{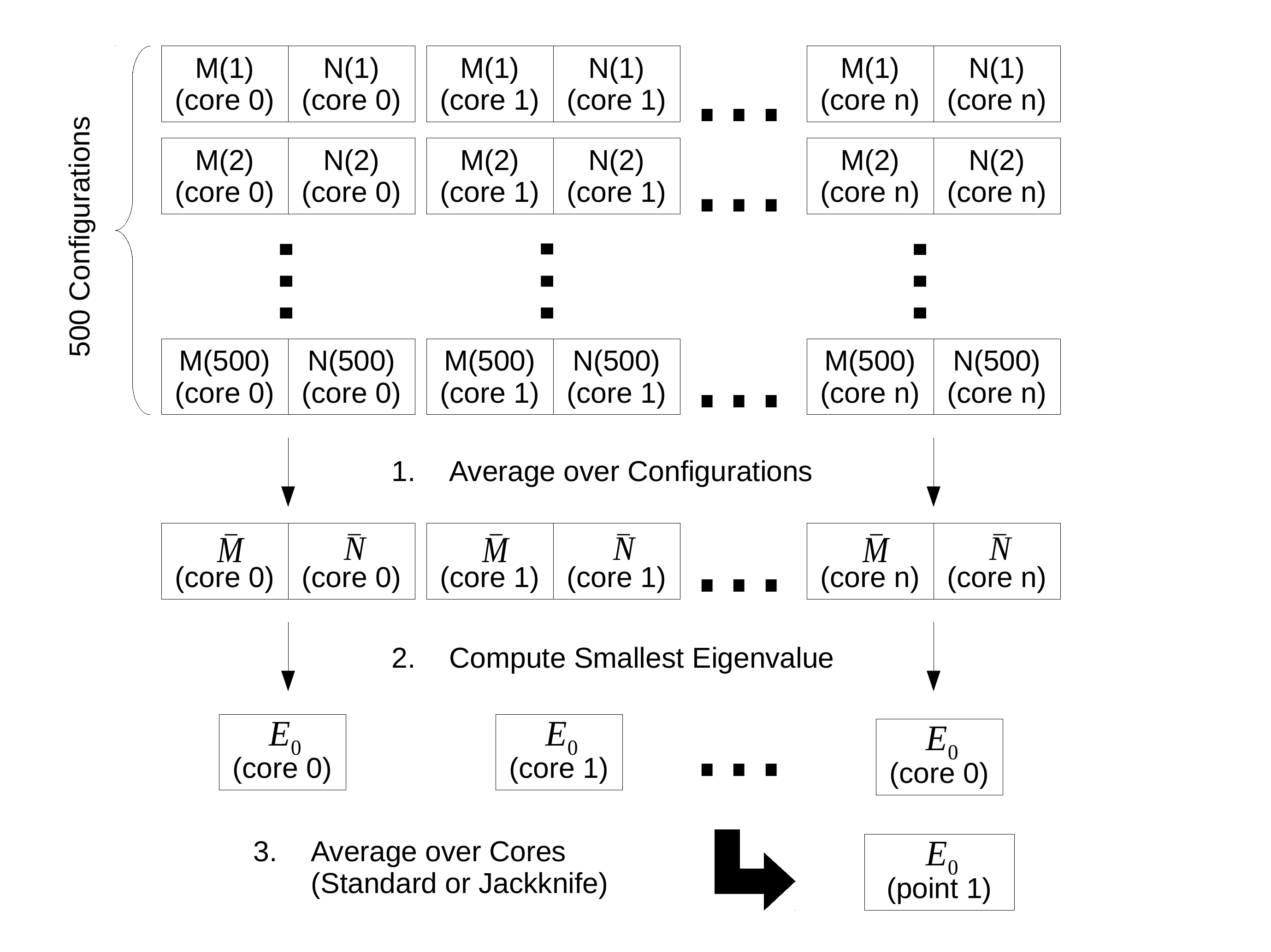}}
\end{center}
\caption[Implementation of online algorithm, part 1]{Steps \#1 - \#3 for the implementation of the online algorithm}
\end{figure}

\begin{figure}
\begin{center}
\makebox[\textwidth][c]{\includegraphics[width=\textwidth]{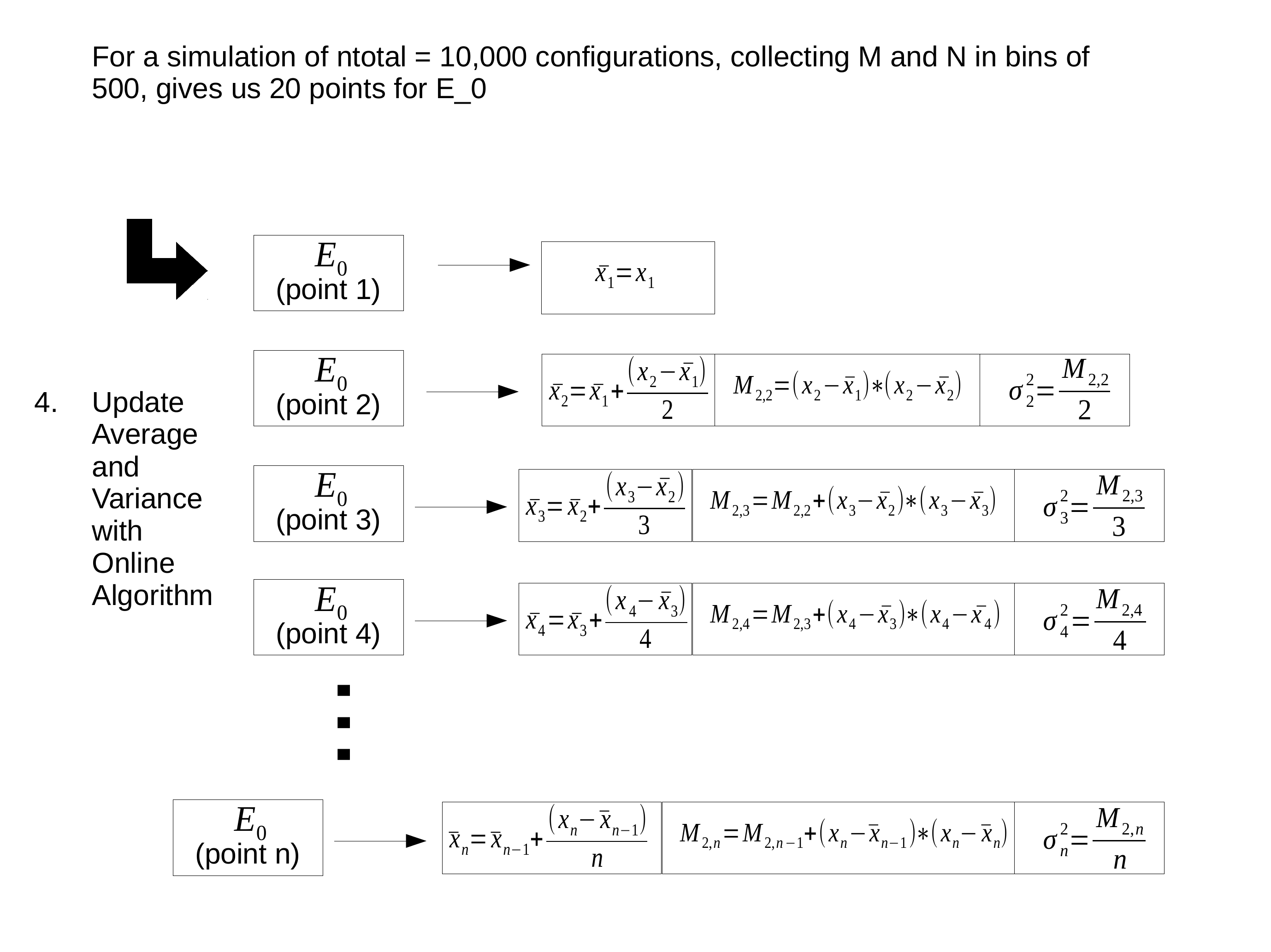}}
\end{center}
\caption[Implementation of online algorithm, part 2]{Step \#4 for the implementation of the online algorithm}
\end{figure}

In Fig. [3.12], we show a comparison of this method with the conventional jackknife analysis. We plot the estimates of the Coulomb energy shift in $^4$He, computed perturbatively (COUL) and via direct calculation (EC), as a function of the number of Monte Carlo steps. On the top graph, both estimates are computed via a jackknife analysis over the different CPUs. On the bottom graph, the same perturbative result is plotted, but the EC estimate is averaged via the online algorithm.

\begin{figure}
\begin{center}
\makebox[\textwidth][c]{\includegraphics[width=0.9\textwidth]{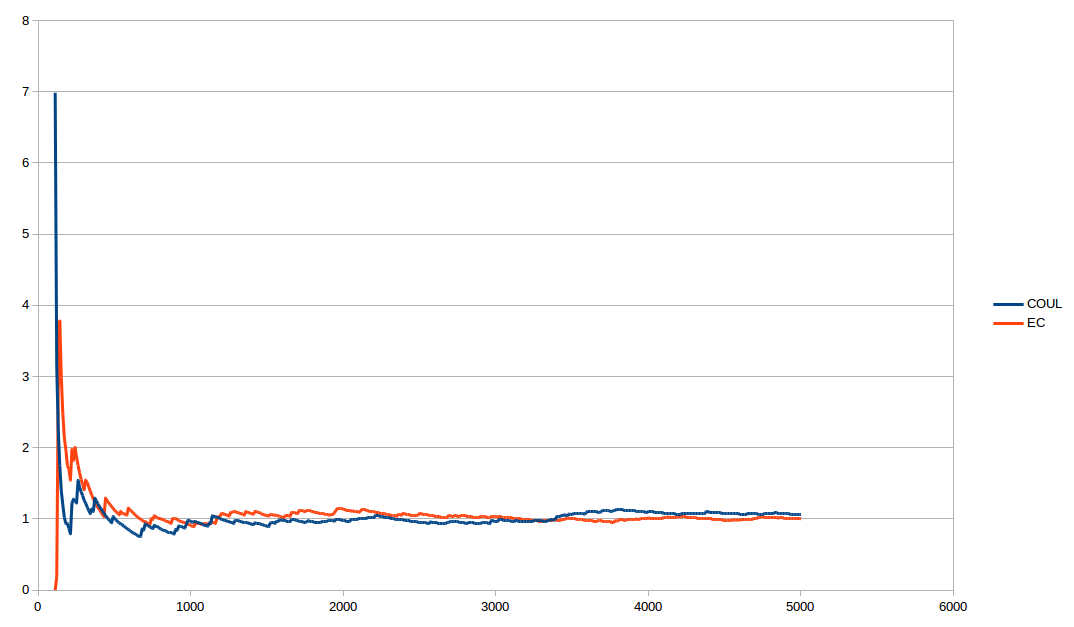}}
\makebox[\textwidth][c]{\includegraphics[width=0.9\textwidth]{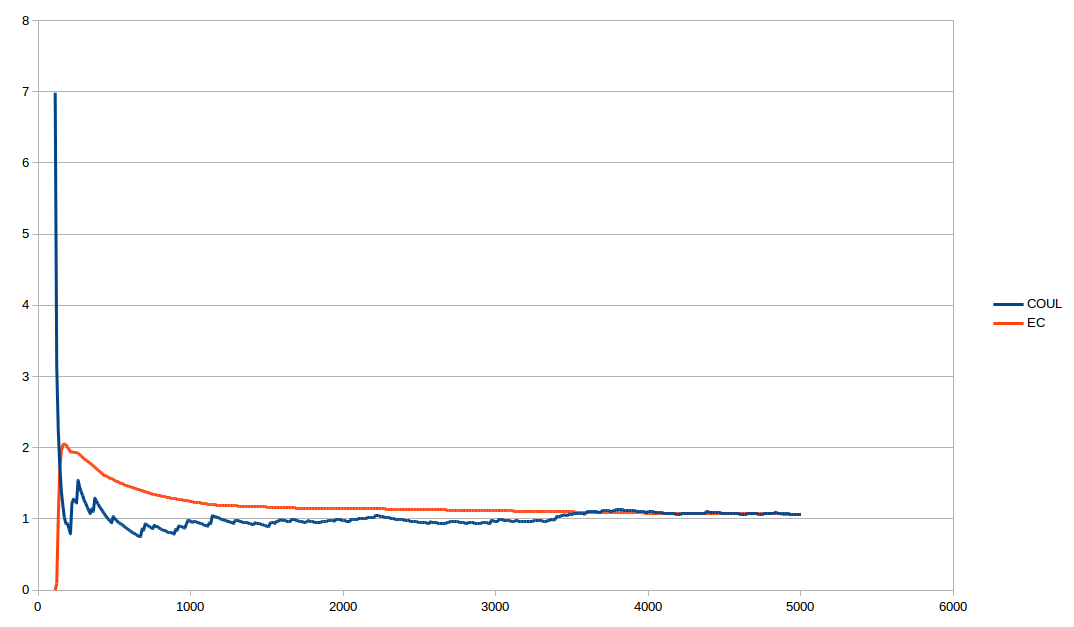}}
\caption[Perturbative and EC Coulomb energies, jackknife vs online]{Plots of the perturbative Coulomb energy shift in $^4$He computed with a jackknife analysis (in blue), versus the eigenvector continuation estimate for the Coulomb energy shift (orange). On the top plot, the EC estimate was done with a jackknife analysis, and on the bottom, it was done using the online algorithm.}
\end{center}
\end{figure}

%%%%%%%%%%%%%%%%%%%%%%%%%%%%%%%%%%%%%%%%%%%%%%%%%%%%%%%%%%%%%%%%%%

\section{Implementation and Computational Cost}

For the most part, Monte Carlo results detailed in this work were done using a modified version of the MCLEFT code \cite{Lu:2018een}, developed by B. Lu (MSU). The modifications were done in Fortran 95, and consisted of a module file containing all the necessary subroutines.

In these codes, the bulk of the work is in generating the auxiliary field configurations, and propagating the initial single-particle wavefunctions through Euclidean time via transfer matrices. These single particle wavefunctions have a spin index, isospin index, and in the modified code, an EC order index. From this we, expect the computational cost of the eigenvector continuation code to be directly proportional to the EC order.

At the middle timestep, we insert various operators to perturbatively calculate various quantities, like energies from higher order Chiral forces, isospin symmetry breaking terms, Galilean invariance restoration terms, etc. These involve forming Slater determinants out of the propagated single-particle states. Due to the size of these determinants, we only store their logarithm. To extract the EC estimates for the energy, we use the BLAS routine DSYGV. Since our projected matrix is a Hamiltonian matrix, we only keep the real part, since the imaginary parts should trend to zero.

\begin{figure}
\begin{center}
\makebox[\textwidth][c]{\includegraphics[width=0.8\textwidth]{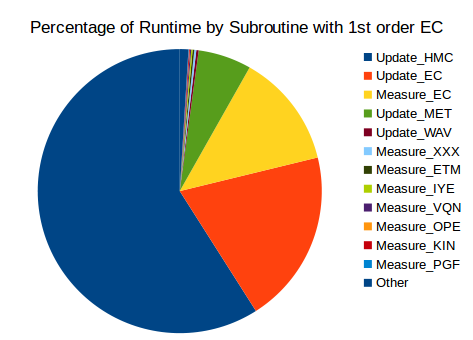}}
\makebox[\textwidth][c]{\includegraphics[width=0.8\textwidth]{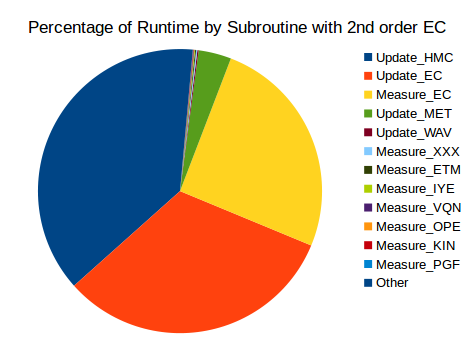}}
\caption[\textit{gprof} profiling of MCLEFT code with 1st/2nd order EC]{\textit{gprof} breakdown of the computational time spent in major subroutines in the MCLEFT code. At 1st order, the overall time spent in the EC subroutines is about 32\%, and at 2nd order, that increases to 56\%}
\end{center}
\end{figure}

In Fig. [3.13], we show the results of profiling this code with \textit{gprof} for a fixed number of MC steps. UPDATE routines involve generation of the auxiliary fields and propagation of the wavefunctions through time. MEASURE routines involve the computation of observables using the wavefunctions at the center time-step. Subroutines with the suffix EC are routines used in the eigenvector continuation method. From here, we see a rapid growth in the overall time used by the EC method by the time that we have reached 2nd order.

\begin{table}
\centering
\begin{tabular}{|c|c|c|c|}
\hline
Routine & 1st Order EC & 2nd Order EC & 3rd Order EC \\
\hline
Update\_HMC & 67.79 & 66.37 & 67.52 \\
Update\_EC & 22.74 & 58.29 & 69.96 \\
Measure\_EC & 14.93 & 46.09 & 132.53 \\
Update\_MET & 6.96 & 6.76 & 6.85 \\
Update\_WAV & 0.31 & 0.27 & 0.33\\
Measure\_XXX & 0.26 & 0.25 & 0.25 \\
Measure\_ETM & 0.22 & 0.21 & 0.21 \\
Measure\_IYE & 0.18 & 0.17 & 0.17 \\
Measure\_VQN & 0.15 & 0.13 & 0.18 \\
Measure\_OPE & 0.03 & 0.02 & 0.03 \\
Measure\_KIN & 0.12 & 0.12 & 0.12 \\
Measure\_PGF & 0.06 & 0.02 & 0.02 \\
Other & 1.12 & 2.62 & 4.90 \\
\hline
Total & 114.63 & 180.08 & 280.08 \\
\hline
\end{tabular}
\caption[\textit{gprof} profiling results]{Tabulated results from \textit{gprof}}
\end{table}

This rapid scaling is in part due to the exponential growth in the number of combinations you can form the eigenvectors into subspaces with. If we compute $N$ eigenvectors, and form a subspace out of $m$ of them, there are
\begin{equation}
\text{\# of Choices} = \binom{N}{m} = \frac{N!}{m!(N-m)!}
\end{equation}
\noindent number of choices, which escalates for large $N$. However, if we only pick a particular set of the $N$ eigenvectors to form the subspace out of, the computation time spent can be reduced.

We now show how the EC algorithm scales as a function of the lattice volume. We run the code for  fixed number of Monte Carlo steps (100 in this case), a fixed number of particles ($N_f$ = 4), and a fixed size of the lattice in the temporal direction ($L_{t,in} = 20$,$L_{t,out} = 20$). Here, we expect the primary term to be proportional to the lattice volume $L^3$. We take our fitting function to be a cubic:
\begin{equation}
T(L) = aL^3 + bL^2 + cL
\end{equation}
\noindent with the y-intercept set to zero. This fit is done using a least-squares regression.

\begin{figure}
\begin{center}
\makebox[\textwidth][c]{\includegraphics[width=\textwidth]{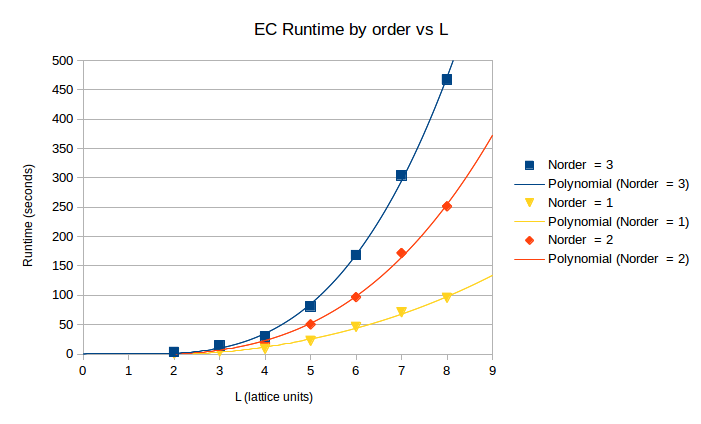}}
\caption[Run-time of the EC method vs L]{Plot of the computational cost/time used by the EC method, as a function of the length of the edges the lattice L. This was done for a fixed number of particles (N=4), and  a fixed number of timesteps (Ltinner = 20, Ltouter = 20).}
\end{center}
\end{figure}

\begin{table}
\centering
\begin{tabular}{|c|c|c|c|c|}
\hline
EC Order & $a$ (cubic term) & $b$ (quadratic term) & $c$ (linear term) & $R^2$ \\
\hline
1st & 0.0729667392 & 1.4274807737 & -3.8672850799 & 0.9972665119 \\
2nd & 0.6035009969 & -0.7136869805 & -1.0474019912 & 0.9990421919 \\
3rd & 1.4201400218 & -4.4940652917 & 4.021016857 & 0.9995114091 \\
\hline
\end{tabular}
\caption[Polynomial time scaling of EC algorithm vs L]{Coefficients of best polynomial fits to computational cost of the EC algorithm, as a function of the size of the lattice L. Since there are $L^3$ points, we expect the leading term to be proportional to $L^3$. We don't expect other terms, but present them for completeness.}
\end{table}

We now look at the scaling of the algorithm with respect to the number of particles $N_f$. Like before, we fix the number of Monte Carlo steps (100), the spatial size of the lattice ($L = 4$), and the size of the lattice in the temporal direction ($L_{t,in} = 20$,$L_{t,out} = 20$). This time, we expect a small term proportional to $N_f^2$, and a larger term proportional to $N_f$. We then take our fitting function to be a quadratic:
\begin{equation}
T(N_f) = aN_f^2 + bN_f
\end{equation}
\noindent with the intercept set to zero as well.

\begin{figure}
\begin{center}
\makebox[\textwidth][c]{\includegraphics[width=\textwidth]{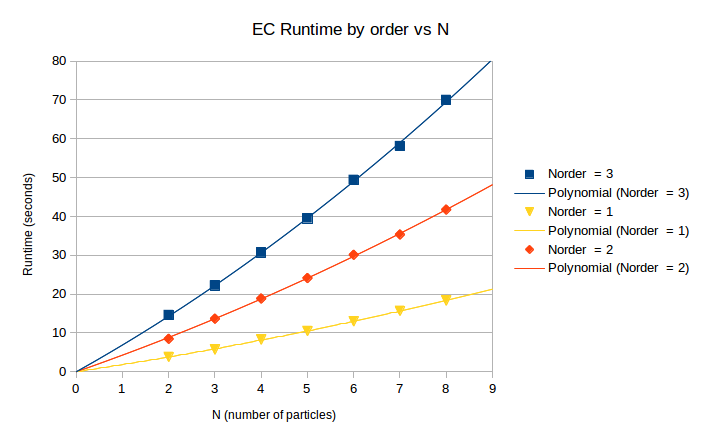}}
\caption[Run-time of the EC method vs N]{Plot of the computational cost/time used by the EC method, as a function of the number of particles in the system. This was done for fixed lattice volume (L=4), and a fixed number of timesteps (Ltinner = 20, Ltouter = 20).}
\end{center}
\end{figure}

\begin{table}
\centering
\begin{tabular}{|c|c|c|c|}
\hline
EC Order & $a$ (quadratic term) & $b$ (linear term) & $R^2$ \\
\hline
1st & 0.0649337121 & 1.7797578463 & 0.9999319128 \\
2nd & 0.1325730519 & 4.1616152597 & 0.9999292763 \\
3rd & 0.255174513 & 6.6464975649 & 0.9999057979 \\
\hline
\end{tabular}
\caption[Polynomial time scaling of EC algorithm vs $N_f$]{Coefficients of best polynomial fits to computational cost of the EC algorithm, as a function of the number of particles in the system $N_f$. We expect a small term proportional to $L^2$, from the matrix elements, and a large linear term, since we keep track of single-particle wavefunctions.}
\end{table}

%%%%%%%%%%%%%%%%%%%%%%%%%%%%%%%%%%%%%%%%%%%%%%%%%%%%%%%%%%%%%%%%%%
%%%%%%%%%%%%%%%%%%%%%%%%%%%%%%%%%%%%%%%%%%%%%%%%%%%%%%%%%%%%%%%%%%
%%%%%%%%%%%%%%%%%%%%%%%%%%%%%%%%%%%%%%%%%%%%%%%%%%%%%%%%%%%%%%%%%%
\chapter{Applications to Nuclear Systems}

In this chapter, we discuss all the applications of eigenvector continuation to nuclear systems that have been investigated so far. We start with the initial investigations into the Bose-Hubbard model and infinite neutron matter. Then, we discuss the main project of computing the contribution of the Coulomb interaction to the ground state energy in light nuclei. Then, we look at a much older model of nuclear structure; the Lipkin-Meshkov-Glick Hamiltonian.

%%%%%%%%%%%%%%%%%%%%%%%%%%%%%%%%%%%%%%%%%%%%%%%%%%%%%%%%%%%%%%%%%%

\section{Bose-Hubbard Model}

The Bose-Hubbard model in three dimensions \cite{Gersch:1963qcm} consists of a system of identical bosons on a three-dimensional cubic lattice. The Hamiltonian has a hopping term, controlling the nearest-neighbour hopping of each boson, a potential term proportional to U controlling the pairwise interaction between bosons on the same site, and a chemical potential $\mu$, and it has the form
\begin{equation}
H = -t\sum_{\langle \textbf{n},\textbf{n'} \rangle} a^\dagger (\textbf{n'}) a(\textbf{n}) + \frac{U}{2}\sum_{\textbf{n}} \rho(\textbf{n}) \left[ \rho(\textbf{n}) - \textbf{1} \right] - \mu \sum_{\textbf{n}} \rho(\textbf{n})
\end{equation}
where $a^\dagger (\textbf{n})$ and $a(\textbf{n})$ are the creation and annihilation operators for bosons at the lattice site $\textbf{n}$, the first summation is over nearest-neighbour pairs $\langle \textbf{n},\textbf{n'} \rangle$, and $\rho(\textbf{n}) = a^\dagger (\textbf{n}) a(\textbf{n})$ is the density operator.

\begin{figure}
\begin{center}
\makebox[\textwidth][c]{\includegraphics[width=0.7\textwidth]{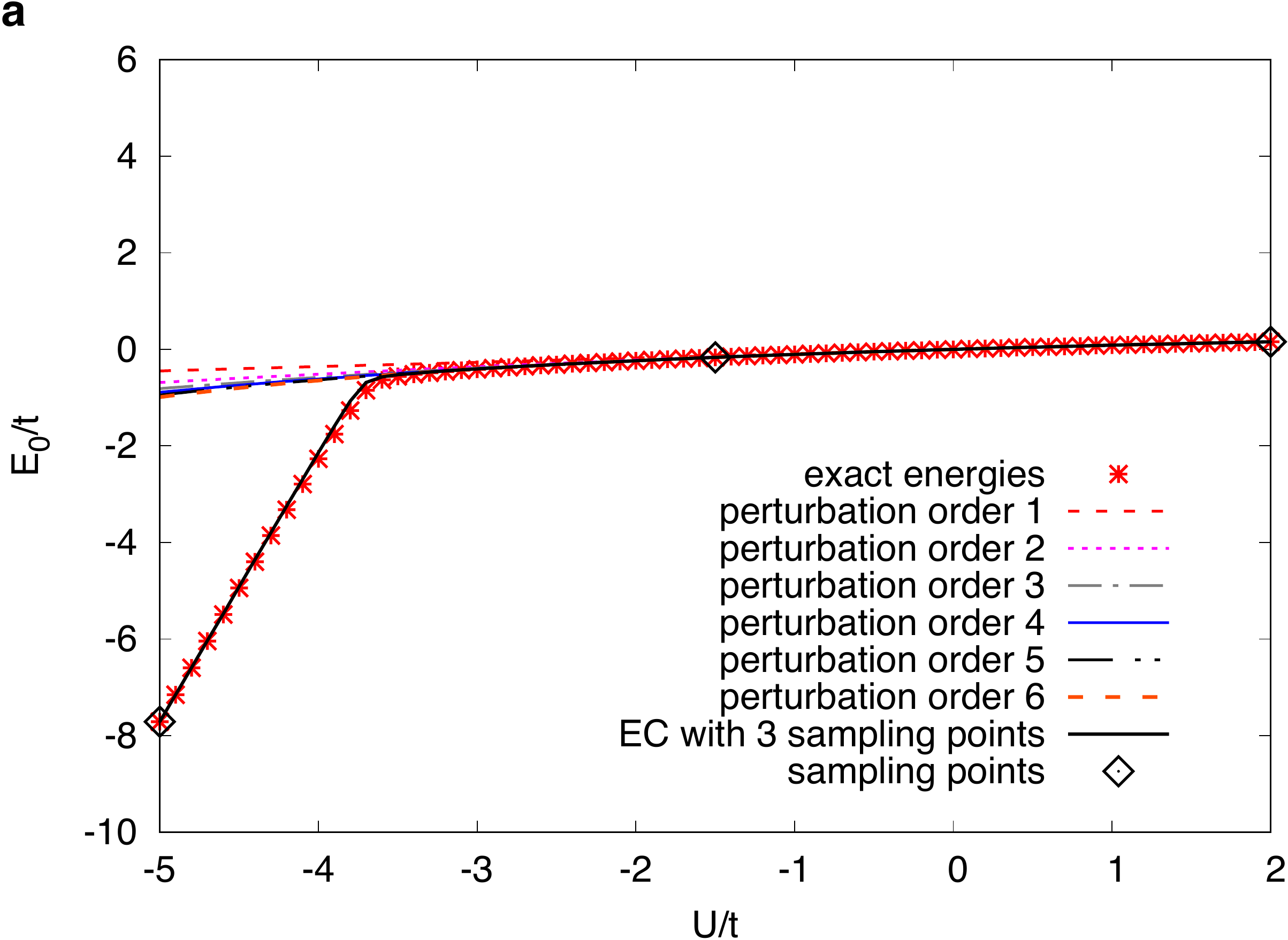}}
\caption[Bose-Hubbard model results with perturbation theory]{Ground state energy for the Bose-Hubbard model, computed by exact diagonalization and by perturbation theory.}
\end{center}
\end{figure}

We consider a system of four particles on a $4\times4\times4$ lattice. To illustrate why we chose this problem, we first try using perturbation theory to find the ground state energy eigenvalue $E_0$ in units of the hopping parameter $t$. In Fig. [4.1], we show the ground state energy $E_0/t$, computed directly using exact diagonalization and up to 6th order in perturbation theory, as functions of $U/t$. For the perturbative results, we expanded in a power series about the point $U/t = 0.0$. We see immediately that there is a critical point at about $U/t = -3.8$, where perturbation theory fails to converge. We note that while perturbation theory seems to converge to the wrong value, at higher orders, it does indeed diverge at higher orders. This divergence is due to an avoided level crossing, where the ground state eigenvalue is merging with the 1st excited state eigenvalue.

It makes some sense that perturbation theory fails, since the physical nature of the ground state wavefunction is changing dramatically across this critical point. It is a Bose gas for $U/t > 0$, a weakly-state for $ -3.8 < U/t < 0$, and a strongly-bound cluster for $U/t < -3.8$.

Eigenvector continuation can enter the picture, once we realise that even though the eigenvector is changing in a linear space of huge dimension, the path it traces out only moves significantly in a small number of orthogonal directions.

\begin{figure}
\begin{center}
\makebox[\textwidth][c]{\includegraphics[width=0.7\textwidth]{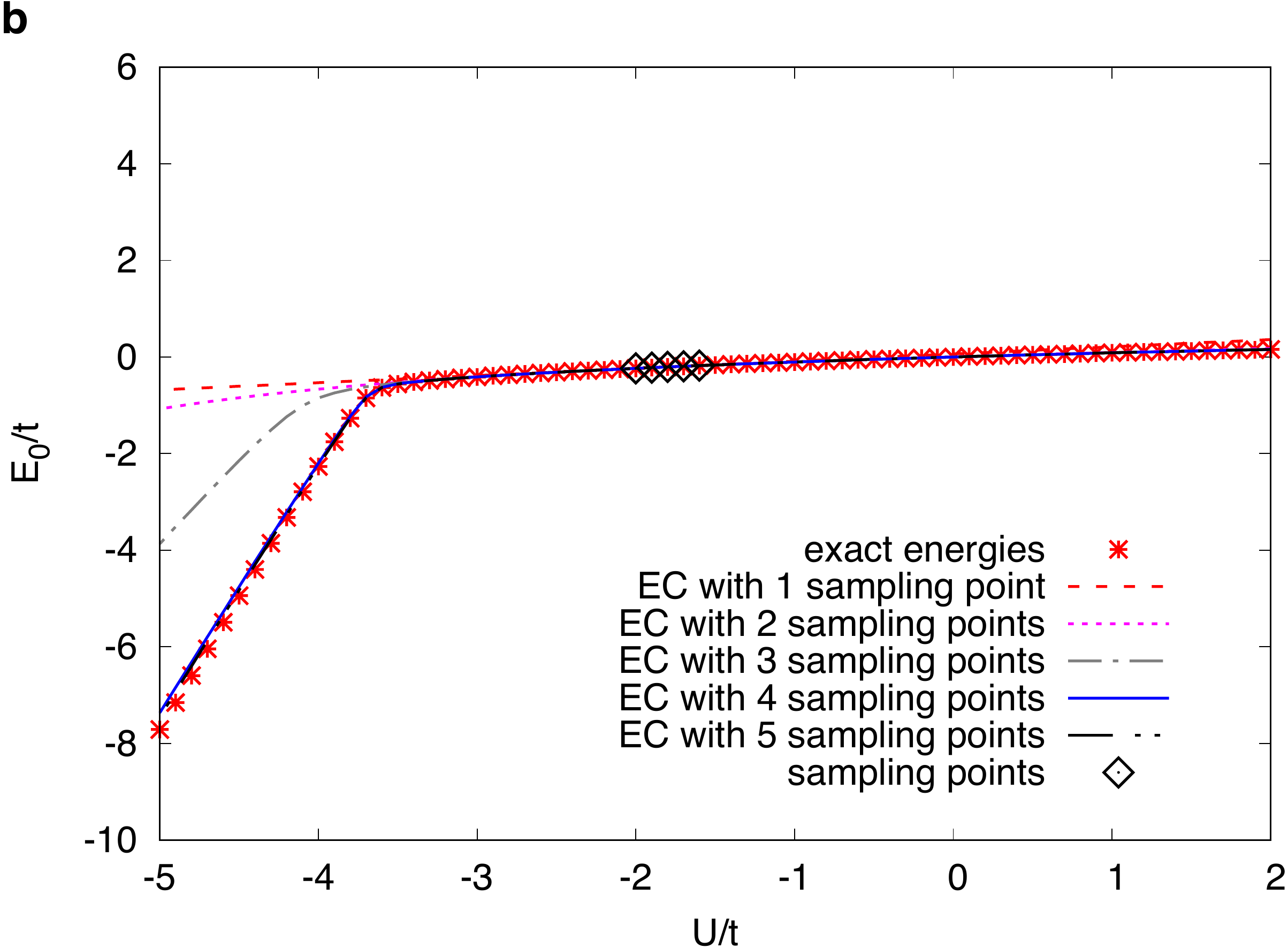}}
\caption[Bose-Hubbard model results with eigenvector continuation]{Ground state energy of the Bose-Hubbard model, done with up to 5th order eigenvector continuation (5 training vectors.)}
\end{center}
\end{figure}

To illustrate this, we consider 3rd order eigenvector continuation, in which we take three sampling points $U/t = -5.0, -1.5, 2.0$. Using the ground state wavefunctions for each coupling, and using the procedure from Section 2.2, we can extract the EC estimate for $E_0$. This result is shown also in Fig. [4.1].

At first glance, it may appear as though we got lucky. After all, we sampled values of the coupling from each of the three "physically-different" regions, so it seems obvious that an expansion using linear combinations of these three wavefunctions would cover the full range of couplings. However, we can sample couplings from entirely within one of the regions, say from the region $ U/t = -2.0 $ to$ -1.6$. If we now consider eigenvector continuation with up to 5 vectors, from the couplings $U/t = -2.0, -1.9, -1.8, -1.7, -1.6$, we get the results in Fig. [4.2]. We see that even in this case, at high enough order, the EC result is able to capture the kink in $E_0$ at the point $U/t \approx -3.8$.

\begin{figure}	% FILLER PLOT
\begin{center}
\makebox[\textwidth][c]{\includegraphics[width=0.8\textwidth]{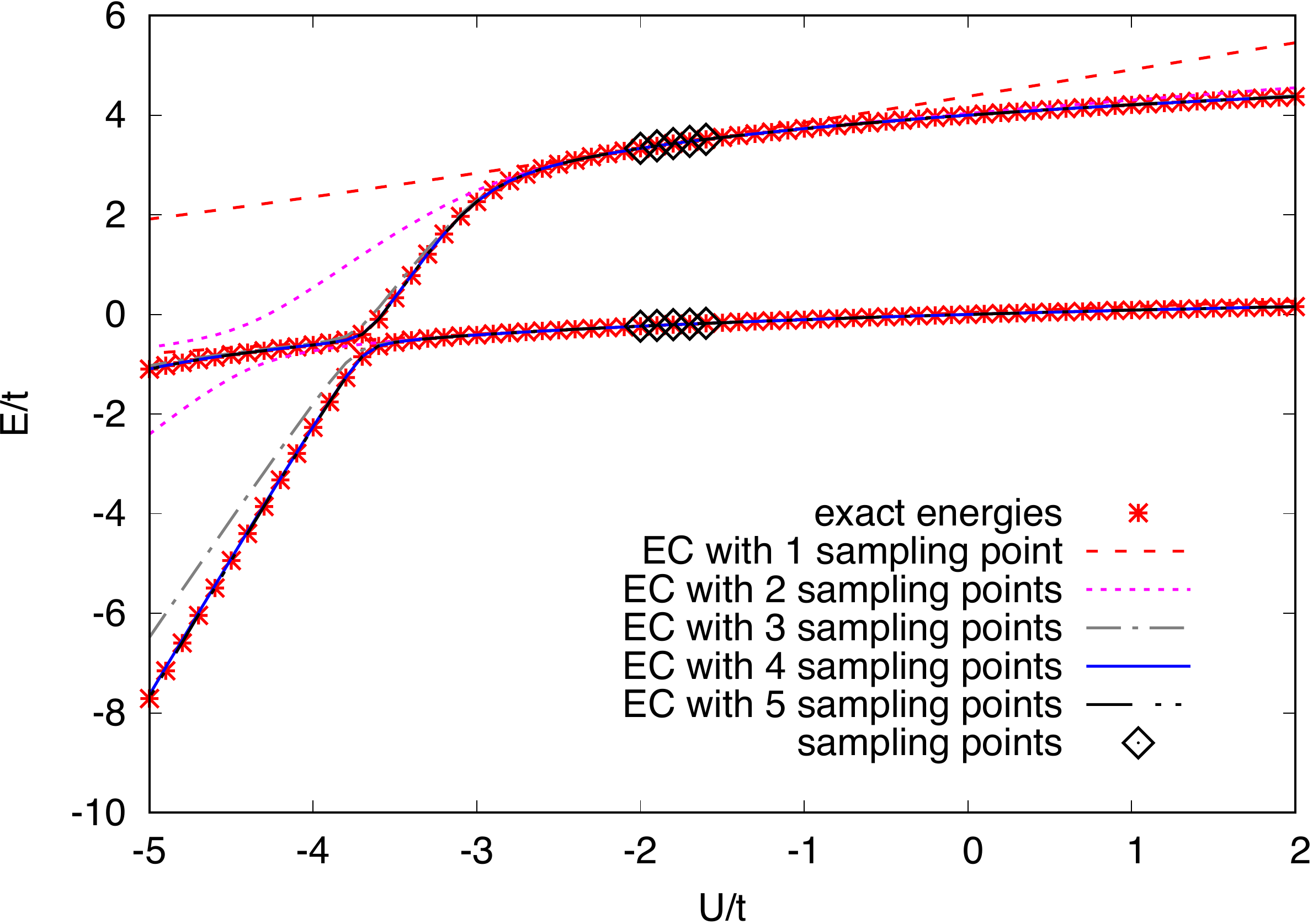}}
\caption[Lowest two states of Bose-Hubbard model, with EC results]{Computation of the lowest two energies in the Bose-Hubbard model. Here, the lowest two eigenvectors at each training coupling were kept and used to construct the EC subspace.}
\end{center}
\end{figure}

In Section 3.1, we discussed how including multiple eigenvectors at each sampling point can accelerate the convergence of the EC results. To demonstrate this, we now wish to compute the two lowest, even-parity states of the Bose-Hubbard system. These results are shown in Fig. [4.3]. We can now clearly see the avoided level crossing. The couplings used are still $U/t = -2.0, -1.9, -1.8, -1.7, -1.6$, however, we now take the two lowest eigenvectors at each value for the coupling, and use these to form our EC subspace. Comparing with the results of only including one vector per sampling point, we see that the convergence has been improved; the third order result is much closer, and the fourth/fifth order results are in quite good agreement with the exact energies.

%%%%%%%%%%%%%%%%%%%%%%%%%%%%%%%%%%%%%%%%%%%%%%%%%%%%%%%%%%%%%%%%%%
\section{Simulations of Neutron Matter}

The second system we studied was the system of pure neutron matter. The objective was to demonstrate the use of eigenvector continuation to a full many-body Quantum Monte Carlo simulation. We considered systems of six or fourteen neutrons at leading order (LO) in chiral effective field theory. While there exist computationally-friendly lattice actions, as described in \cite{Elhatsari:2017aic,Elhatisari:2016nds}, the more demanding lattice action \cite{Hjorth-Jensen:2017acn} was instead chosen to serve as a better test for the method.

Since we are considering only neutrons, we can simplify the lattice action by reducing the number of independent contact interactions from two to one. In the following equations, we use $a_i\left(\textbf{n}\right)$ and  $a^{\dagger}_i\left(\textbf{n}\right)$ to denote the fermion annihilation and creation operators, with spin component $i$ located at the lattice site $\textbf{n}$. We use the shorthand $a\left(\textbf{n}\right)$ and  $a^{\dagger}\left(\textbf{n}\right)$ to denote the column vector of nucleon components and row vector of nucleon components, respectively. We use a spatial lattice spacing of $a$ = 1/100 MeV$^{-1}$ = 1.97 fm, and a temporal lattice spacing of $a_t$ = 1/150 MeV$^{-1}$ = 1.32 fm. We also set $\hbar = c = 1$.

Our free, non-relativistic lattice Hamiltonian is given by
\begin{equation}
H_{\text{free}}\left( a^{\dagger}, a \right) = \sum_{k=0,1} \frac{\left(-1\right)^k}{2m}\sum_{\textbf{n}}\sum_{l=1,2,3} a^{\dagger}\left(\textbf{n}\right) \left[ a\left(\textbf{n} + k\hat{\textbf{l}}\right) + a\left(\textbf{n} - k\hat{\textbf{l}}\right) \right]
\end{equation}
where $\hat{\textbf{l}} = \hat{\textbf{1}},\hat{\textbf{2}},\hat{\textbf{3}}$ denote the three spatial lattice unit vectors. The nucleon mass $m$ is taken to be 938.92 MeV. 

We use the auxiliary field formalism to include the interactions between the neutrons. We then express the Euclidean time-evolution operator over $L_t$ time steps as a product of transfer matrix operators which depend on the auxiliary field $s$ and the pion field $\pi_0$.  
\begin{equation}
U\left(L_t,g_A^2\right) = \int Ds D\pi_0 \exp \left[ -\mathcal{S}_{SS}\left(s\right) - \mathcal{S}_{\pi_0\pi_0}\left(\pi_0\right) \right] \left\lbrace M^{\left(L_t-1\right)} \cdots M^{\left(0\right)} \right\rbrace
\end{equation}
\noindent For convenience, we have explicitly listed the dependence of this operator on the square of the axial vector coupling $g_A^2$. The quadratic part of the auxiliary field action is
\begin{equation}
\mathcal{S}_{SS}\left(s\right) = \frac{1}{2} \sum_{\textbf{n},n_t} s^2\left(\textbf{n},n_t\right)
\end{equation}
\noindent and the quadratic part of the pion field action is
\begin{align}
\mathcal{S}_{\pi_0\pi_0}\left(\pi_0\right) &= \frac{1}{2}\alpha_t m_\pi^2 \sum_{\textbf{n},n_t} \pi_0^2 \left(\textbf{n},n_t\right) \\
& + \frac{1}{2}\alpha_t \sum_{k=0,1} \left(-1\right)^k \sum_{\textbf{n},n_t}\sum_{l=1,2,3} \pi_0\left(\textbf{n},n_t\right) \left[ \pi_0(\textbf{n} + k\hat{\textbf{l}},n_t) + \pi_0(\textbf{n} - k\hat{\textbf{l}},n_t) \right]
\end{align}
\noindent with the pion mass $m_\pi$ is taken to be 134.98 MeV.

The normal-ordered auxiliary field transfer matrix, for a given time step $n_t$, is given by
\begin{equation}
M^{\left(n_t\right)} = : \exp \left[ -H^{\left(n_t\right)}(a^\dagger, a, s, \pi_0) \alpha_t \right] :
\end{equation}
\noindent The Hamiltonian at the time step $n_t$ is given by
\begin{equation}
H^{\left(n_t\right)}(a^\dagger, a, s, \pi_0) \alpha_t = H_{\text{free}}\left(a^\dagger, a\right) \alpha_t + S_s^{\left(n_t\right)}\left(a^\dagger, a, s\right) + S_{\pi_0}^{\left(n_t\right)}\left(a^\dagger, a, \pi_0\right)
\end{equation}
\noindent where
\begin{align}
&S_s^{\left(n_t\right)}(a^\dagger, a, s) = \sqrt{-C \alpha_t} \sum_{\textbf{n}} s(\textbf{n},n_t)a^\dagger(\textbf{n})a(\textbf{n}) \\
&S_{\pi_0}^{\left(n_t\right)}(a^\dagger, a, \pi_0) = \frac{g_A \alpha_t}{2 f_\pi} \sum_{\textbf{n}} \sum_{l=1,2,3} \frac{1}{2} [ \pi_0(\textbf{n} + k\hat{\textbf{l}},n_t) + \pi_0(\textbf{n} - k\hat{\textbf{l}},n_t) ] a^\dagger(\textbf{n}) \sigma_l a(\textbf{n})
\end{align}
\noindent Here, C is the coupling of the contact interaction, and is taken to be $-4.5\times10^{-5}$ MeV$^{-2}$. This is different from the value used in \cite{Hjorth-Jensen:2017acn}; however, this value was chosen to reproduce a more realistic equation of state for neutron matter at the densities being probed in our simulations. $\sigma_l$ denote the Pauli spin matrices. $f_\pi$ is the pion decay constant, which we take to be 92.2 MeV. $g_A$ is the axial-vector coupling, taken as 1.29. After integrating over the pion fields, we obtain a one-pion exchange potential that is quadratic in $g_A$. 

\begin{figure}
\begin{center}
\makebox[\textwidth][c]{\includegraphics[width=0.8\textwidth]{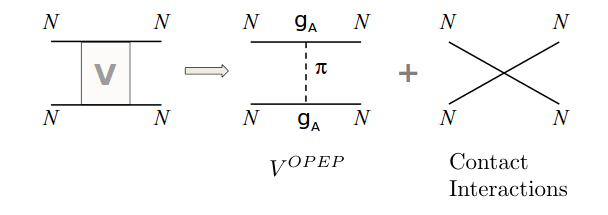}}
\caption[LO chiral forces used in neutron matter simulations]{LO chiral forces used in neutron matter simulations. Included interactions include a contact term and one-pion exchange}
\end{center}
\end{figure}

We performed simulations of six and fourteen neutrons in a 4$^3$ box, extracting the ground state wavefunctions. In physical units, L = 7.9 fm, and the corresponding number densities are 0.012 fm$^3$ for six neutrons and 0.028 fm$^3$ for fourteen neutrons. Our initial and final states are taken as free Fermi gas wavefunctions, which we denote $|\Phi\rangle$. We use Euclidean time projection to produce the ground states for given values of $g_A^2$, which we denote $|\Phi,n_t,g_A^2\rangle$.
\begin{equation}
|\Phi,n_t,g_A^2\rangle = U\left(n_t,g_A^2\right)|\Phi\rangle
\end{equation}
\noindent For large $n_t$, $|\Phi,n_t,g_A^2\rangle$ will asymptotically approach the ground state.  

First, we present the results of direct calculation without eigenvector continuation. In these calculations, we perform Monte Carlo simulations for the pion and auxiliary fields, and calculate the ratio
\begin{equation}
r\left(L_t\right) = \frac{\langle \Phi | U\left(L_t,g_A^2\right) | \Phi \rangle}{\langle \Phi | U\left(L_t - 1,g_A^2\right) | \Phi \rangle}
\end{equation}
\noindent in the large $L_t$ limit. This ratio is converted into an energy by
\begin{equation}
E(L_t) = -\log\lbrace r(L_t) \rbrace / a_t \approx E_0 + c e^{-\delta E L_t a_t}
\end{equation}

In Fig. [4.5], we show the direct result for the calculation of six neutrons. The red open squares show the lattice results versus projection time, including the one-$\sigma$ error bars reflecting the stochastic errors in the simulation. From the asymptotic form in Eqn. [4.13], the best fit is shown as a red, solid line, and the limits of the one-standard-deviation error bars are shown as red dashed lines. In Fig. [4.6], we show the results of the direct calculation for fourteen neutrons. The red open squares show the lattice results versus projection time, including one-standard-deviation error bars indicating stochastic errors. The best fit is shown as a red solid line, and the limits of the one-standard-deviation error bands are shown as red dashed lines. Due to large sign oscillations, it is not possible to do simulations at large projection times. This is reflected in the large uncertainties of the ground state energy extrapolations.

\begin{figure}
\begin{center}
\makebox[\textwidth][c]{\includegraphics[width=0.9\textwidth]{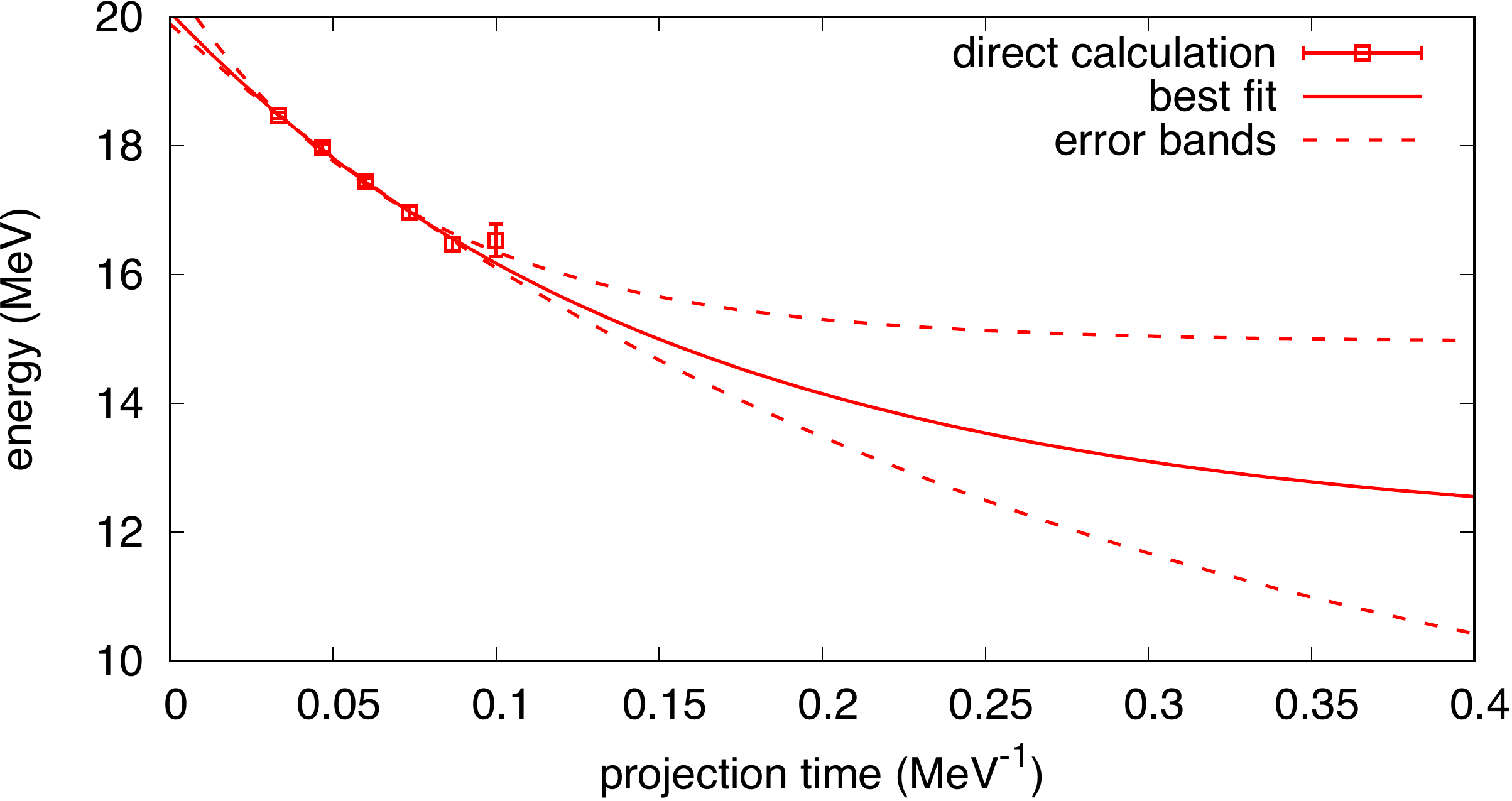}}
\caption[Direct calculation of six neutrons (L = 8 fm)]{Direct calculation of the ground state energy for six neutrons, shown as a function of Euclidean time. The red, open squares are the lattice results, and the solid red line is the best fit showing the exponential decay of the signal. The dashed red line is the 1$\sigma$ error band for the fit.}
\end{center}
\end{figure}

\begin{figure}
\begin{center}
\makebox[\textwidth][c]{\includegraphics[width=0.9\textwidth]{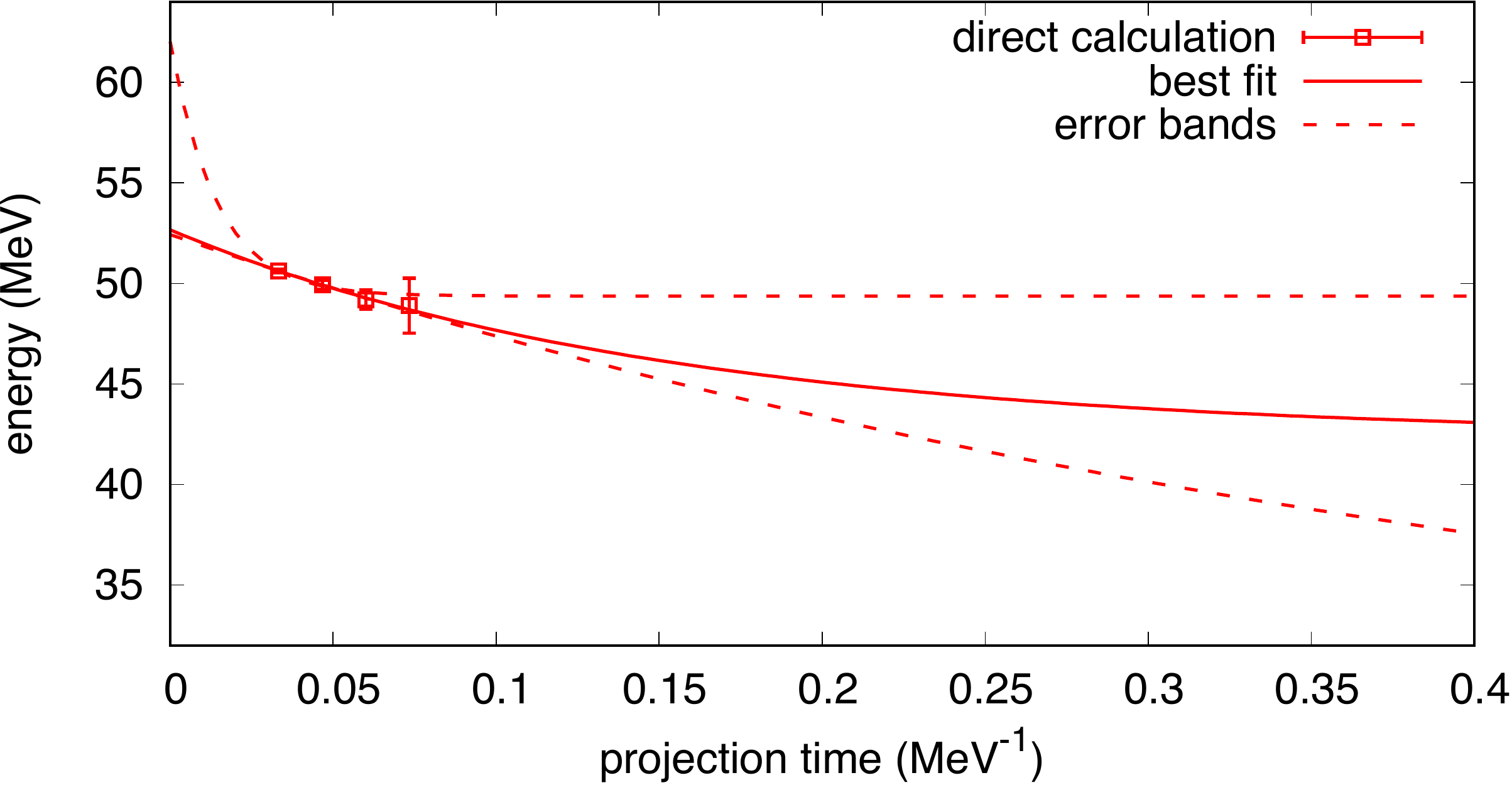}}
\caption[Direct calculation of fourteen neutrons (L = 8 fm)]{Direct calculation of the ground state energy for fourteen neutrons, shown as a function of Euclidean time. The red, open squares are the lattice results, and the solid red line is the best fit showing the exponential decay of the signal. The dashed red line is the 1$\sigma$ error band for the fit.}
\end{center}
\end{figure}

For the eigenvector continuation, we compute the inner products
\begin{equation}
N_{i,i'}(n_t) = \langle \Phi,n_t,c_i' | \Phi,n_t,c_i \rangle
\end{equation}
\noindent for sampling values $g_A^2 = c_1, c_2, c_3$, where $c_1 = 0.25$, $c_2 = 0.60$, and $c_3 = 0.95$. We also calculate the matrix elements of the full transfer matrix for the target value $c_\odot = 1.29^2 = 1.66$,
\begin{equation}
M_{i,i'}(n_t) = \langle \Phi,n_t,c_i' | U(1,c_\odot) | \Phi,n_t,c_i \rangle
\end{equation}
\noindent To extract the EC energy, we solve the generalized eigenvalue problem $N^{-1/2} M N^{-1/2}$, finding the largest eigenvalue $\lambda(n_t)$, and using the relation
\begin{equation}
E(n_t) = -\log\lbrace \lambda(n_t) \rbrace / a_t \approx E_0 + c e^{-\delta E (2 n_t + 1) a_t}
\end{equation}

In these calculations, we consider seven different choices for the EC subspace. For the first three, we consider the 1-D subspaces formed by a single vector $| \Phi,n_t,c_i \rangle$ for $i=1,2,3$. The next three consist of the 2-D subspaces formed by taking two vectors $| \Phi,n_t,c_i \rangle$ and $| \Phi,n_t,c_i' \rangle$ for $i \neq i'$. Lastly, we consider the full 3-D subspace formed from all three vectors.

The EC results for six neutrons are shown in Fig. [4.7]. The ground state energy is shown versus projection time using sampling data $g_A^2 = c_1, c_2, c_3$, where $c_1 = 0.25$, $c_2 = 0.60$, and $c_3 = 0.95$. We show results for one, two and three vectors. The errors are estimated using a jackknife analysis of the Monte Carlo data. In order to decrease the sensitivity to stochastic noise, we have used a common $\delta E$ for all cases. We have also imposed a variational constraint that adding more vectors to the EC subspace will not increase the energy. The EC results for fourteen neutrons are shown in Fig. [4.8], with the same choices for $c_1, c_2, c_3$.

\begin{figure}
\begin{center}
\makebox[\textwidth][c]{\includegraphics[width=0.9\textwidth]{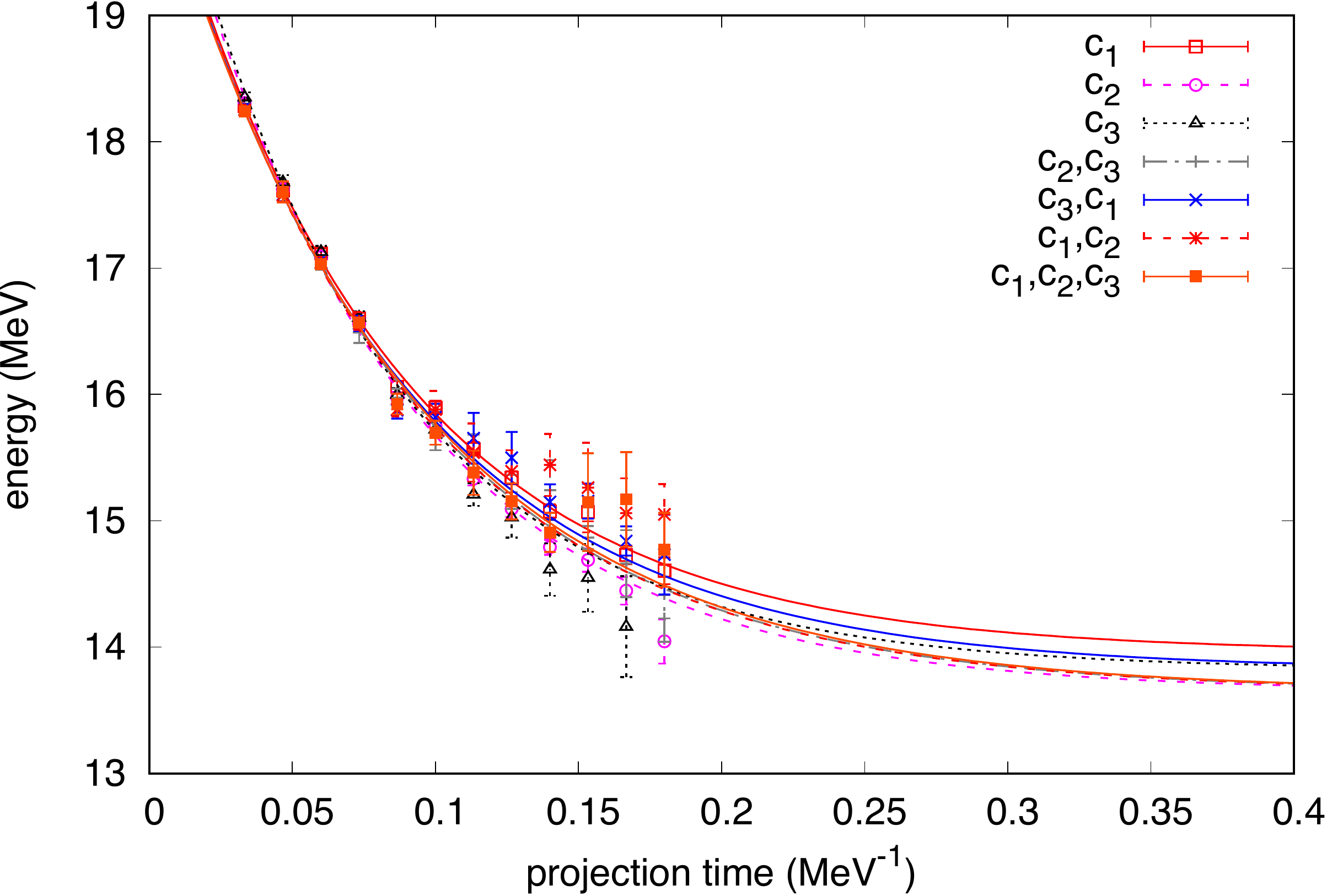}}
\caption[Eigenvector Continuation with six neutrons (L = 8 fm)]{Eigenvector continuation results for six neutrons, including up to three vectors in the training subspace. Open squares show the EC estimates for the ground state energy for a particular set of vectors used to construct the subspace. The lines show the best fits of these results.}
\end{center}
\end{figure}

\begin{figure}
\begin{center}
\makebox[\textwidth][c]{\includegraphics[width=0.9\textwidth]{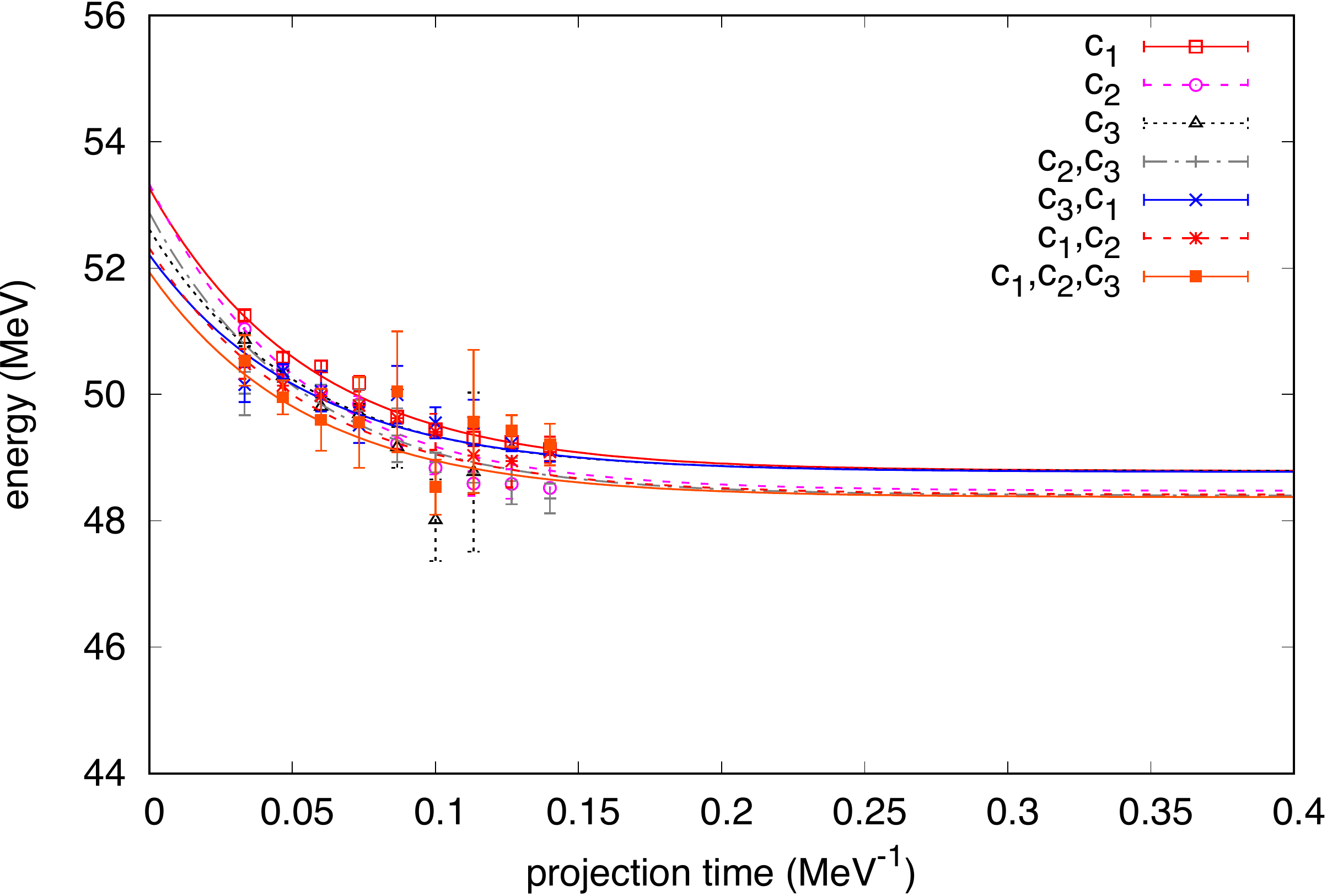}}
\caption[Eigenvector Continuation with fourteen neutrons (L = 8 fm)]{Eigenvector continuation results for fourteen neutrons, including up to three vectors in the training subspace. Open squares show the EC estimates for the ground state energy for a particular set of vectors used to construct the subspace. The lines show the best fits of these results.}
\end{center}
\end{figure}

We summarize all of the results in Table [4.1]. For reference, the nuclear densities present in each simulation are provided. These are computed by $\rho_N = N/V$, where $N$ is the number of neutrons in the box, and $V$ is the volume of the box, in units of fm$^{-3}$. In both cases, we see that the estimates for the error have improved by a factor of 10 to 20.

\begin{table}
\begin{center}
\begin{tabular}{c | c | c}
\hline \hline
$g^2_A$ Values & $E_0$ for 6 neut. (MeV) & $E_0$ for 14 neut. (MeV) \\ \hline
$c_1$ & 13.8(1) & 48.9(4) \\
$c_2$ & 13.6(2) & 48.4(5) \\
$c_3$ & 13.6(2) & 48.9(6) \\
$c_2$,$c_3$ & 13.6(2) & 48.1(6) \\
$c_3$,$c_1$ & 13.6(2) & 48.9(6) \\
$c_1$,$c_2$ & 13.6(2) & 48.0(6) \\
$c_1$,$c_2$,$c_3$ & 13.6(2) & 48.0(6) \\
\hline
direct calc. & 12$\left(^{+3}_{-4}\right)$ & 42$\left(^{+7}_{-15}\right)$ \\
\hline \hline
$\rho_N$ & 0.012 fm$^{-3}$ & 0.028 fm$^{-3}$ \\ \hline
\end{tabular}
\end{center}
\caption[Summary of neutron matter results]{Summary of the neutron matter results, for 6 and 14 neutrons, compared to the direct Monte Carlo calculations. For reference, the relevant nuclear densities are provided.}
\end{table}

%%%%%%%%%%%%%%%%%%%%%%%%%%%%%%%%%%%%%%%%%%%%%%%%%%%%%%%%%%%%%%%%%%
\section{Coulomb Interaction in $^{12}$C and $^{16}$O}

The non-perturbative simulation of the Coulomb interaction in light nuclei is the main project described in this work. For most nuclear simulations, the Coulomb interaction is handled perturbatively. However, due to its repulsive nature, it leads to sign problems for all but the lightest nuclei.

We want to use eigenvector continuation to circumvent this problem, by adding the Coulomb interaction directly to the transfer matrix via auxiliary fields, and getting ground state wavefunctions for various couplings. Then, we can build the EC estimate for the energy shift due to the Coulomb interaction using the method described in Section 2.4.4.

On the lattice, the Coulomb interaction has the form
\begin{equation}
V(|\textbf{n}-\textbf{n}'|) = \frac{\alpha_{em}}{\text{max}( 0.5,|\textbf{n}-\textbf{n}'|)}
\end{equation}
\noindent where $\alpha_em = 1/137$ is the fine structure constant, and $|\textbf{n}-\textbf{n}'$ is the distance between lattice sites. The denominator has the max function to prevent the potential taking an infinite value at zero distance. The value at the origin is chosen to be $V(0) = 2\alpha_{em}$. While this approximation may seem incorrect, it doesn't affect the p-p scattering phase shifts or any of the low-energy nuclear observables.

To add this interaction into the transfer matrix, we define a new auxiliary field $s_1$,
\begin{equation}
M(s,s_1,t) = : \exp \lbrace -H_{\text{free}}\alpha_t + \sqrt{-\alpha_{em} \alpha_t} \sum_{\vec{n}} s_1(\vec{n},n_t) \rho(\vec{n},n_t)  \rbrace :
\end{equation}
\noindent where $H_{\text{free}}$ consists of the kinetic energy term, and $\rho(\vec{n},n_t)$ is the density operator at the lattice site $(\vec{n},n_t)$. The $s_1$ auxiliary field is pulled from the distribution 
\begin{equation}
P = \exp \left( -\mathcal{S} \right) = \exp \left( - \sum_{\vec{n},\vec{n}'} s_1(\vec{n},n_t) V^{-1}(\vec{n} - \vec{n}') s_1(\vec{n}',n_t) \right)
\end{equation}
\noindent following the prescription in section 2.4.4.

\begin{figure}	% FILLER PLOT
\begin{center}
\makebox[\textwidth][c]{\includegraphics[width=0.8\textwidth]{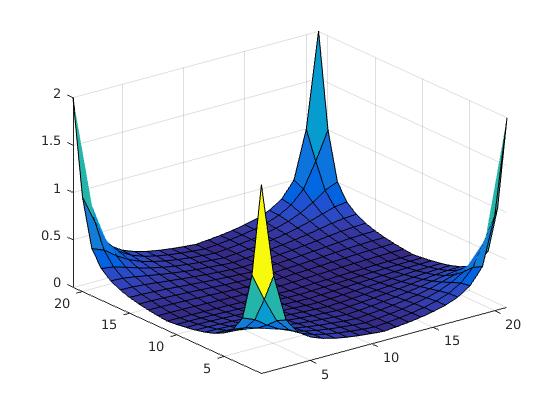}}
\caption[Slice of the Coulomb potential]{Values of the Coulomb potential on the $n_z$=0 plane. The value at the origin is chosen to be $V(0) = 2\alpha_{em}$. This is large enough to give good results.}
\end{center}
\end{figure}

This calculation was implemented in the MCLEFT code \cite{Lu:2018een}, which uses the shuttle algorithm to update the auxiliary fields and propagate the wavefunctions using the transfer matrices. Two new modules were written, \textit{eigenvectorContinuation} and \textit{eigenvectorContinutationAuxiliary}, which were designed to compute the ground state wavefunctions for various couplings for this $s_1$ field. Then, the eigenvector continuation technique could be applied using the interactions in the MCLEFT code as a baseline. Also, the MCLEFT code calculates the Coulomb energy perturbatively, so that was used as a benchmark for the method.

We start with $^{12}$C, computed on an $L=5$ lattice, with lattice spacings of $a = 1/100$ $\text{ MeV}^{-1}$ and $a_t = 1/1000 \text{ MeV}^{-1}$. We first motivate the need for the eigenvector continuation technique, by looking at the ground state energy of $^{12}$C with Coulomb included via direct calculation. The Coulomb interaction has a noticeable sign problem in larger nuclei, so we tune the fine structure constant to be four times its physical value, $C_\odot = 4\alpha_{em}$. Results for this is shown in Fig. [4.10]. 

\begin{figure}	% FILLER PLOT
\begin{center}
\makebox[\textwidth][c]{\includegraphics[width=\textwidth]{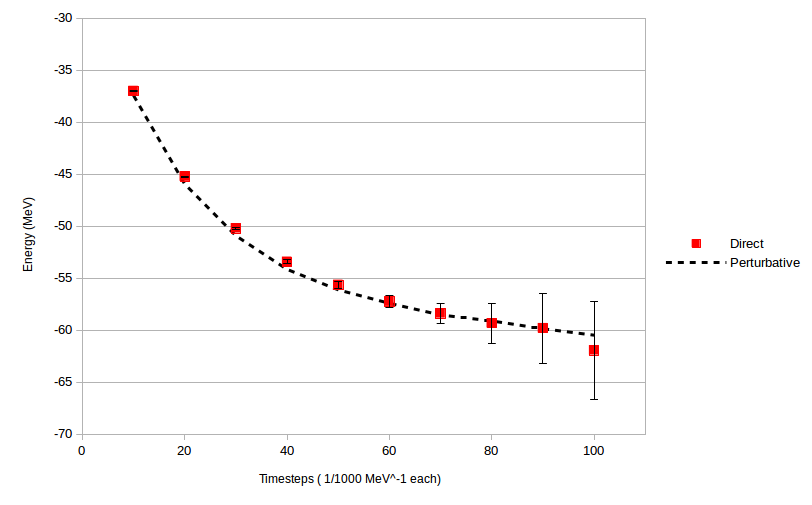}}
\caption[Direct calculation of Coulomb energy shifts for $^{12}$C]{Direct calculation of Coulomb energy shifts for $^{12}$C, at $C_\odot = 4\alpha_{em}$. The black dotted line shows the perturbative results, and the orange squares show the results of the direct calculation. We see that for large projection times the sign problem begins to show up, before we have reached a proper plateau.}
\end{center}
\end{figure}

We use perturbative results as a reference value. In general, perturbative results will not be available, or if they are, they may not be reliable. In the case of the Coulomb interaction, perturbation theory works well enough to estimate the energy that it can be used as a benchmark. These are shown with the black dotted line.

Ideally, we would compute this energy for several projection times, and show a nice exponential decay that we could fit the true, infinite projection time ground state energy to. However, we see the rapid onset of the Monte Carlo sign problem, which causes large oscillations in the amplitudes at large projection time. This causes the statistical errors to explode, making large $\tau$ extrapolations virtually impossible.

Like we did for the neutron matter system, we turn to eigenvector continuation to save us. We can take a set of wavefunctions for small couplings, small enough that the sign problem doesn't bother us, and use them to extrapolate to couplings that would be otherwise inaccessible.

We now show the results of this EC calculation for $^{12}$C, up to 3rd order. The couplings used in the EC subspace are $C_{EC} = 0\alpha_{em}, 2.0\alpha_{em}, \text{ and } 4.0\alpha_{em}$, and the target coupling is $C_\odot = 4\alpha_{em}$. This is the same target coupling used in the direct calculation, so that we can easily compare the accuracy and stability of the results. These EC results are plotted in Fig. [4.11].

\begin{figure}	% FILLER PLOT
\begin{center}
\makebox[\textwidth][c]{\includegraphics[width=\textwidth]{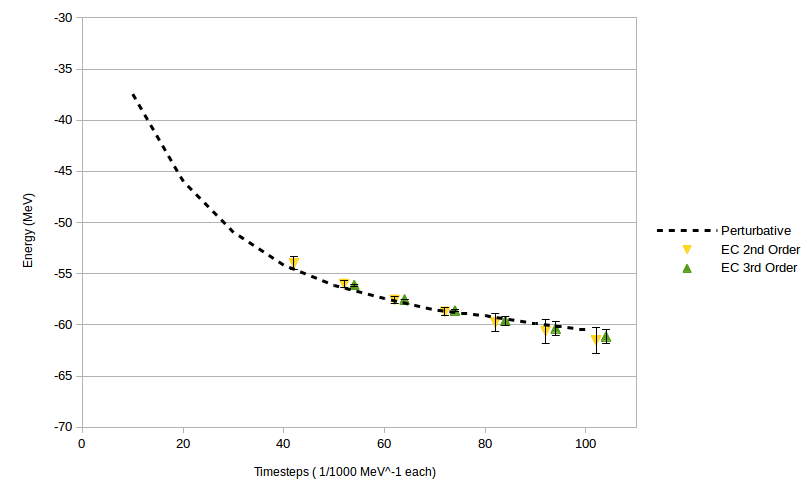}}
\caption[EC Calculation of Coulomb energy shifts in $^{12}$C]{EC Calculation of Coulomb energy shifts in $^{12}$C. Again, we show the perturbative results as a reference. Early on, the EC norm matrix is very singular, so data is not present before $L_t = 40$. We note that the 2nd and 3rd order results are computed at the same projection times; they are only separated on the plot for visual clarity.}
\end{center}
\end{figure}

In yellow, we show the 2nd order results, computed using the wavefunctions at coupling $C_{EC} = 2.0\alpha_{em}, \text{ and } 4.0\alpha_{em}$. In green are the 3rd order results, computed with the wavefunctions at all three couplings $C_{EC} = 0\alpha_{em}, 2.0\alpha_{em}, \text{ and } 4.0\alpha_{em}$. For small projection times, we ran into numerical instability in the norm matrix. This is due to the fact that all the wavefunctions start in the same initial state; thus, for small projection times, the wavefunctions haven't gotten the chance to change much. It takes until $L_t = 40$ ($\tau = 7.88\text{ fm}$) for the 2nd order results to become stable, and $L_t = 50$ ($\tau = 9.85\text{ fm}$) for the 3rd order results to become stable. We now show these two sets of data on the same plot, in Fig. [4.12], to better compare the methods. The values on this plot are summarized in Table [4.2].

\begin{figure}	% FILLER PLOT
\begin{center}
\makebox[\textwidth][c]{\includegraphics[width=\textwidth]{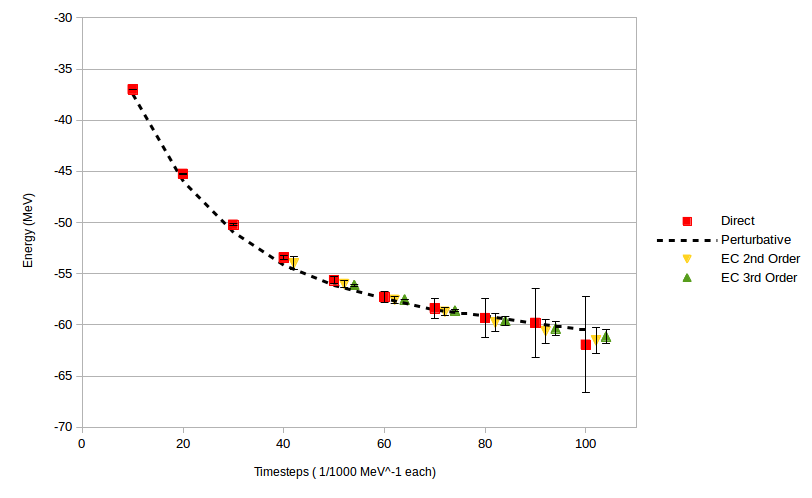}}
\caption[Comparison of EC and direct calculations of Coulomb energy shifts in $^{12}$C]{Combined plot showing the direct results and the EC estimates of Coulomb energy shifts in $^{12}$C. We can immediately see the improvement gained by the EC method, and the overall resilience to the sign problem. We note that the three points in each pack are computed at the same projection times; they are only separated on the plot for visual clarity.}
\end{center}
\end{figure}

\begin{table}
\begin{center}
\begin{tabular}{| c | c | c | c | c | c |}
\hline
Timesteps & $\tau$ (fm) & Perturbative & Direct Calc. & 2nd Order EC & 3rd Order EC \\ \hline
10 & 1.97 & -37.448(51) & -37.012(20) & - & - \\
20 & 3.94 & -45.964(52) & -45.239(43) & - & - \\
30 & 5.91 & -50.934(59) & -50.209(100) & - & - \\
40 & 7.88 & -54.164(63) & -53.435(177) & -53.982(633) & - \\
50 & 9.85 & -56.138(77) & -55.648(335) & -56.002(376) & -56.111(95) \\
60 & 11.82 & -57.428(72) & -57.268(571) & -57.553(321) & -57.515(45) \\
70 & 13.79 & -58.546(84) & -58.397(936) & -58.716(398) & -58.606(130) \\
80 & 15.76 & -59.120(99) & -59.341(1900) & -59.788(897) & -59.613(474) \\
90 & 17.73 & -59.872(105) & -59.814(3352) & -60.601(1169) & -60.343(680) \\
100 & 19.70 & -60.490(99) & -61.944(4708) & -61.527(1316) & -61.133(686) \\
\hline
\end{tabular}
\end{center}
\caption[Summary of $^{12}$C results for the Coulomb interaction]{Summary of the Coulomb results in $^{12}$C, computed via direct calculation, and by eigenvector continuation. Dashed entries are where a singular norm matrix prevented stable calculations.}
\end{table}

In Section 3.2, we mentioned that it is often easier to break the computation of the EC inserted operator matrix $M$ into two pieces
\begin{equation}
M = M_0 + C_\odot M_{EC}
\end{equation}
\noindent so that we could perform the eigenvector simulations in the post-processing stage, for arbitrary target coupling.

In Fig. [4.13], we show the results of tuning the target Coulomb strength $C_\odot = x \alpha_{em}$ in the range $0.0 < x < 10.0$. We show the perturbative results again as a reference. Using perturbation theory, we expect this energy correction to be linear in the coupling $E_0 \propto \alpha_{em}$. Thus, any deviations from this would be hints of higher-order terms.

\begin{figure}	% FILLER PLOT
\begin{center}
\makebox[\textwidth][c]{\includegraphics[width=0.84\textwidth]{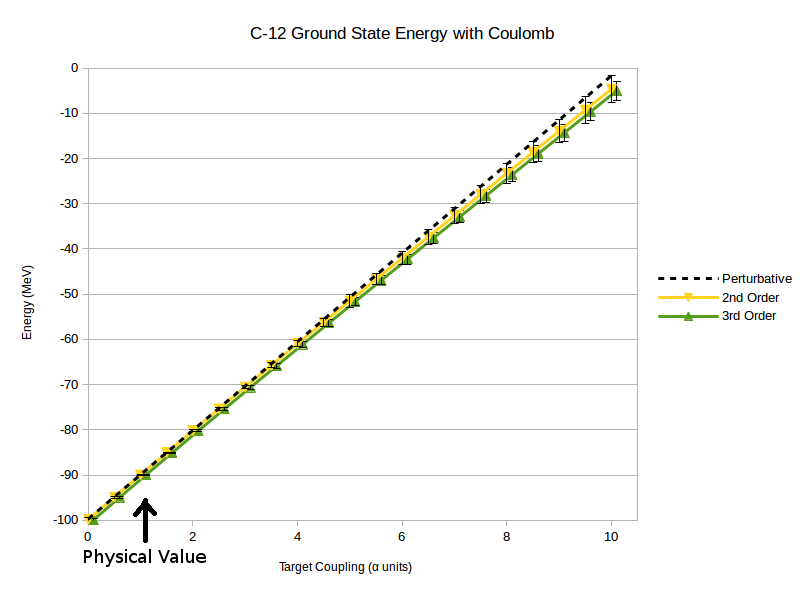}}
\makebox[\textwidth][c]{\includegraphics[width=0.84\textwidth]{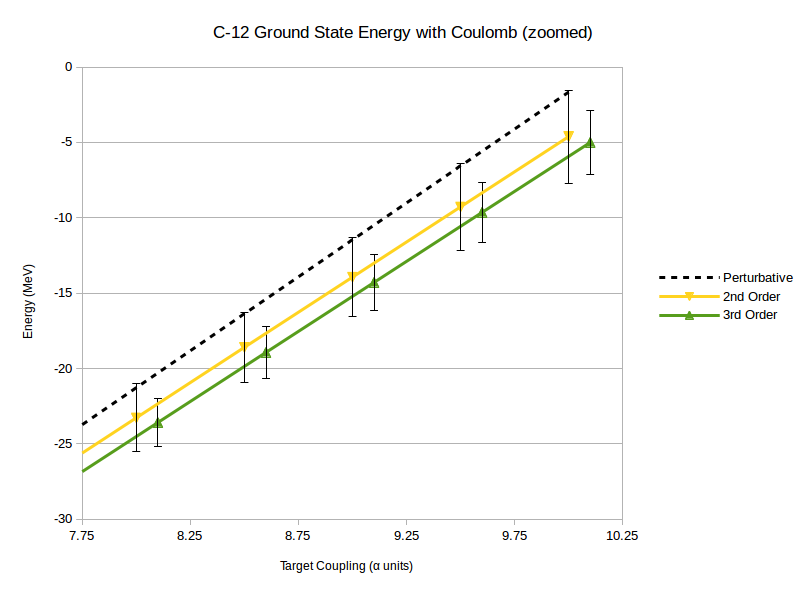}}
\caption[Computation of Coulomb energy shifts in $^{12}$C for arbitrary target coupling.]{Computation of Coulomb energy shifts in $^{12}$C for arbitrary target coupling. $C_\odot$ was varied in the range $0.0\alpha_{em} \leq C_\odot \leq 10.0\alpha_{em}$. Shown in the bottom figure is a zoom of larger coupling region, showing a clear deviation from the perturbative results.}
\end{center}
\end{figure}

We now show a few other tests that we performed to check the validity of the simulations. In Fig [4.14], we show the convergence of the $^{12}$C EC ground state energy estimate as a function of the Monte Carlo step. This is a test any Monte Carlo calculation should check first; a good result should be converged after a suitably large number of trials. We see that here; after 7,000 steps or so, the result doesn't vary much, and the error estimate from the Online algorithm seems to be approaching a steady value.

\begin{figure}	% FILLER PLOT
\begin{center}
\makebox[\textwidth][c]{\includegraphics[width=0.8\textwidth]{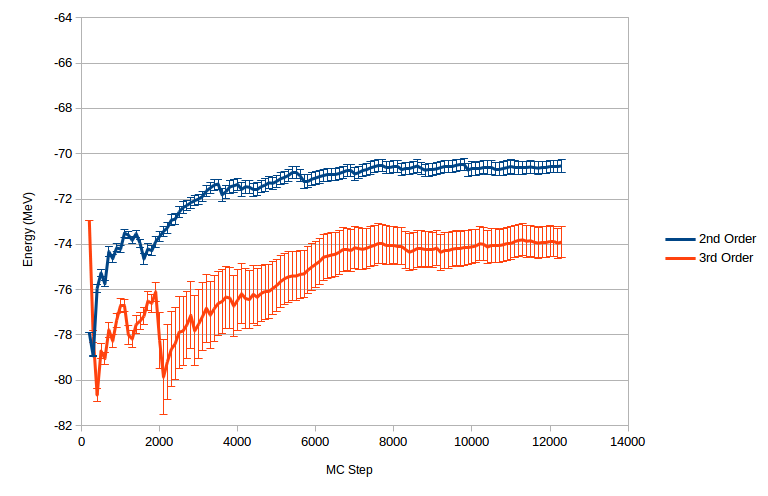}}
\caption[Value of $^{12}$C ground state energy vs MC steps]{Plot of the EC estimate for the ground state energy of $^{12}$C, as a function of the MC step. Shown are also the 1$\sigma$ error bars at each step. These calculations were done at a target coupling $C_\odot = 3 \alpha_{\text{em}}$. We see that after ~7000 steps, the results are more-or-less converged, though the 3rd order results experience some numerical instability.}
\end{center}
\end{figure}

We now move on to simulations of $^{16}$O. Here, the number of particles is starting to get high enough that simulations at large projection times become even more problematic. In Figs. [4.15 - 4.17], we show plots like the $^{12}$C plots; the results from direct calculation of the Coulomb energy shifts, and a result for performing the eigenvector continuation method. We see that for large projection times, the sign problem once again kills us, though, it is much more sever this time around. Also, the use of the eigenvector continuation has not only stabilized the results, but has also reduced the errors rather dramatically. The values on these plots are summarized in Table [4.3].

\begin{figure}	% FILLER PLOT
\begin{center}
\makebox[\textwidth][c]{\includegraphics[width=\textwidth]{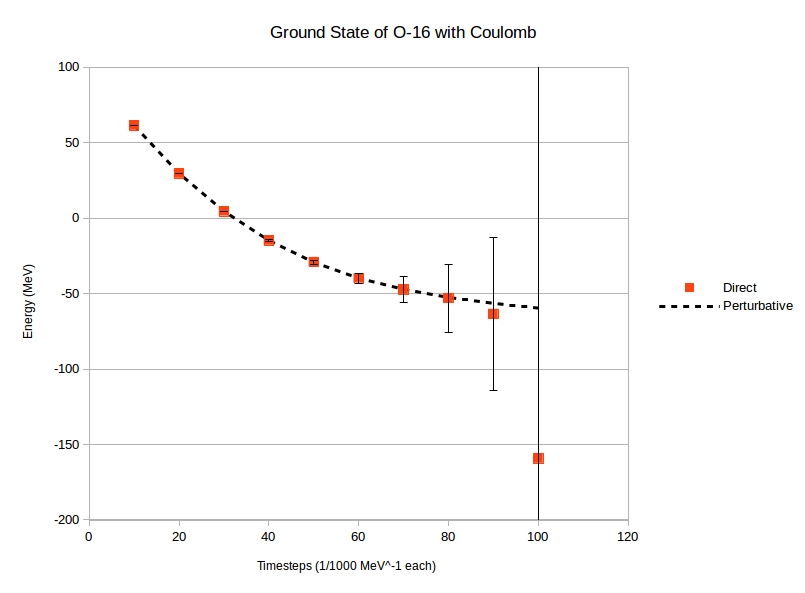}}
\caption[Direct calculation of Coulomb energy shifts for $^{16}$O]{Direct calculation of Coulomb energy shifts for $^{16}$O}
\end{center}
\end{figure}

\begin{figure}	% FILLER PLOT
\begin{center}
\makebox[\textwidth][c]{\includegraphics[width=\textwidth]{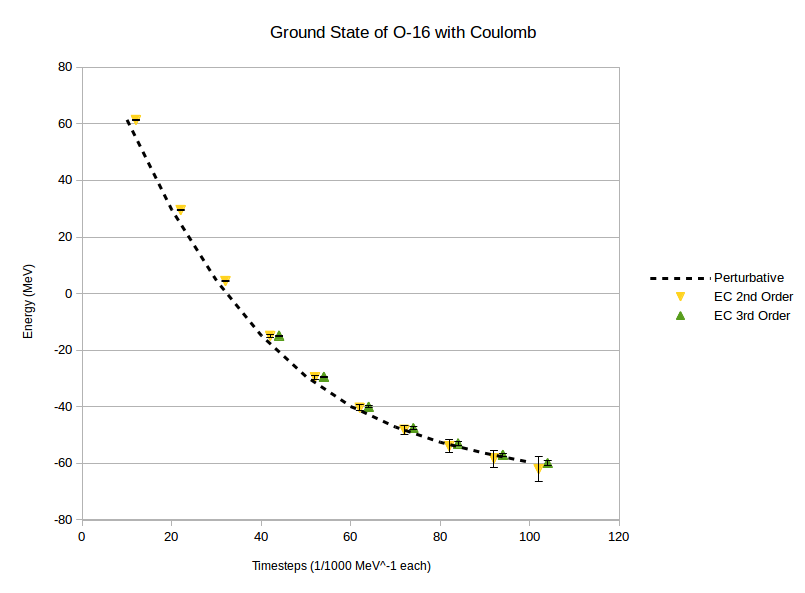}}
\caption[EC Calculation of Coulomb energy shifts in $^{16}$O]{EC Calculation of Coulomb energy shifts in $^{16}$O. For $L_t = 10-30$, the 3rd order results were too unstable to extract reliable data.}
\end{center}
\end{figure}

\begin{figure}	% FILLER PLOT
\begin{center}
\makebox[\textwidth][c]{\includegraphics[width=\textwidth]{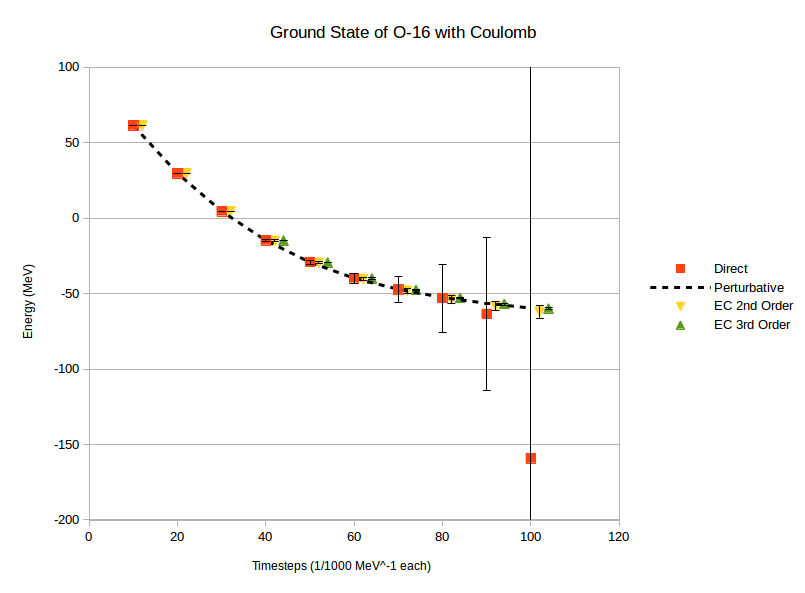}}
\caption[Comparison of EC and direct calculations of Coulomb energy shifts in $^{16}$O]{Combined plot showing the direct results and the EC estimates of Coulomb energy shifts in $^{16}$O.}
\end{center}
\end{figure}

\begin{table}
\begin{center}
\begin{tabular}{| c | c | c | c | c | c |}
\hline
Timesteps & $\tau$ (fm) & Perturbative & Direct Calc. & 2nd Order EC & 3rd Order EC \\ \hline
10 & 1.97 & 61.406(89) & 61.408(28) & 61.35(18) & - \\
20 & 3.94 & 29.60(10) & 29.579(73) & 29.51(19) & - \\
30 & 5.91 & 4.53(10) & 4.54(17) & 4.43(28) & - \\
40 & 7.88 & -14.81(12) & -14.74(50) & -14.94(48) & -14.87(17) \\
50 & 9.85 & -29.31(12) & -29.12(125) & -29.56(69) & -29.43(18) \\
60 & 11.82 & -39.82(13) & -39.93(346) & -40.34(113) & -40.02(38) \\
70 & 13.79 & -47.16(13) & -47.25(833) & -48.11(167) & -47.59(49) \\
80 & 15.76 &  -52.54(15) & -53.02(2254) & -53.92(236) & -52.98(58) \\
90 & 17.73 & -56.44(15) & -63.55(5091) & -58.35(302) & -57.00(66) \\
100 & 19.70 & -59.60(15) & -159.27(426912) & -62.04(446) & -56.96(83) \\
\hline
\end{tabular}
\end{center}
\caption[Summary of $^{16}$O results for the Coulomb interaction]{Summary of the Coulomb results in $^{16}$O, computed via direct calculation, and by eigenvector continuation. Dashed entries are where a singular norm matrix prevented stable calculations.}
\end{table}

\begin{figure}	% FILLER PLOT
\begin{center}
\makebox[\textwidth][c]{\includegraphics[width=0.84\textwidth]{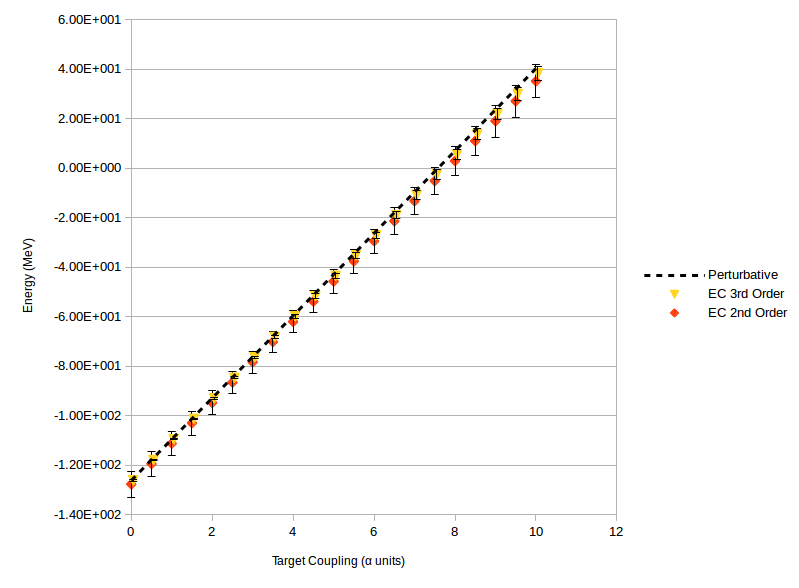}}
\makebox[\textwidth][c]{\includegraphics[width=0.84\textwidth]{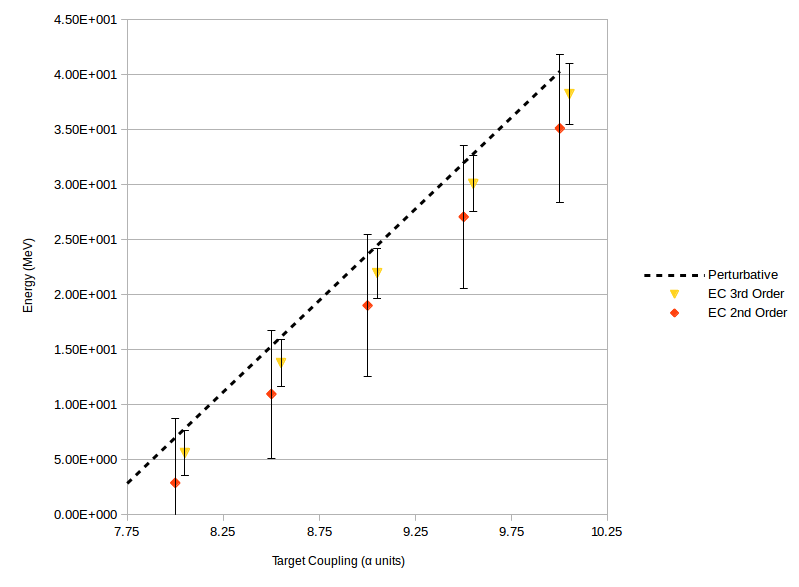}}
\caption[Computation of Coulomb energy shifts in $^{16}$O for arbitrary target coupling.]{Computation of Coulomb energy shifts in $^{16}$O for arbitrary target coupling. $C_\odot$ was varied in the range $0.0\alpha_{em} \leq C_\odot \leq 10.0\alpha_{em}$. Shown in the bottom figure is a zoom of larger coupling region.}
\end{center}
\end{figure}

We now repeat the same consistency checks that we performed for the $^{12}$C simulations. In Fig. [4.19], we show the convergence of the $^{16}$O EC ground state energy estimate as a function of the Monte Carlo step. It takes a bit longer to reach a stable value, but it does appear to be converged.

\begin{figure}	% FILLER PLOT
\begin{center}
\makebox[\textwidth][c]{\includegraphics[width=\textwidth]{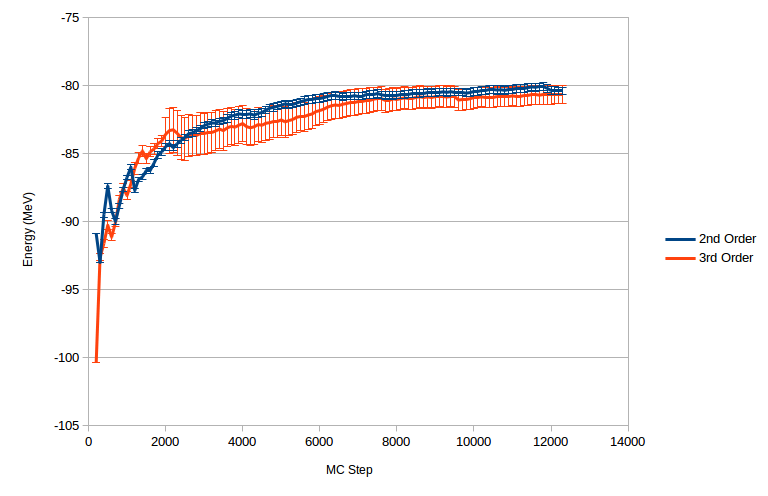}}
\caption[Value of $^{16}$O ground state energy vs MC steps]{Plot of the EC estimate for the ground state energy of $^{16}$O, as a function of the MC step. Shown are also the 1$\sigma$ error bars at each step. These calculations were done at a target coupling $C_\odot = 3 \alpha_{\text{em}}$. }
\end{center}
\end{figure}

%%%%%%%%%%%%%%%%%%%%%%%%%%%%%%%%%%%%%%%%%%%%%%%%%%%%%%%%%%%%%%%%%%
\section{EC Solutions of the Lipkin Hamiltonian} 

\begin{wrapfigure}{r}{0.5\linewidth}
\centering
\includegraphics[width=\linewidth]{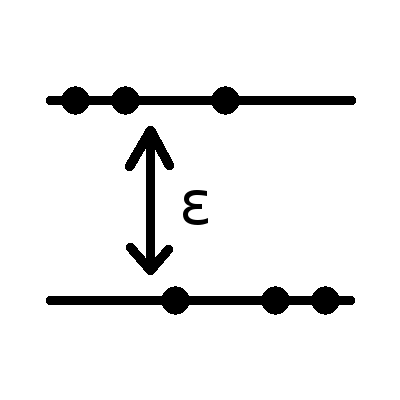}
\caption[Basic Schematic of the Lipkin Model]{Basic schematic of the Lipkin model. There are N particles distributed among two angular momentum levels, $m = \pm 1/2$, separated by some gap $\epsilon$.}
\end{wrapfigure}

The Lipkin-Meshkov-Glick (or just Lipkin) model \cite{Lipkin1965:vmb} is a model that was developed to test many-body approximation methods and numerical techniques. It was designed to both be exactly solvable, yet to have non-trivial behaviour.

The model consists of N fermions distributed between two levels, each assumed to be N-fold degenerate. Each particle is assumed to be spin-1/2, and the overall wavefunction $|J, m\rangle$ is given by a total angular momentum $J$ and projection of the angular momentum onto the z-axis $m$. The interaction is written in terms of angular momentum operators
\begin{equation}
H= \epsilon J_z + \frac{V}{2} ( J_{+}^2 + J_{-}^2 ) + \frac{W}{2} ( J_+ J_- + J_- J_+ )
\end{equation}
where $J_z$, $J_+$ and $J_-$ act according to ($\hbar = 1$)
\begin{align}
J_z | J, m \rangle &= m | J, m \rangle \nonumber \\
J_+ | J, m \rangle &= \sqrt{J(J+1) - m(m+1)} | J, m+1 \rangle \\
J_- | J, m \rangle &= \sqrt{J(J+1) - m(m-1)} | J, m-1 \rangle \nonumber
\end{align}

The first term is a count of the single particle energies, while the second and third term are what leads to interesting behaviours. The second term, or the V term, is responsible for moving two particles from one level to the other, thus changing m by two units. The third term, or the W term, allows swapping an upper-level particle and a lower-level particle.

This terms that allow mixing are responsible for a quantum phase transition. This feature of the Lipkin model has been well-studied \cite{Arias:2019kvz} \cite{2018arXiv181003517W} \cite{Co:2018ixu} \cite{2018IJQI...1650029Y} \cite{2018PhRvE..97a2112S}  \cite{2018PhRvA..97a2115H}, and has been used as a testing ground for studying these phase transitions.

While the dimension of this problem may seem large, we note that none of the operators in the Hamiltonian actually change $J$. This means the Hamiltonian matrix will be block diagonal. There will be $J_{\text{max}} = N/2$ blocks, one per value of $J$, and each block will be dimension $2J + 1$, corresponding with the values of $m$. In this paper, we will only consider the $(N+1) \times (N+1)$ sub-matrix corresponding with $J = J_{\text{max}}$.

The first thing that was checked was how well perturbation theory worked in this model. For this, we used a Matlab code, developed by A. Sarkar (publication in the works). Results for these are shown in Figs. [4.21] and [4.22].

\begin{figure}
\centering
\includegraphics[width=0.8\linewidth]{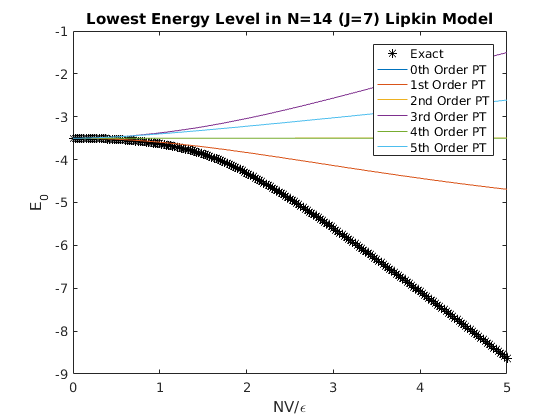}
\caption[Pert. theory results for N=14 Lipkin Model]{Results of the N=14 particle system using perturbation theory, up to 5th order. It is clear that perturbation theory diverges quickly for couplings far from 0.}
\end{figure}

\begin{figure}
\centering
\includegraphics[width=0.8\linewidth]{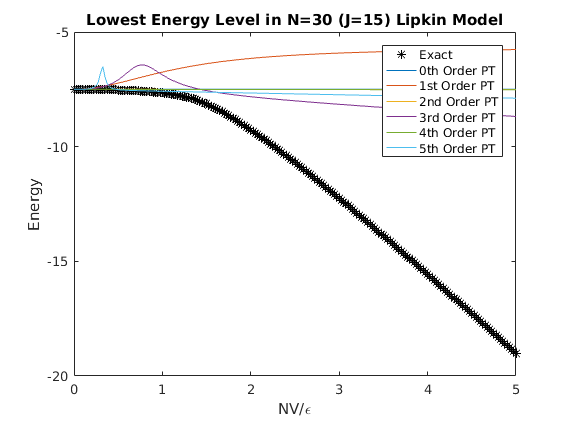}
\caption[Pert.theory results for N=30 Lipkin Model]{Results of the N=30 particle system using perturbation theory, up to 5th order. Convergence here is worse than the 14 particle system.}
\end{figure}

We now demonstrate the use of the eigenvector continuation technique to this model. Using a short Matlab code, we computed the first five eigenvectors of the Lipkin Hamiltonian, treating $V$ as our tunable parameter, for five different values of V; $V = 0.1, 0.2, 0.3, 0.4, 0.5$. For now, we take $W = 0$.

In each order of EC, which we call $\mu$, we formed a projection operator by taking the span of the first five eigenvectors for $\mu$ of the couplings, resulting in a $5\mu$-dimensional operator. Then, we projected the Hamiltonian $H(V')$ into this subspace, resulting in a new matrix $\tilde{H}$. We then diagonalized $\tilde{H}$; the lowest 5 eigenvalues of this matrix we call the $\mu$-th order EC estimates. 

\begin{figure}
\centering
\includegraphics[width = \linewidth]{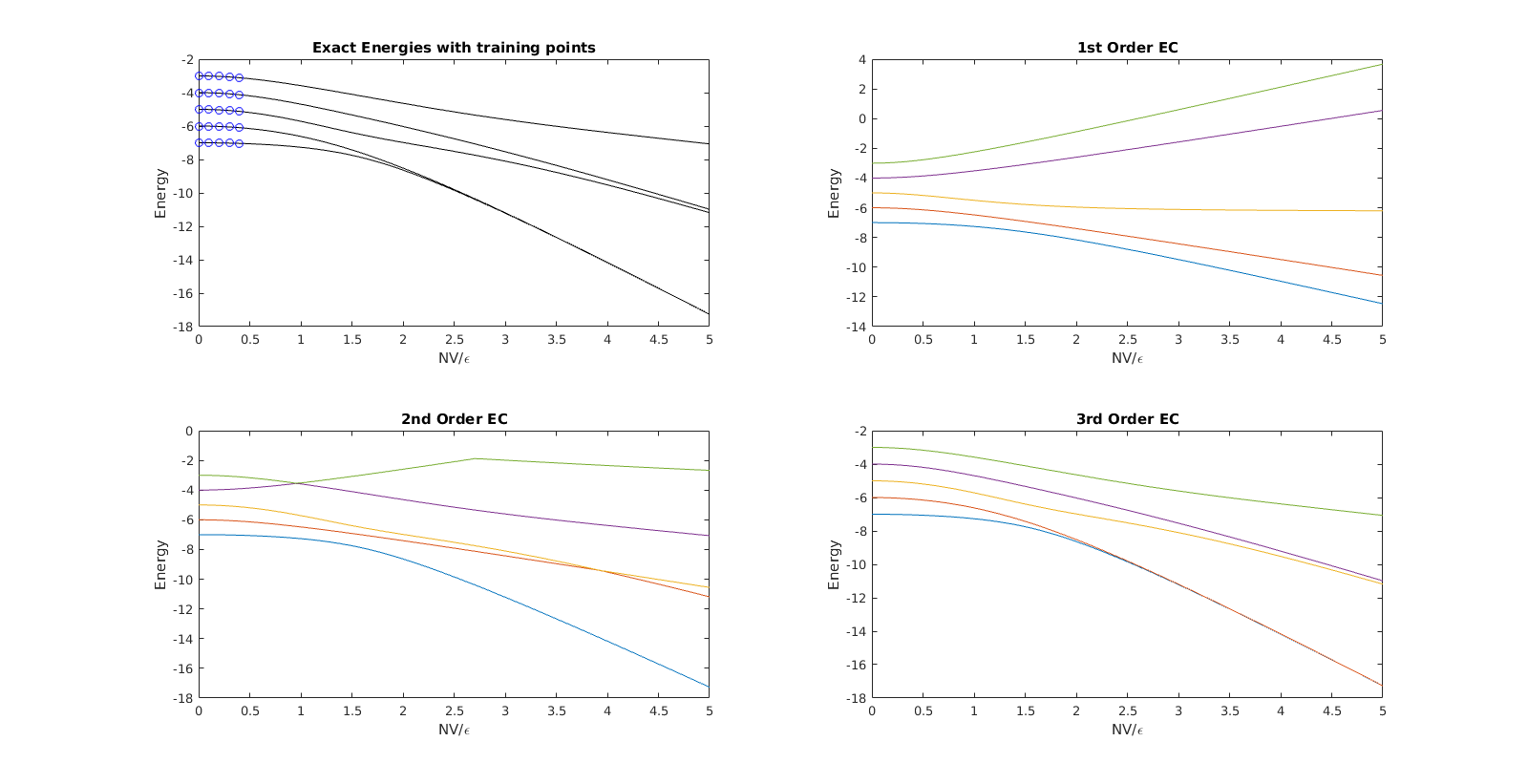}
\caption[EC estimates of the lowest 5 energy levels of N=14 Lipkin Model]{EC Estimates of the five lowest energy levels of the N=14 Lipkin model. The exact energies, obtained via diagonalizing $H$, are shown on the top left. At first order and second order, the estimates are not great, but the third order estimate (and beyond) are extremely good.}
\end{figure}

\begin{figure}
\centering
\includegraphics[width = \linewidth]{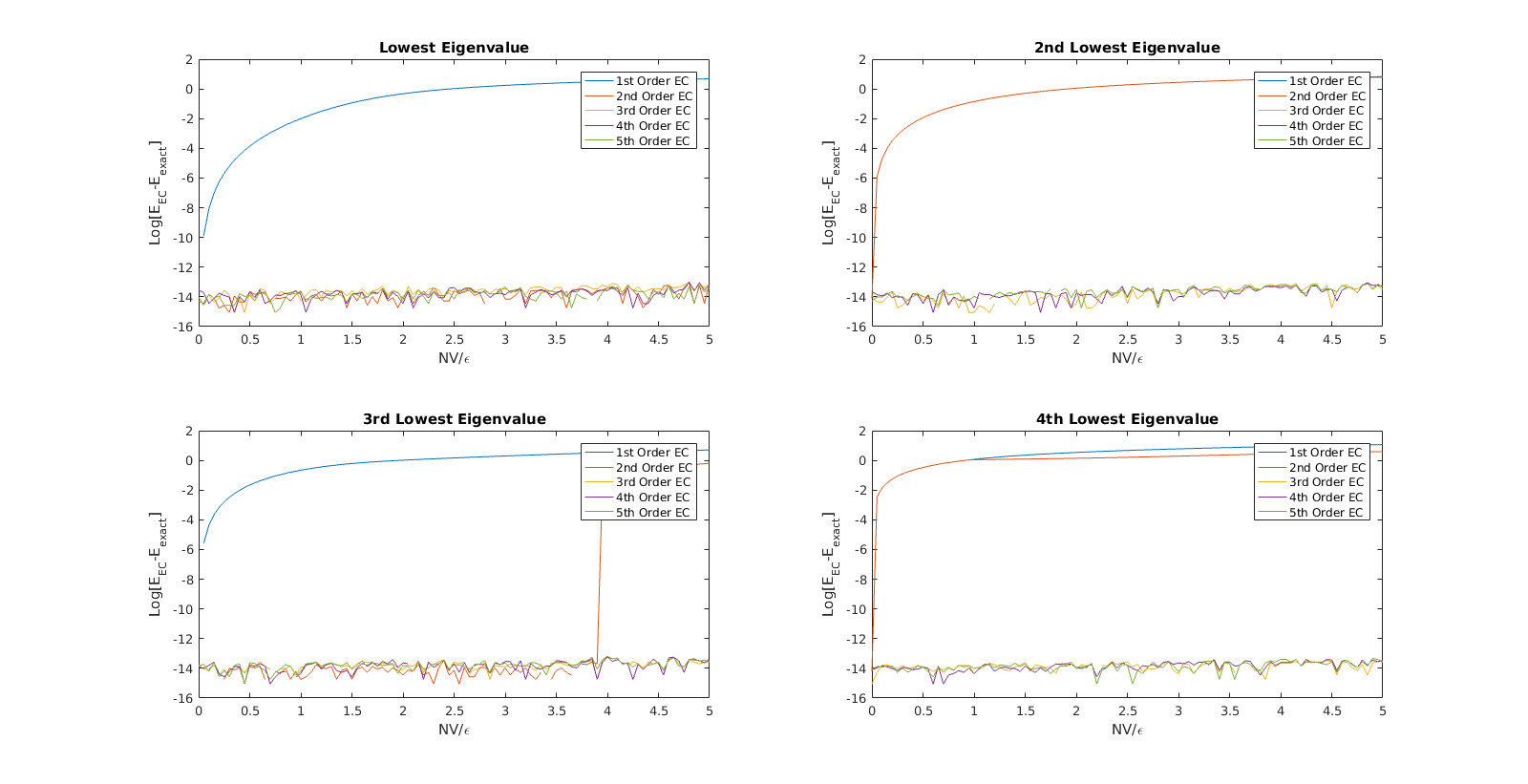}
\caption[Residuals from N=14 Lipkin model EC energy estimates]{Error between the EC estimates and the exact eigenvalues for N=14 Lipkin model, shown on a log scale. The four lowest eigenvalues are shown, each calculated up to 5th order. $10^{-14}$ is about the machine level of precision, so better errors could be obtained by increasing precision.}
\end{figure}

\begin{figure}
\centering
\includegraphics[width = \linewidth]{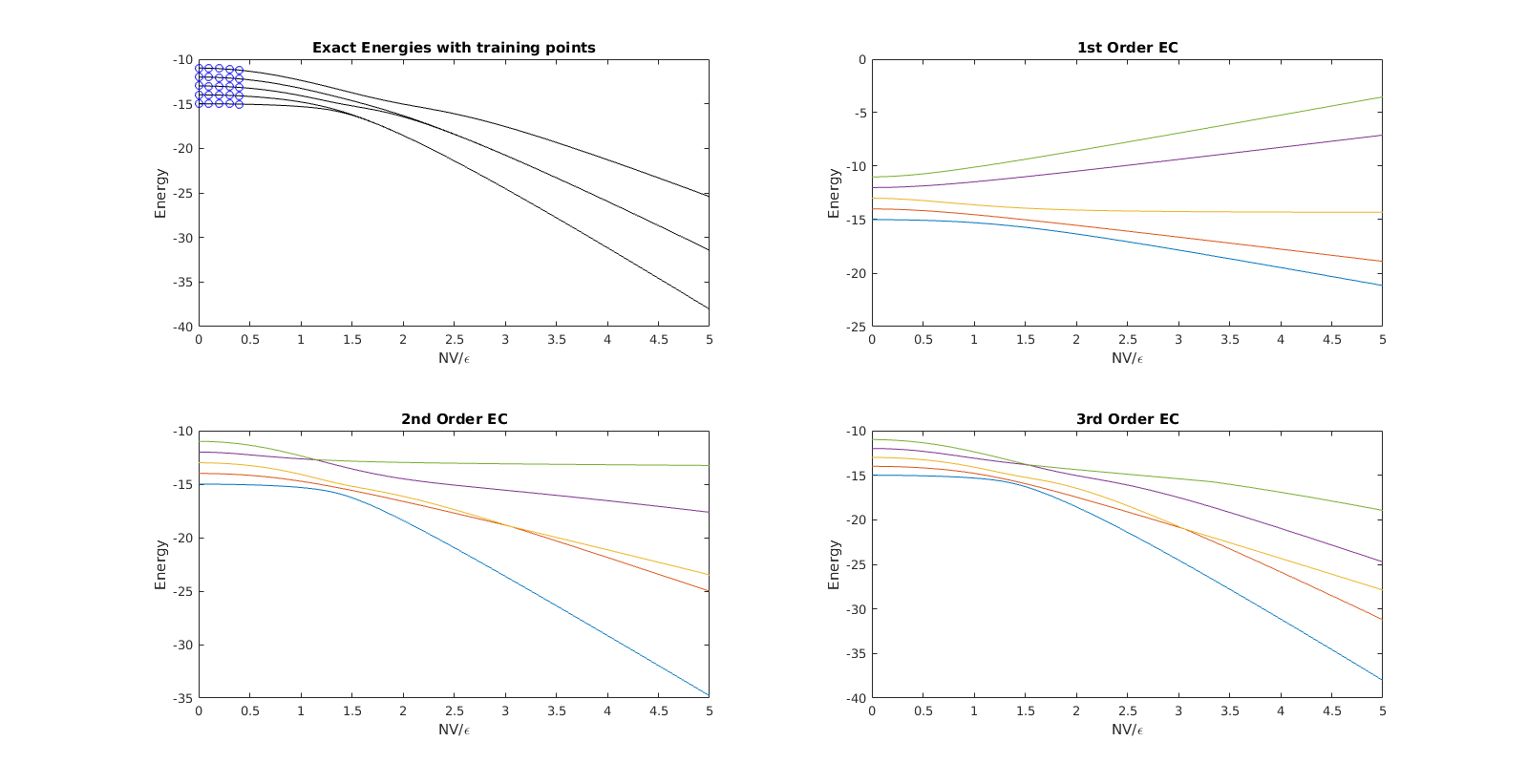}
\caption[EC estimates of the lowest 5 energy levels of N=30 Lipkin Model]{EC Estimates of the five lowest energy levels of the N=30 Lipkin model. The exact energies, obtained via diagonalizing $H$, are shown on the top left. The results converge much more slowly, and the improvements at higher orders are noticeable.}
\end{figure}

\begin{figure}
\centering
\includegraphics[width = \linewidth]{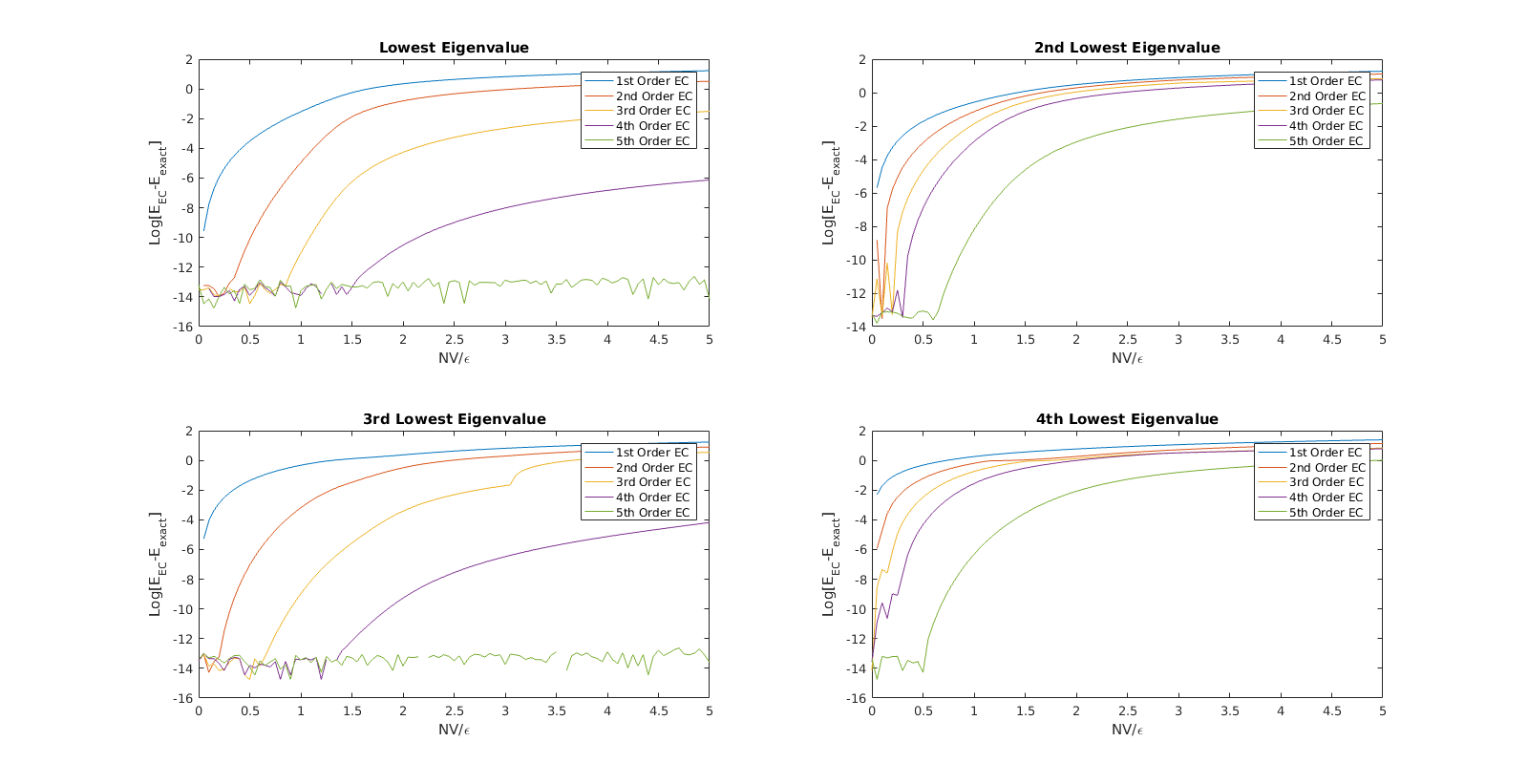}
\caption[Residuals from N=14 Lipkin model EC energy estimates]{Error between the EC estimates and the exact eigenvalues for N=30 Lipkin model, shown on a log scale. The four lowest eigenvalues are shown, each calculated up to 5th order. The order-by-order convergence is more pronounced with this larger number of particles.}
\end{figure}

\begin{figure}
\centering
\includegraphics[width = \linewidth]{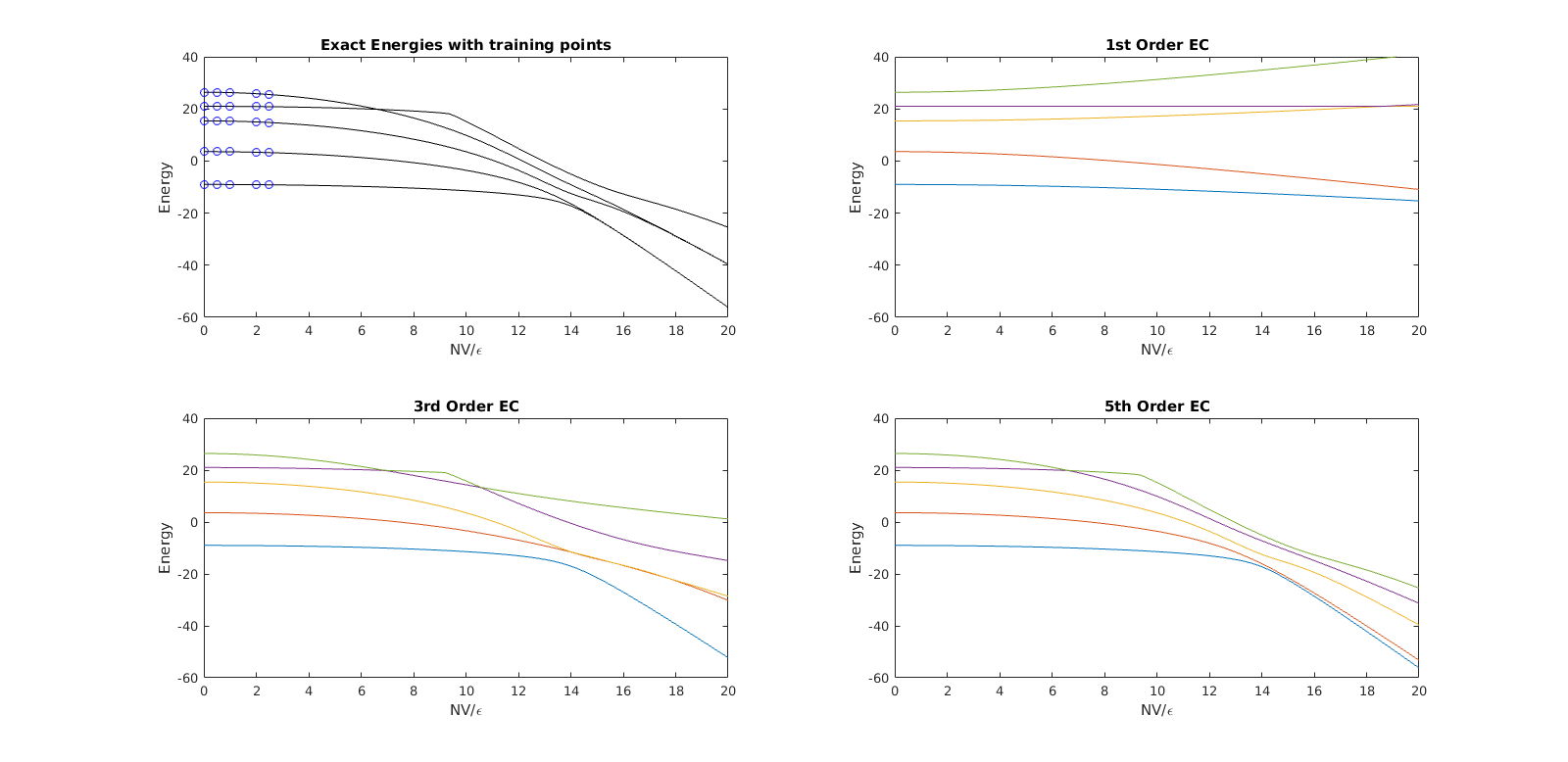}
\caption[EC estimates of the lowest 5 energy levels of N=30 Lipkin Model, $W=0.4$]{EC Estimates of the five lowest energy levels of the N=30 Lipkin model, with $W=0.4$. The exact energies, obtained via diagonalizing $H$, are shown on the top left. The 1st, 3rd, and 5th order results are shown, because convergence is overall slower than the results from $W=0$}
\end{figure}

\begin{figure}
\centering
\includegraphics[width = \linewidth]{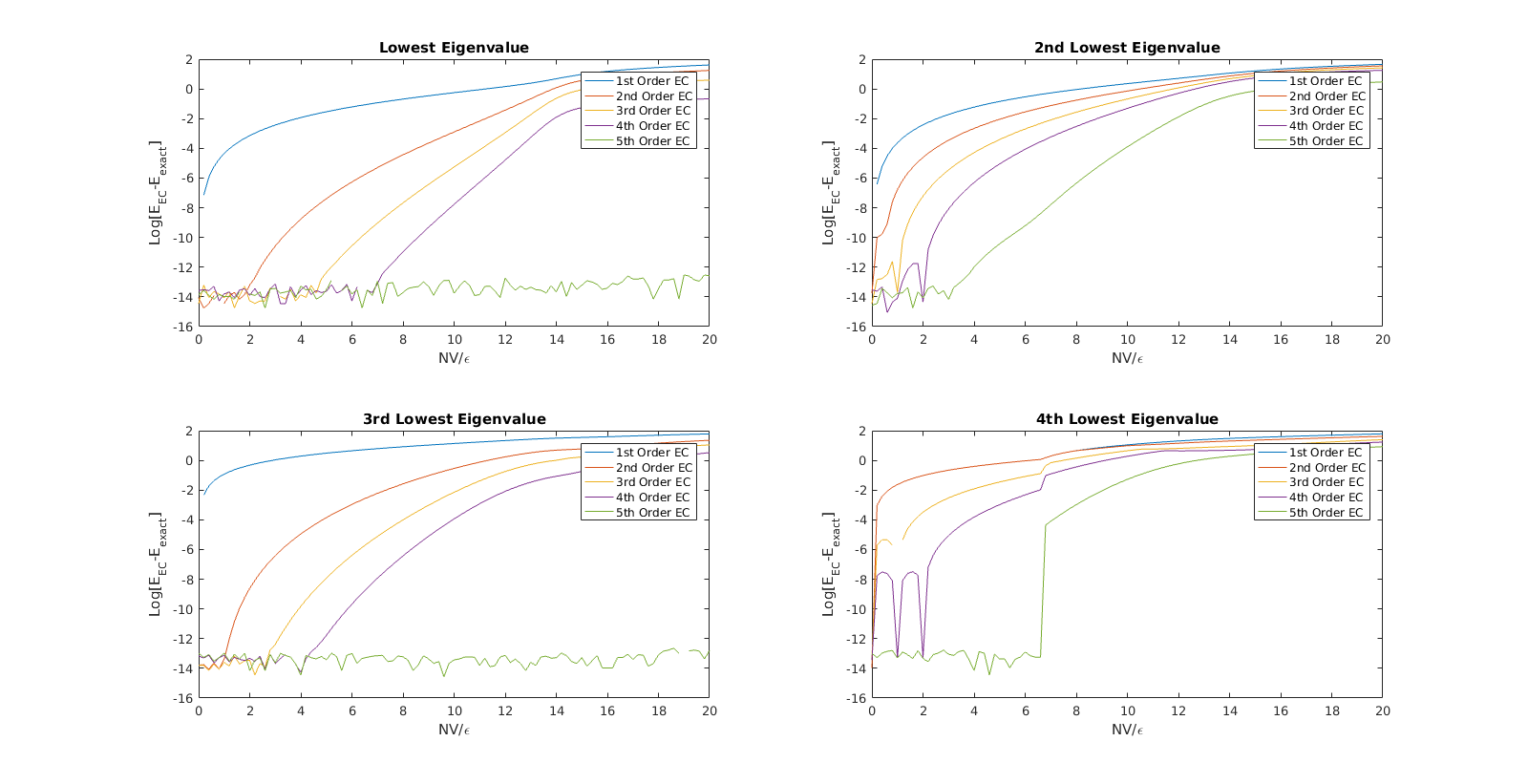}
\caption[Residuals from N=30 Lipkin model, $W=0.4$, EC energy estimates]{Error between the EC estimates and the exact eigenvalues for N=30 Lipkin model, with W = 0.4, shown on a log scale. The four lowest eigenvalues are shown, each calculated up to 5th order. While the residuals are larger, the order-by-order convergence is still present.}
\end{figure}

In Figures [4.23] and [4.25], we show the results of these calculations. We show for reference the exact energies, shown by the solid black lines in the top-left plot. On this top-left plot, we also show the points that were used in the construction of the EC subspace. The other three plots on each figure show the 1st, 2nd, and 3rd order results of the EC method. We see that while the 1st and 2nd order results are close for couplings near the training region, they tend to not behave for other couplings. However, when we reach 3rd order, we see almost perfect agreement, down to the level of precision in the calculation. We computed the 4th and 5th order estimates for the energies, but did not bother to plot them because we had already reached the numerical precision threshold.

To illustrate this agreement better, we show the residuals between the EC estimates for the energy and the exact values
\begin{equation}
R = \log_{10} \left( | E_{j,\text{EC}} - E_{j,\text{exact}} | \right)
\end{equation}
plotted in Figures [4.24] and [4.26]. These are organized by eigenvalue, to show the order-by-order convergence in the results. $10^{-14}$ is as good as Matlab can do by default, so this put a limit on how accurately our results could be. For the $N=14$ system, the convergence was already below the machine threshold by the third order calculations. For the $N=30$ system, these errors were much larger, allowing us to see the improvements in each order of EC.

Now, to really test the method, we turn the W term on, taking a value of $W = 0.4$. This leads to a more interesting energy level spectrum, with level crossings and the like. We still consider $N=30$. In Fig. [4.27], we show the first five eigenvalues, at first, third, and fifth order in EC. This is because the convergence isn't as quick as before. Looking at the residuals, shown in Fig. [4.28], it is clear that the fifth order results are not quite converged enough.

Having demonstrated that the EC technique works, we will now push it to the strong coupling or large $N$ limit. Looking at the exact spectrum, we see a very tight pinch on the spectrum around $C = 1.0$. This is signalling a branch point is very close to the real axis, which would cause the convergence of the EC method to dramatically slow down (if it works at all.)

\begin{figure}	% FILLER PLOT
\begin{center}
\makebox[\textwidth][c]{\includegraphics[width=0.8\textwidth]{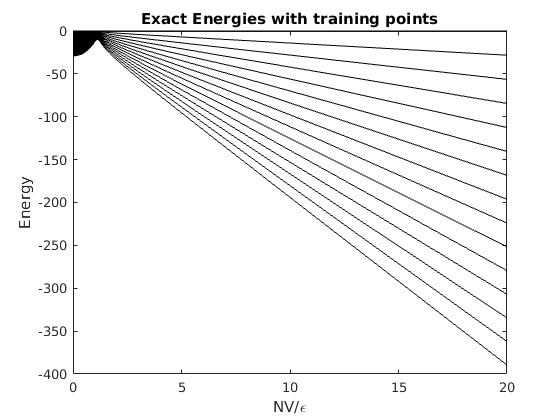}}
\caption[Spectrum of N=1000 Lipkin Model]{Plotted here are the first 20 energy levels for the N=1000 Lipkin Model. Due to the overall energy scale, we have subtracted out the lowest energy level. We note that the energy levels pinch together at around $C = 1.0$, suggesting a nearby branch point.}
\end{center}
\end{figure}

We now apply the EC method, up to 15th order, to this system. In Figs. [4.30] and [4.31], we show the odd EC estimates (1st order to 11th order) for the first five eigenvalues, with the lowest eigenvalues subtracted out. On the top-left of each, we show the exact spectrum, with the lowest eigenvalue also subtracted out, and the training points in blue circles.

\begin{figure}	% FILLER PLOT
\begin{center}
\makebox[\textwidth][c]{\includegraphics[width=\textwidth]{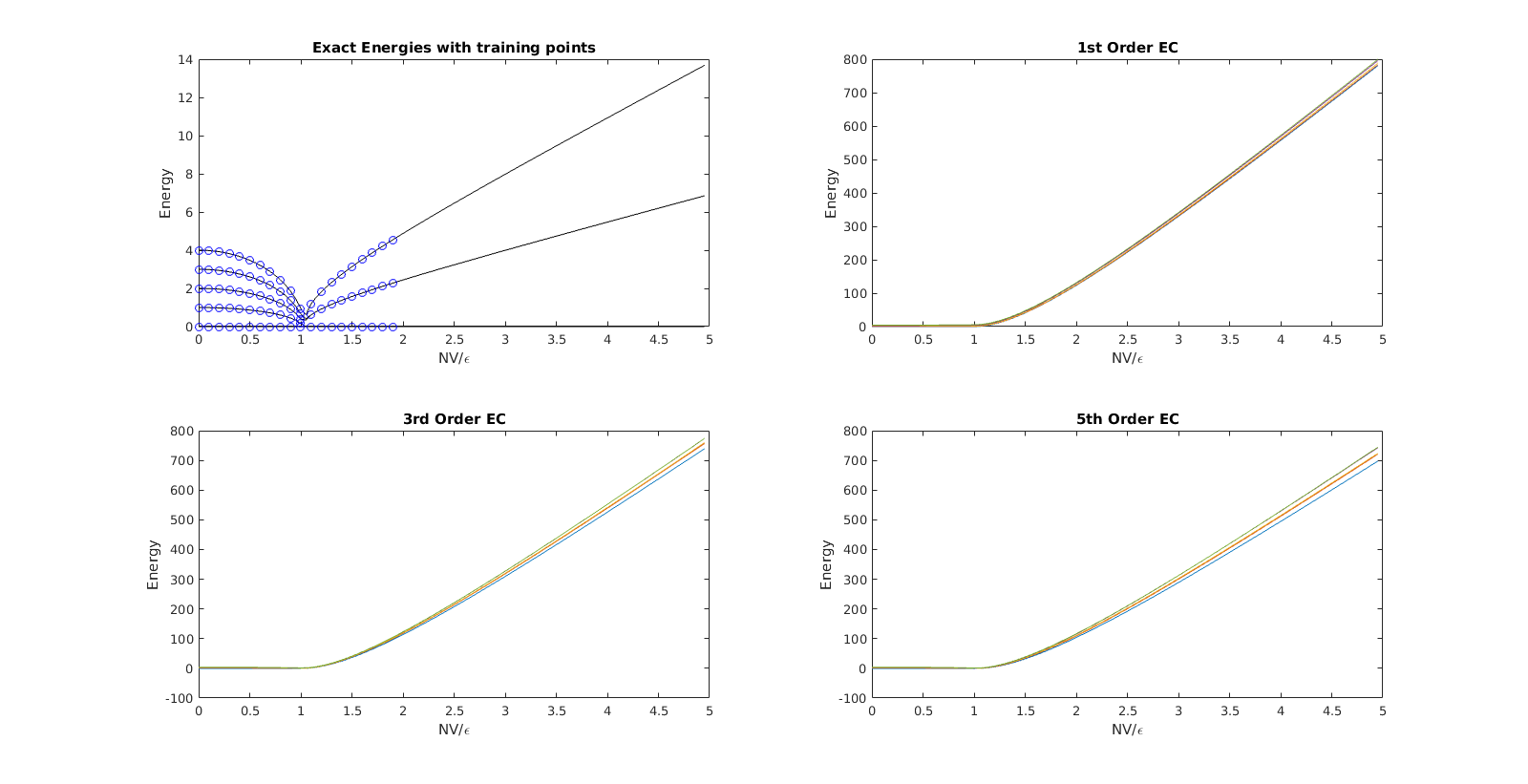}}
\caption[1st, 3rd, and 5th order EC estimates for N=1000 Lipkin model]{EC estimates for the lowest 5 eigenvalues of the N=1000 Lipkin model. We show the energy with the lowest ground state eigenvalue subtracted out, to show the kink present in the spectrum. The EC results diverge rapidly past the branch point at $C = 1$.}
\end{center}
\end{figure}

\begin{figure}	% FILLER PLOT
\begin{center}
\makebox[\textwidth][c]{\includegraphics[width=\textwidth]{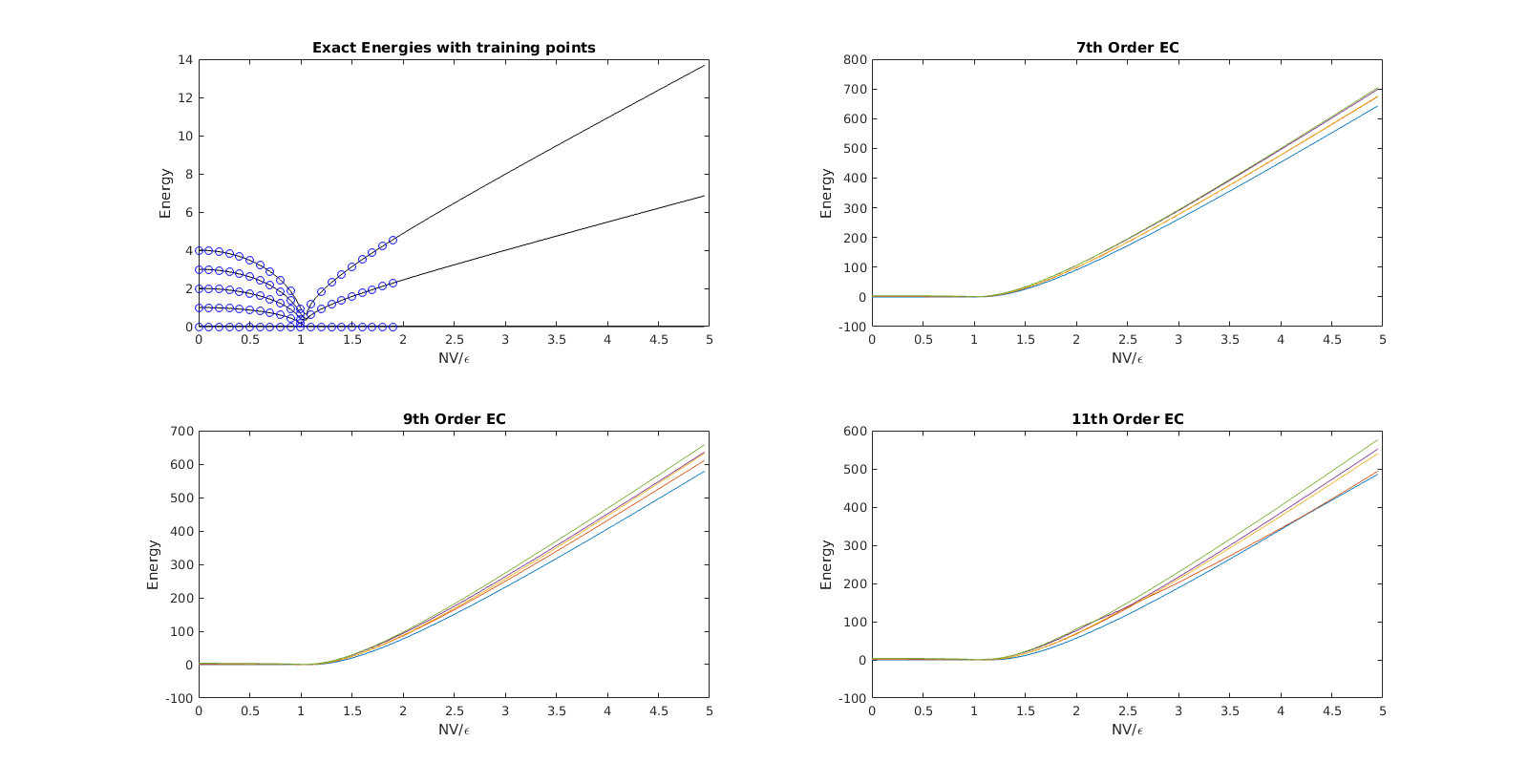}}
\caption[7th, 9th, and 11th order EC estimates for N=1000 Lipkin model]{EC estimates for the lowest 5 eigenvalues of the N=1000 Lipkin model. We show the energy with the lowest ground state eigenvalue subtracted out, to show the kink present in the spectrum. The EC results diverge rapidly past the branch point at $C = 1$.}
\end{center}
\end{figure}

\begin{figure}	% FILLER PLOT
\begin{center}
\makebox[\textwidth][c]{\includegraphics[width=\textwidth]{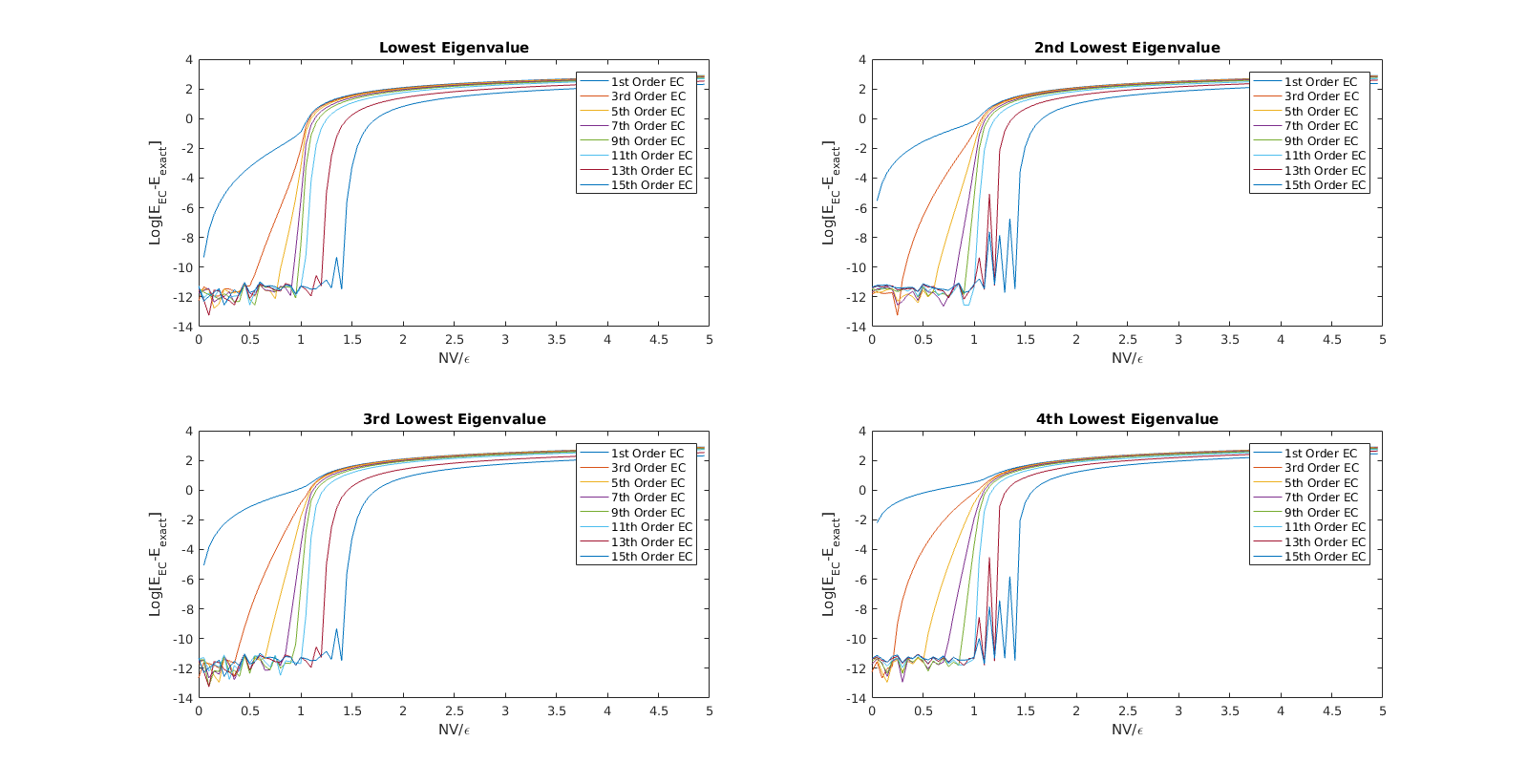}}
\caption[Residuals for the first 4 eigenvalues of the N=1000 Lipkin model]{Residuals for the first 4 eigenvalues for the N=1000 Lipkin model. The first 15 EC orders are shown for each eigenvalue.}
\end{center}
\end{figure}

We note that the EC method is not able to extrapolate past this kink in the spectrum. While the order-by-order estimates for the eigenvalues before that point get better and better, they all go bad by the time they reach the branch point.

Turning on the W term, we can move the location of the kink to around $C = 3$. Like before, we consider up to 15th order EC, showing the results of the first five energy levels for odd EC orders in Figs. [4.33] and [4.34], with the corresponding residuals in Figs. [4.35].

\begin{figure}	% FILLER PLOT
\begin{center}
\makebox[\textwidth][c]{\includegraphics[width=\textwidth]{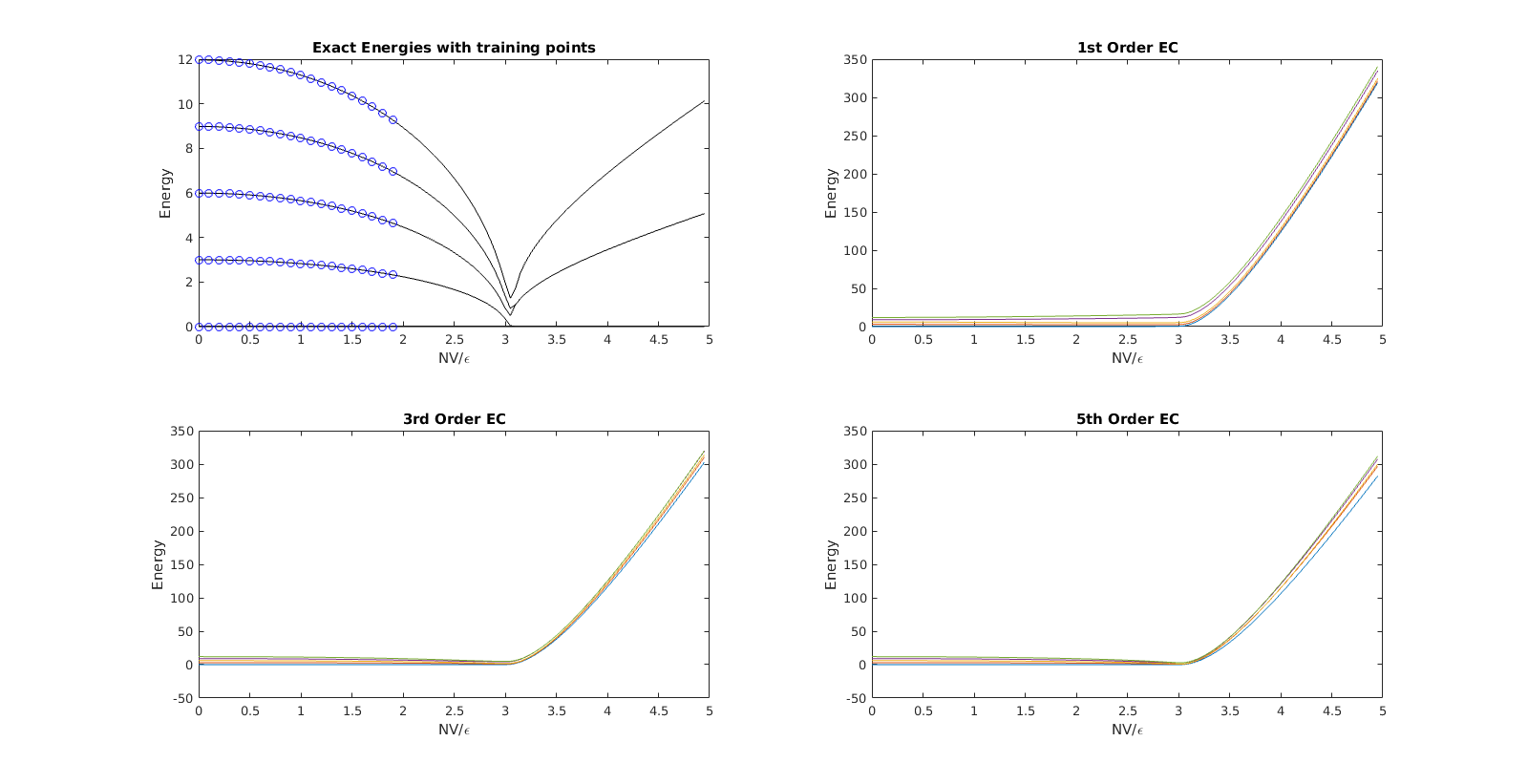}}
\caption[1st, 3rd, and 5th order EC estimates for N=1000 Lipkin model]{EC estimates for the lowest 5 eigenvalues of the N=1000 Lipkin model, with non-zero W. We show the energy with the lowest ground state eigenvalue subtracted out, to show the kink present in the spectrum. The EC results diverge rapidly past the branch point at $C = 3$.}
\end{center}
\end{figure}

\begin{figure}	% FILLER PLOT
\begin{center}
\makebox[\textwidth][c]{\includegraphics[width=\textwidth]{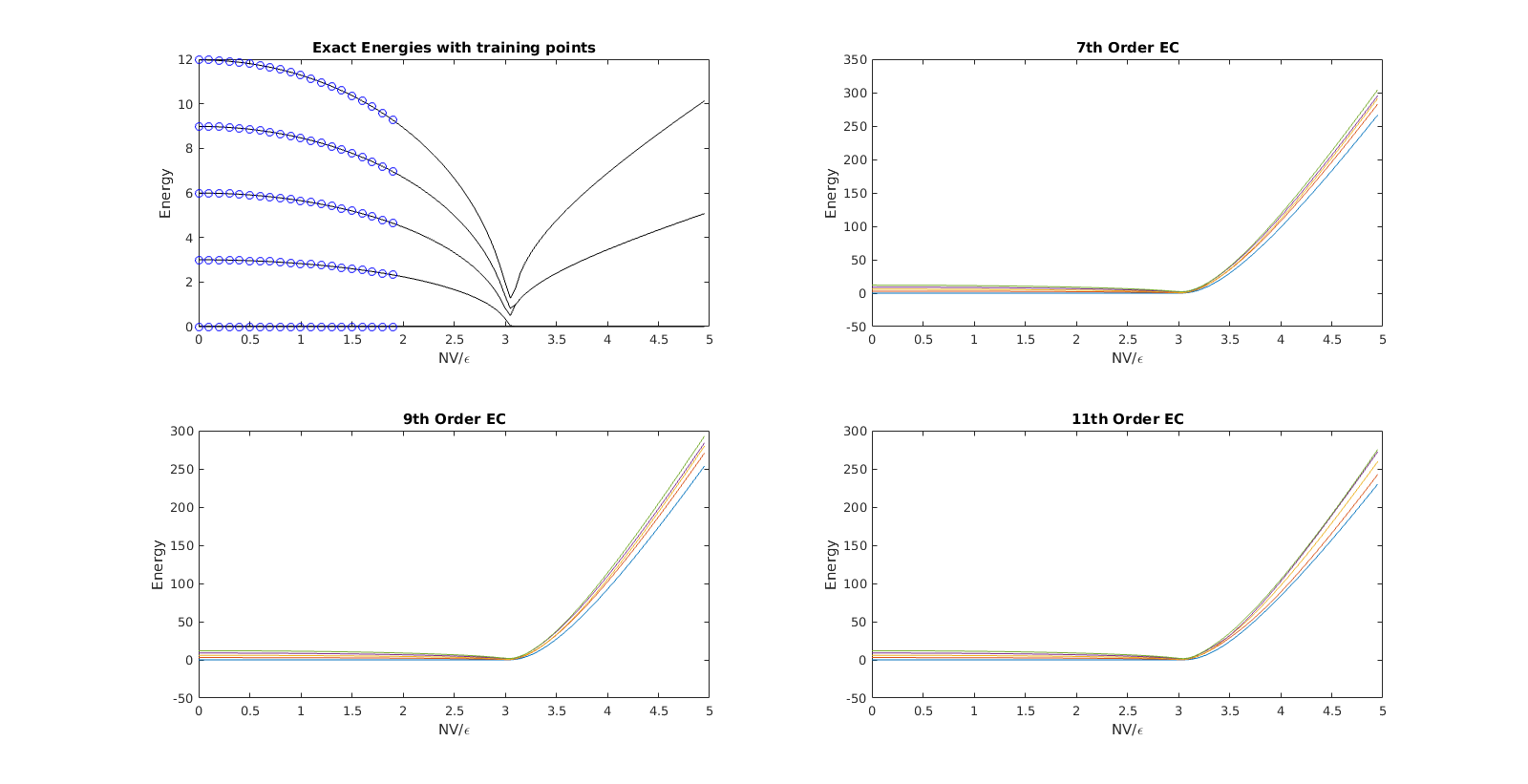}}
\caption[7th, 9th, and 11th order EC estimates for N=1000 Lipkin model]{EC estimates for the lowest 5 eigenvalues of the N=1000 Lipkin model, with non-zero W. We show the energy with the lowest ground state eigenvalue subtracted out, to show the kink present in the spectrum. The EC results diverge rapidly past the branch point at $C = 3$.}
\end{center}
\end{figure}

\begin{figure}	% FILLER PLOT
\begin{center}
\makebox[\textwidth][c]{\includegraphics[width=\textwidth]{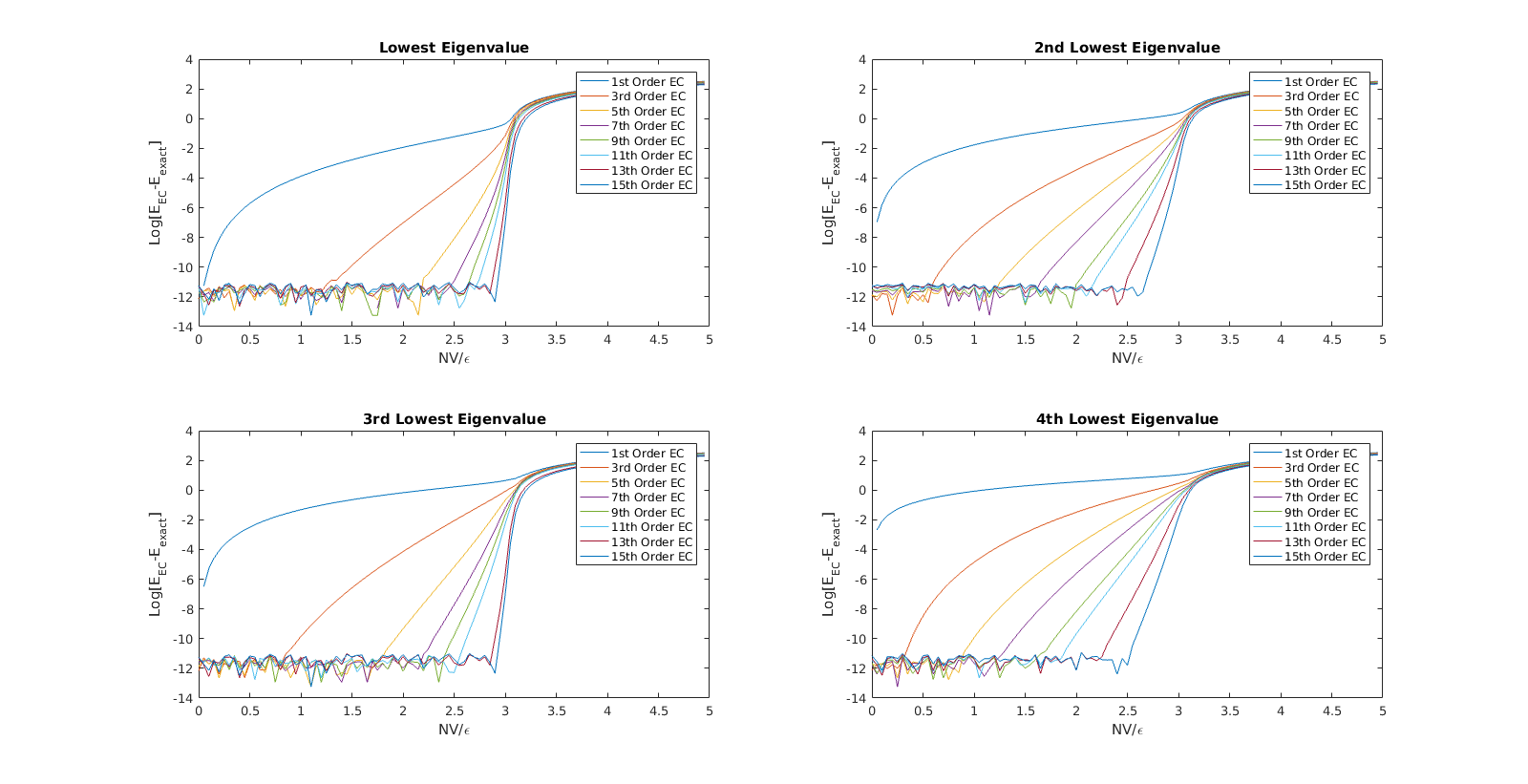}}
\caption[Residuals for the first 4 eigenvalues of the N=1000 Lipkin model, with non-zero W]{Residuals for the first 4 eigenvalues for the N=1000 Lipkin model, with non-zero W. The first 15 EC orders are shown for each eigenvalue.}
\end{center}
\end{figure}

%%%%%%%%%%%%%%%%%%%%%%%%%%%%%%%%%%%%%%%%%%%%%%%%%%%%%%%%%%%%%%%%%%
%%%%%%%%%%%%%%%%%%%%%%%%%%%%%%%%%%%%%%%%%%%%%%%%%%%%%%%%%%%%%%%%%%
%%%%%%%%%%%%%%%%%%%%%%%%%%%%%%%%%%%%%%%%%%%%%%%%%%%%%%%%%%%%%%%%%%
\chapter{Concluding Remarks}

In this work, we have demonstrated a new technique for evaluating the extremal eigenvalues and eigenvectors at points inaccessible to direct calculations. This technique relies on forming a low-dimensional space out of the eigenvectors in some coupling region, and then projecting the Hamiltonian at some target coupling into this space. Using analytic function theory, we showed that this projected Hamiltonian can be used to determine the analytically continued eigenvalues and eigenvectors at the target coupling.  

As an application of this method, we considered four lattice systems. First, we considered the Bose-Hubbard model, which consists of $N$ bosons on a lattice, with a Hamiltonian consisting of a hopping term, and pair-wise interaction, and a chemical potential. This system has a phase transition which cannot be captured easily in perturbation theory, when the system goes from a weakly bound Bose gas, to a strongly bound cluster state. Using eigenvector continuation, we showed that this phase transition could be captured. The second system studied was pure neutron matter, done at leading order in Chiral EFT for a system of six and fourteen neutrons. In direct calculations, the sign problem prevented accurate results from being obtained, but by the use of eigenvector continuation,  we could obtain accurate, large projection time estimates for the energies. Next, we looked at the non-perturbative inclusion of the Coulomb interaction into our lattice calculations. Since the Coulomb interaction is repulsive, it has a sign problem for large systems. We showed once again that the direct calculations yielded poor results, and that the use of the eigenvector continuation technique could dramatically help the determinations of the energy. Finally, we looked at the Lipkin model, which consists of N fermions distributed into two N-fold degenerate angular momentum levels. This levels are coupled, and allowed to mix, giving interesting phase transitions. Like in the Bose-Hubbard model, we showed that perturbation theory failed to capture this phase transition, and that the eigenvector continuation method did a good job at reproducing the energy levels.

In the future, we have two primary avenues to continue this work. One is on the mathematical side, looking at proving the convergence of this method, or providing error bounds on the estimates of the eigenvalues. The other direction this work could continue is in the applications to other systems; ones with more impact, like in QCD or in scattering calculations. The eigenvector continuation technique is very basic, so the number of research areas where it could be applied are quite broad. We would like to continue the search for applications, as we continue the work on the mathematical foundations of the method.

\nocite{*}

\end{doublespace}

%%%%%%  Bibliography %%%%%
%% A bibliography is required. By default it is called, "Bibliography"
%% You may use ÒLITERATURE CITEDÓ, ÒWORKS CITEDÓ or ÒREFERENCESÓ 
%% instead of ÒBIBLIOGRAPHYÓ if that is the convention in your discipline. 
%% To do so, copy and paste your choice into the empty argument 
%% of the following command and remove the "%".
%\renewcommand{\bibname}{}

%% The bibliography may be made using BibTeX.
%% To do so the necessary commands must be entered in the 
%% preamble and here.
%% If the Bibliography is made from scratch,
%% remove the "%" in front of \begin{thebibliography}{???}
%% replacing the ??? with the appropriate entry and 
%% remove the "%" in front of \end{thebibliography} below.
%\begin{thebibliography}{dkframe_dissertation}
\bibliography{dkframe_dissertation}
%\end{thebibliography}
%% In either case, the bibliography is automatically entered
%% in the Table of Contents.
\end{document}